\documentclass[12pt]{article}

\usepackage[a4paper,margin=1in]{geometry}
\usepackage{setspace}
\usepackage{parskip}

\usepackage{newtxtext,newtxmath}
\usepackage[T1]{fontenc}
\usepackage[utf8]{inputenc}
\usepackage{microtype}

\usepackage{amsmath,amsthm}
\usepackage{booktabs}
\usepackage{tabularx, array, multirow}
\usepackage{adjustbox}
\usepackage[table]{xcolor}
\usepackage{bm}

\usepackage{float}
\usepackage{placeins}
\usepackage{graphicx}
\usepackage{tikz}
\usetikzlibrary{calc,positioning}
\usepackage{caption}
\usepackage{subcaption}
\graphicspath{{images/}}

\usepackage[english]{babel}
\usepackage{csquotes}
\usepackage{cite}

\usepackage{longtable,enumitem,listings}
\newcolumntype{L}[1]{>{\raggedright\arraybackslash}p{#1}}
\newcolumntype{Y}{>{\raggedright\arraybackslash}X}
\lstdefinestyle{promptjson}{basicstyle=\ttfamily\small,breaklines=true,breakatwhitespace=false,columns=fullflexible,keepspaces=true,showstringspaces=false,frame=none,aboveskip=0.4em,belowskip=0.4em}
\newcommand{\parencite}[1]{\cite{#1}}
 
\usepackage{titling}
\pretitle{\centering\Large\bfseries}
\posttitle{\par\vskip 0.5em}
\preauthor{\centering\small}
\postauthor{\par}
\predate{\centering\normalsize}
\postdate{\par}

\usepackage{titlesec}
\titleformat{\section}[block]{\large\bfseries}{\thesection}{0.7em}{}
\titleformat{\subsection}[block]{\large\bfseries}{\thesubsection}{1em}{}
\titleformat{\subsubsection}[block]{\normalsize\bfseries\itshape}{\thesubsubsection}{1em}{}       

\newcommand{\keywords}[1]{%
  \par\noindent\textbf{Keywords: } #1
}

\usepackage[
  colorlinks=true,
  linkcolor=black,
  citecolor=black,
  urlcolor=black
]{hyperref}

\newcommand{\DataRepositoryLink}{%
  \href{https://1drv.ms/f/c/d282f76e12dba942/IgAYXRao6IG2RYrE4FPltqSgASFc2-Lh6UMGv--3AIMIamM?e=RlC24i}{Anonymous Data Repository}}%

\newcommand{\CodeArchiveLink}{%
  \href{https://anonymous.4open.science/r/Social_Science_of_LLMs-D28B/data/}%
  {Anonymous Code Repository}%
}

\usepackage{fancyhdr}
\title{Mapping the Emerging Social Science of Large Language Models}

\author{Zhanzhan Zhao\textsuperscript{1,2,\textdagger,*}, Yi Yang\textsuperscript{2,\textdagger}, Xiao Jia\textsuperscript{2,\textdagger}, Zeyun Dong\textsuperscript{3}, and Chenzhang Wang\textsuperscript{4}}
\date{}

\begin{document}
\maketitle
\begin{center}\small
\textsuperscript{1}School of Humanities and Social Science, The Chinese University of Hong Kong, Shenzhen\\
\textsuperscript{2}School of Artificial Intelligence, The Chinese University of Hong Kong, Shenzhen\\
\textsuperscript{3}School of Computer Science and Technology, Xidian University\\
\textsuperscript{4}School of Mathematics, The University of Edinburgh\\[4pt]
\textsuperscript{\textdagger}Zhanzhan Zhao, Yi Yang, and Xiao Jia contributed equally to this work.\\
\textsuperscript{*}Corresponding author: Zhanzhan Zhao (\href{mailto:zhanzhanzhao@cuhk.edu.cn}{zhanzhanzhao@cuhk.edu.cn})
\end{center}
\noindent\textbf{Article type:} Review\\
\textbf{Short title:} Social science of large language models

\begin{abstract}
As large language models (LLMs) blur the boundary between cognition and computation, they reshape knowledge production and collective life. We combine a curated review of 198 papers with computational mapping of 47,719 formally published papers from five databases. Text embeddings, clustering, and topic modeling organize the curated literature into 13 subcategories across three domains. LLM as Social Minds examines mind-like reasoning and socially grounded behavior in individual models. LLM Societies explores how interactions among agents produce collective behavior and social organization. LLM--Human Interactions examines how people perceive and engage with LLMs, and how interaction reshapes human thinking, creativity, and relationships. Structural topic modeling of the larger corpus traces how research attention shifts over time and differs across publication venues. We present a reproducible framework that identifies the field's underlying structure and enables its continuous refinement as new studies emerge. The review also identifies a gap in current understanding: the mechanisms underlying many of LLMs' observed cognitive and social behaviors remain unclear. Human-like performance can then be mistaken for understanding, obscuring the norms, biases, and structures of power that LLMs reproduce. We call for shared benchmarks and robustness checks across the three domains to build a reliable evidence base for understanding LLMs' social behaviors and informing their design and use.
\end{abstract}

\keywords{large language models; social science; taxonomy; literature review; human--AI interaction; multi-agent systems; trust; bias; institutions; mind attribution}

\section{Introduction}

In the seminal work \textit{The Sciences of the Artificial}, Herbert Simon \cite{simon2019sciences} advanced the view that artifacts differ fundamentally from both natural phenomena and abstract formal systems. They are purpose-driven constructions, designed to meet human ends, yet their meaning and behavior are defined by the environments in which they are embedded. Artifacts thus occupy a liminal ontological space: their functions cannot be fully understood apart from the social, cognitive, and institutional contexts that give them significance. Large language models (LLMs) are paradigmatic examples of such artifacts. Conceived as computational instruments for text prediction, they have rapidly become general-purpose infrastructures that not only execute linguistic tasks but also manifest patterns of interaction and interpretation that invite comparison with psychological, social, and institutional processes \parencite{mei2024turing,park2023generative}. To conceptualize LLMs merely as technical systems is therefore insufficient; they should be studied as social artifacts, whose significance emerges precisely at the intersection of engineering design and social embedding \parencite{bail2024can, peter2025benefits}. Accordingly, we use the term social science of LLMs to denote systematic inquiry that takes LLMs themselves as social objects of analysis, with attention to the behavioral, interactive, and institutional patterns they generate \parencite{xu2024ai}.

This perspective situates the social science of LLMs within a longer intellectual lineage. In sociology and psychology, artifacts have long been treated as windows into cognition and culture: from Vygotsky’s account of tools as mediators of thought, to Latour’s actor--network theory in which technologies shape human associations, to recent cognitive science accounts of distributed cognition \parencite{hutchins1995cognition, latour2005reassembling, vygotsky1978mind}. LLMs extend this tradition in unprecedented ways. They provoke questions about mind-like attributions when their outputs mimic intentional reasoning \parencite{kosinski2024evaluating, strachan2024testing, lehr2025kernels, hu2025generative}. They generate novel social dynamics when multiple models interact, negotiate, or coordinate \parencite{li2023camel, park2023generative, akata2025playing}. And they are transforming human practices by altering productivity, learning, governance, and communication \parencite{chen2024large, noy2023experimental}. These three levels provide complementary perspectives on the field, but the distribution of research across them, their overlaps, and their development over time remain less well characterized. In response, this study combines substantive synthesis with reproducible computational mapping to clarify how the literature is organized and changing.

Existing syntheses of adjacent AI-and-society literatures include narrative reviews (broad, agenda-setting, e.g. Bommasani and colleagues \cite{bommasani2021opportunities}; Xu and colleagues \cite{xu2024ai}; Bail \cite{bail2024can}), domain-specific reviews (e.g. education, Yan and colleagues \cite{yan2024practical}; law, Lai and colleagues \cite{lai2024large}; medicine, Haltaufderheide and Ranisch \cite{haltaufderheide2024ethics}), and task-focused syntheses (e.g. persuasion, Rogiers and colleagues \cite{rogiers2024persuasion}; misinformation, Kuntur and colleagues \cite{kuntur2024under}; intelligent agents, Xi and colleagues \cite{xi2025rise}; Wang and colleagues \cite{wang2024survey}). Recent related work includes Liu's examination of conceptual and methodological tensions in the sociology of AI \cite{liu2026sociology} and Chen and colleagues' synthesis of AI agent behavior across individual, multi-agent, and human--agent settings \cite{chen2026agentbehavior}. Our three-domain taxonomy is developed through computational clustering, topic interpretation, and full-text synthesis of the 198-paper corpus. It organizes the substantive review of LLMs as social artifacts; mapping the wider literature then examines how research across these domains is distributed and changing. This combination makes the proposed organization explicit and open to reassessment as the field develops~\parencite{fasce2023taxonomy, reimers2019sentence}.

We address three questions: What themes organize the social science of LLMs? How consistently do computational analyses and author classifications support these categories, and where do they overlap? How does research attention vary across publication periods and disciplinary venues? Study 1 combines a full-text review of 198 curated papers with computational clustering and topic modeling. Authors interpret the resulting themes and representative papers to name the domains, then compare this organization with independent author and LLM classifications. These comparisons identify both shared structure and boundary-spanning work. Study 2 extends the mapping to 47,719 formally published papers from five databases, using structural topic modeling to examine thematic patterns over time and across publication venues. Methods and Supplementary Materials describe the analyses and validation procedures.

Our analysis supports a three-domain organization of the reviewed corpus, which we label \emph{LLM as Social Minds}, \emph{LLM Societies}, and \emph{LLM--Human Interactions}. The first domain synthesizes research investigating whether and under what conditions LLMs display mind-like and socially grounded behavior, including reasoning about others’ beliefs, making moral judgments, expressing preferences, and engaging in strategic communication. The second domain brings together work on multi-LLM interactions and emergent collective behaviors, including how role-play and memory-planning architectures give rise to social regularities, norms and institutions, cooperative behaviors, and collective epistemics like information spreading and polarization. The third domain examines how people perceive and trust LLMs, how interaction with large language models alters human thinking, creativity, and collaboration in educational and professional contexts, and how such transformations create both new opportunities and risks that demand adaptive governance. Stability analyses and classification comparisons support this organization while revealing overlap between the domains. Mapping the larger corpus in Study~2 shows that human--LLM interactions account for the largest share of mapped research content. Over time, attention shifts from broad discussion toward studies of human use and agent societies, while conference publications give greater relative emphasis to model behavior and agent societies than journals. By connecting these domains, the framework helps researchers examine how individual LLM behavior and collective agent dynamics shape one another, and how both influence and are shaped by human society.

\section{Results}
\label{sec:results}

Mapping analyses place the substantive review in the context of the broader literature. A curated corpus of 198 papers, reviewed in full, supports interpretation of the three domains and their subcategories. A second corpus of 47,719 formally published papers supports examination of the field's thematic composition, development over time, and variation across publication venues.

\subsection{Study 1: Computational Taxonomy via Unsupervised Clustering}

\subsubsection{Global Structure of the Corpus}

The curated corpus contains 115 journal articles, 76 conference papers, and seven preprints. Titles and abstracts were represented with MPNet sentence embeddings and compared across K-means solutions with $K=2$--9. Internal indices favored different resolutions, but the three-cluster solution combined the highest Calinski--Harabasz index with the strongest resampling stability. Across 100 draws containing 90\% of the corpus, its mean adjusted Rand index (ARI) was 0.952, compared with 0.721 for $K=2$ and 0.778 for $K=4$. It also remained consistent across random initializations (Fig.~\ref{fig:k-selection-stability}).

\subsubsection{Topic Modeling within Clusters}

The selected clusters contained 95, 52, and 51 papers. Within-cluster topic modeling identified four, four, and five topics, respectively, whose interpretation yielded \emph{LLM as Social Minds}, \emph{LLM Societies}, and \emph{LLM--Human Interactions}. These themes provide the 13 subcategories developed in the review. The cluster projection is presented in Fig.~\ref{fig:kmeans-pca-clusters}, and topic summaries are provided in Figs.~\ref{fig:cluster0}--\ref{fig:cluster2}.

The three-domain taxonomy is summarized in Fig.~\ref{fig:llm-taxonomy}. Sections~\ref{sec:social-minds}--\ref{sec:llm-human-interactions} develop the substantive synthesis within each domain.

\begin{figure}[!htb]
\centering
\begingroup
\setstretch{1}
\definecolor{mindblue}{HTML}{0072B2}
\definecolor{societygreen}{HTML}{009E73}
\definecolor{humanorange}{HTML}{D55E00}
\newcommand{\taxonomyitem}[2]{\vspace{5pt}{\fontsize{9}{10.4}\selectfont\bfseries #1}\par\vspace{1pt}{\fontsize{8.4}{9.6}\selectfont\color{black!75}#2}\par}
\begin{adjustbox}{max width=\linewidth,center}
\begin{tikzpicture}[
  every node/.style={font=\sffamily},
  taxonomy root/.style={align=center,text width=100mm,anchor=north,inner sep=0pt},
  taxonomy panel/.style={draw,rounded corners=3pt,line width=0.65pt,inner sep=5pt,anchor=north},
  taxonomy connector/.style={draw=black!30,line width=0.6pt}
]
\node[taxonomy root] (taxonomyroot) at (0,0) {
  {\fontsize{12}{14}\selectfont\bfseries Social Science of LLMs}\\[3pt]
  {\fontsize{9}{10.5}\selectfont\color{black!65}Three domains and 13 subcategories in the curated review}
};
\node[taxonomy panel,draw=mindblue!65,fill=mindblue!3] (minds) at (-54mm,-23mm) {
\begin{minipage}[t][94mm][t]{46mm}
\raggedright\setlength{\parskip}{0pt}
{\fontsize{10}{11.5}\selectfont\bfseries\color{mindblue}LLM as Social Minds}\par\vspace{3pt}
{\fontsize{8.5}{10}\selectfont Individual model behavior}\par\vspace{4pt}
{\color{mindblue!35}\hrule height 0.5pt}\vspace{2pt}
\taxonomyitem{Human-Like Reasoning and Theory of Mind}{Reasoning about beliefs and intentions}
\taxonomyitem{Personality and Demographic Bias}{Personality profiles and social stereotypes}
\taxonomyitem{Political and Moral Judgment}{Values, moral evaluation, and cultural variation}
\taxonomyitem{Preference Alignment, Persuasion, and Deception}{Preference consistency and strategic influence}
\end{minipage}
};
\node[taxonomy panel,draw=societygreen!65,fill=societygreen!3] (societies) at (0mm,-23mm) {
\begin{minipage}[t][94mm][t]{46mm}
\raggedright\setlength{\parskip}{0pt}
{\fontsize{10}{11.5}\selectfont\bfseries\color{societygreen}LLM Societies}\par\vspace{3pt}
{\fontsize{8.5}{10}\selectfont Interacting model-based agents}\par\vspace{4pt}
{\color{societygreen!35}\hrule height 0.5pt}\vspace{2pt}
\taxonomyitem{Behavioral Games of LLM Agents}{Strategic choices and repeated interaction}
\taxonomyitem{Collective Intelligence and Emergent Dynamics}{Information aggregation, consensus, and polarization}
\taxonomyitem{Group Decision-Making of LLM Agents}{Leadership, norms, and institutional rules}
\taxonomyitem{Scalable LLM-Based Social and Network Simulation}{Populations, networks, and social dynamics}
\end{minipage}
};
\node[taxonomy panel,draw=humanorange!65,fill=humanorange!3] (interactions) at (54mm,-23mm) {
\begin{minipage}[t][94mm][t]{46mm}
\raggedright\setlength{\parskip}{0pt}
{\fontsize{10}{11.5}\selectfont\bfseries\color{humanorange}LLM--Human Interactions}\par\vspace{3pt}
{\fontsize{8.5}{10}\selectfont Human perceptions and practices}\par\vspace{4pt}
{\color{humanorange!35}\hrule height 0.5pt}\vspace{2pt}
\taxonomyitem{Perception and Trust}{Anthropomorphism, reliance, and adoption}
\taxonomyitem{Simulated Empathy and Support}{Emotional support and companionship}
\taxonomyitem{Generative Work and Collaboration}{Productivity, teamwork, and human agency}
\taxonomyitem{Creative Partnership}{Originality, authorship, and creative control}
\taxonomyitem{Education and Assessment}{Learning, AI literacy, and assessment}
\end{minipage}
};
\draw[taxonomy connector] (taxonomyroot.south) -- (0,-18mm);
\draw[taxonomy connector] (-54mm,-18mm) -- (54mm,-18mm);
\draw[taxonomy connector] (-54mm,-18mm) -- (minds.north);
\draw[taxonomy connector] (0,-18mm) -- (societies.north);
\draw[taxonomy connector] (54mm,-18mm) -- (interactions.north);
\end{tikzpicture}
\end{adjustbox}
\endgroup
\captionsetup{font=small}
\caption{\textbf{Three-domain taxonomy of the social science of LLMs.} The 13 subcategories summarize the full-text review and within-cluster topic interpretation of the 198-paper curated corpus. They describe the substantive scope of each domain; the domains also overlap. The complete subcategory reference inventory appears in Table~\ref{tab:corpus-reference-coverage}. These subcategories are distinct from the STM topics estimated in the larger corpus (Fig.~\ref{fig:stm-topics-domains}).}
\label{fig:llm-taxonomy}
\end{figure}

\subsubsection{Expert-Informed Taxonomy and Human--Machine Alignment}

Author classifications based on full texts corresponded with K-means domains for 77.78\% of papers. An LLM applying the same codebook to titles and abstracts corresponded with K-means for 75.25\% and with authors for 85.86\% (Table~\ref{tab:classification-correspondence}). Allowing an LLM-assigned secondary domain increased correspondence with K-means to 87.88\%. Secondary categories were assigned to 87 papers, most commonly connecting \emph{LLM as Social Minds} and \emph{LLM--Human Interactions}. Disagreements concentrated near semantic boundaries: ten low-consensus papers had lower classification correspondence, silhouette values, and centroid margins than the remaining papers. The domains thus organize the literature while accommodating substantive overlap.

\subsection{Study 2: Field-Scale Thematic Structure}

\begin{figure}[!htbp]
    \centering
    \includegraphics[width=\textwidth]{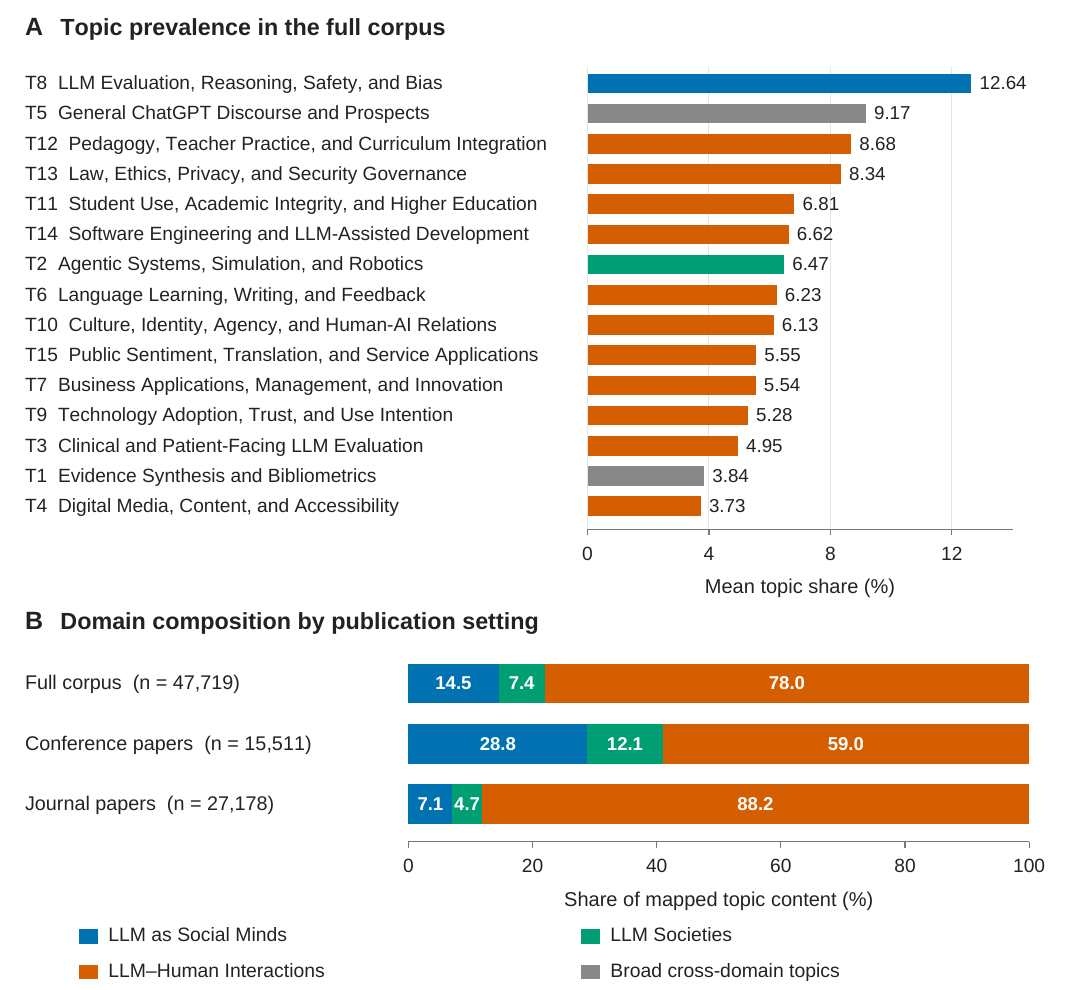}
    \caption{\textbf{Research topics and domain composition in the field-scale corpus.} (A) Mean expected prevalence of all 15 STM topics across 47,719 papers. Colors indicate domain correspondence; Topics~1 and 5 are broad cross-domain topics. (B) Domain composition of mapped topic content in the full corpus, conference and proceedings-series papers, and journal papers. Other venue types are included in the full corpus. Percentages in (B) are normalized over the 13 domain-mapped STM topics, excluding Topics~1 and 5; they describe topic content, not percentages of papers. These STM topics are distinct from the 13 curated-review subcategories in Fig.~\ref{fig:llm-taxonomy}. Citation- and venue-score subsets are shown in Fig.~\ref{fig:venue-subsets}.}
    \label{fig:stm-topics-domains}
\end{figure}

The field-scale corpus was assembled from five databases through bibliographic processing, substantive eligibility screening, and publication-venue classification. Structural topic models with 10--30 topics were compared using coherence and exclusivity; the combined criterion favored 15 topics. The most prevalent were \emph{LLM Evaluation, Reasoning, Safety, and Bias} (12.64\%), \emph{General ChatGPT Discourse and Prospects} (9.17\%), \emph{Pedagogy, Teacher Practice, and Curriculum Integration} (8.68\%), and \emph{Law, Ethics, Privacy, and Security Governance} (8.34\%; Fig.~\ref{fig:stm-topics-domains}A).

Thirteen of the 15 STM topics mapped onto the three domains. These topics were estimated in the larger corpus and are distinct from the 13 subcategories of the curated review: one mapped to \emph{LLM as Social Minds}, one to \emph{LLM Societies}, and eleven to \emph{LLM--Human Interactions}. Two broad topics, concerning evidence synthesis and general ChatGPT discourse, remained cross-domain. Among the mapped topics, \emph{LLM--Human Interactions} represented 78.02\%, \emph{LLM as Social Minds} 14.53\%, and \emph{LLM Societies} 7.44\%. These are shares of modeled topic content, normalized over the 13 mapped topics, rather than mutually exclusive percentages of papers. Human--LLM research extended across education, healthcare, business, law, media, software development, technology adoption, and cultural relations.

\begin{figure}[!htbp]
    \centering
    \includegraphics[width=\textwidth]{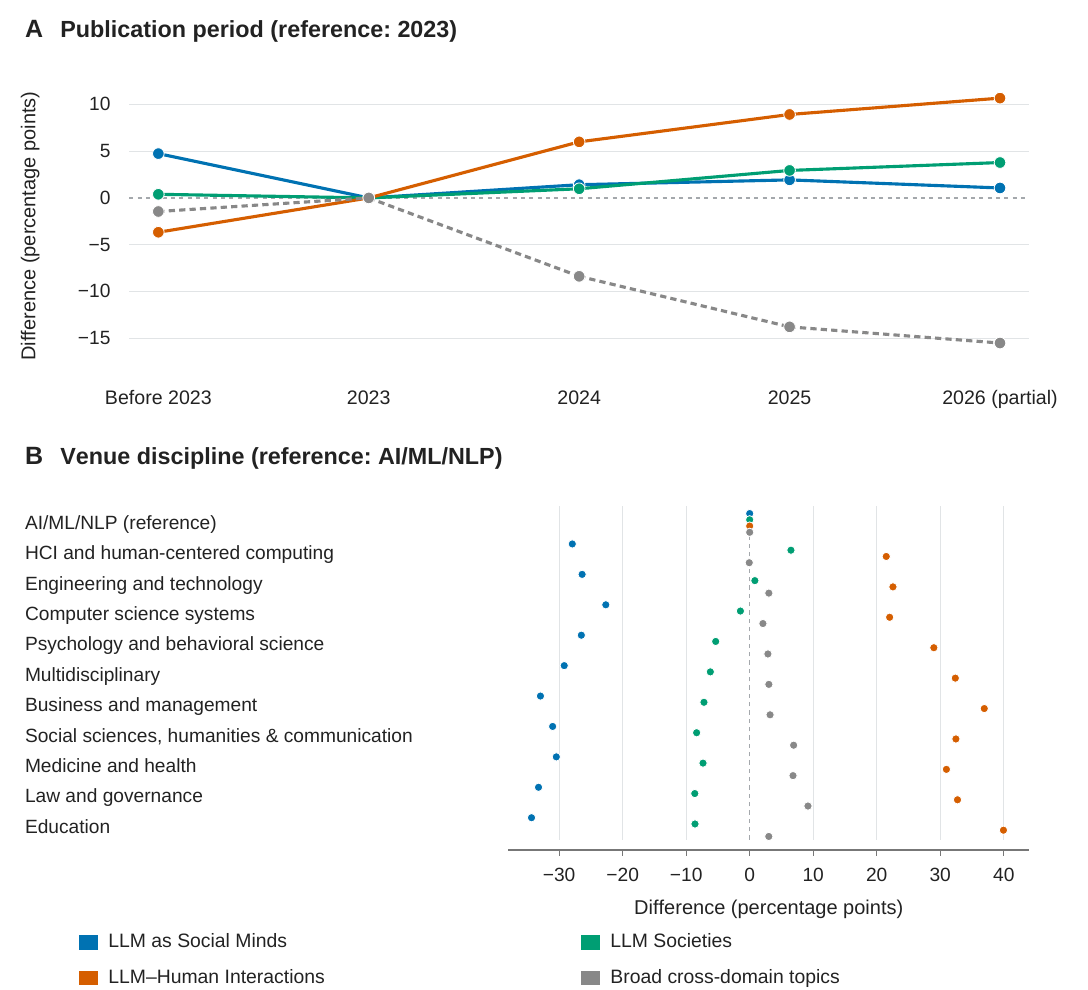}
    \caption{\textbf{Temporal and disciplinary differences in research attention.} (A) Model-adjusted differences in expected STM topic content relative to 2023, holding venue discipline constant. The pre-2023 category combines 2020--2022; 2026 includes publications through 23 July only. (B) Differences across venue disciplines relative to AI/ML/NLP venues, holding publication period constant. Points show coefficient estimates, with domain estimates obtained by summing the coefficients of their mapped topics. The two broad cross-domain topics are shown separately. Positive and negative values indicate higher and lower expected topic content than the reference category. Lines in (A) connect point estimates; uncertainty intervals are not displayed.}
    \label{fig:stm-temporal-venue-patterns}
\end{figure}

The corpus expanded from 4,128 papers in 2023 to 11,133 in 2024 and 18,190 in 2025. The 14,091 papers published through 23 July 2026 cover only part of that year. Accounting for venue discipline, the expected topic content associated with \emph{LLM--Human Interactions} increased relative to 2023 by 5.98 percentage points in 2024, 8.90 in 2025, and 10.64 in partial 2026. Broad cross-domain discussion decreased over the same periods. \emph{LLM Societies} increased by 0.98, 2.93, and 3.77 points (all $p<0.001$), while \emph{LLM as Social Minds} remained modestly above its 2023 level. The pattern suggests increasing specialization in how LLMs are used, experienced, and governed, alongside growth in multi-agent research (Fig.~\ref{fig:stm-temporal-venue-patterns}A).

Venue patterns also distinguished the domains (Fig.~\ref{fig:stm-temporal-venue-patterns}B). After accounting for publication period, \emph{LLM as Social Minds} was concentrated in artificial intelligence, machine learning, and natural language processing (AI/ML/NLP) venues, with expected topic content 22.65--34.35 percentage points lower in other venue categories. \emph{LLM Societies} was 6.50 points higher in human--computer interaction (HCI) and human-centered computing venues than in AI/ML/NLP venues ($p<0.001$) and 0.81 points higher in engineering and technology venues ($p=0.017$), but lower elsewhere.

The overall distribution also varied between journals and conferences (Fig.~\ref{fig:stm-topics-domains}B). \emph{LLM as Social Minds} and \emph{LLM Societies} together represented 40.95\% of mapped content among 15,511 conference and proceedings-series papers, compared with 11.83\% among 27,178 journal papers. Human--LLM interactions remained the largest domain in both groups. These descriptive contrasts show how the picture of the field varies across publication settings. Detailed citation- and venue-score comparisons are provided in Fig.~\ref{fig:venue-subsets}; the venue scores are LLM estimates, not official JIF or CORE rankings.

\subsubsection{Sensitivity and coverage}

Full-text reassessment confirmed eligibility for 288 of 300 sampled records (96\%); ten were ineligible and two indeterminate. The field-scale corpus recovered 162 of the 198 curated papers (81.82\%), with similar recovery across domains. Repeating the curated analysis for the 148 recovered papers at or above the 50th citation percentile retained the three-domain interpretation and yielded 79.05\% correspondence with author classifications. Full screening, stability, classification, and coverage results are provided in the Supplementary Materials.

\FloatBarrier
\section{LLM as Social Minds}
\label{sec:social-minds}
\label{sec:taxonomy}
At the model level, \emph{LLM as Social Minds} examines response patterns that resemble human reasoning, personality, values, and strategic behavior. The term refers to observable, socially meaningful patterns rather than evidence that LLMs possess human-like internal states. The following four subcategories examine these patterns and their dependence on tasks and contexts.

\paragraph{Human-Like Reasoning and Theory of Mind}
A core thread asks whether LLMs reason in ways that resemble humans and whether such resemblance remains stable across tasks and contexts. Studies report human-like performance on reasoning and social-cognitive tasks \parencite{ref4,ref9,ref45,ref46,ref50}, but broader batteries and confidence-calibration tests reveal uneven performance \parencite{ref13,ref42}. Success in one task may therefore not reflect a general or stable reasoning capacity. This tension between behavioral resemblance and contextual instability raises the question of whether models reason like humans or produce human-like responses. Theory of mind (ToM)—the capacity to infer others’ mental states—provides a particularly contested case \parencite{premack1978does,apperly2010mindreaders}. Although some studies report human-comparable ToM performance \parencite{kosinski2023theory,kosinski2024evaluating,ref23,ref7}, adversarial tests and prompt perturbations reveal brittle reliance on cues and heuristics \parencite{strachan2024testing,ullman2023large,shapira2023clever,zhao2025llms,amirizaniani2025mind,ref44,ref96}. Prompted deliberation can improve performance, but may provide external scaffolding rather than reveal underlying cognition \parencite{binz2023using}. Accordingly, human-like behavior should be evaluated across controlled and perturbed task batteries, rather than treated as evidence of human-like underlying cognition \parencite{marchetti2025artificial,ref39,ref49}.

\paragraph{Personality and Demographic Bias}
A second thread examines how LLMs express personality and reproduce demographic bias. Personality studies show that models can emulate targeted profiles, but measured traits vary across instruments, models, elicitation procedures, demographic prompts, and ideological manipulations, sometimes producing unrealistically coherent or unstable profiles \parencite{ref1,ref10,ref60,ref61,ref62,ref65}. Apparent personality is therefore better understood as a context-dependent behavioral pattern than as a stable model disposition. Bias audits similarly identify systematic differences associated with names, occupations, religion, gender, race, ethnicity, and their intersections, alongside broader political and cultural orientations \parencite{kotek2023gender,lu2025cultural,abid2021persistent,siddique2024better,hu2025generative,ref14,ref24,ref28,ref47,ref54,ref57,ref58,ref181,ref183}. Such patterns can persist after tuning, appear under nominally unbiased training, and carry into simulated populations and consequential assessments \parencite{bai2025explicitly,mahesh2024investigating,ref2,ref127,ref190,ref192}. Overall, social bias in LLMs is structured and persistent, but its measured form and magnitude depend strongly on prompts, models, groups, and domains. Reliable evaluation therefore requires repeated sampling, counterfactual tests, intersectional analysis, and explicit checks of reliability, construct validity, and downstream harm \parencite{lee2024large,gallegos2024bias,radaideh2025fairness,torres2024comprehensive,an2025measuring,ref103}.

\paragraph{Political and Moral Judgment}
A third line of work examines political and moral judgment as a property of LLM response behavior. Political audits identify stable ideological tendencies in model responses, but comparative studies of ChatGPT-4, Perplexity, Gemini, and Claude show that their orientations differ substantially across systems \parencite{ref118,rozado2024political}. Moral evaluations similarly find that model judgments vary across model sizes, prompting strategies, moral domains, and the framing of specific dilemmas \parencite{ref31,ref55}. These findings do not imply that LLMs literally hold political beliefs or moral values. Rather, they reveal systematic and socially consequential response profiles whose interpretation depends on the model, evaluation instrument, and context. Political and moral judgment should therefore be assessed through comparative and repeated audits rather than general claims that LLMs are inherently neutral or biased.

\paragraph{Preference Alignment, Persuasion, and Deception}
A fourth line of work examines preference alignment, persuasion, and deception as related capacities through which LLMs adapt their responses to prior commitments, interlocutor characteristics, and strategic incentives. Under controlled conditions, LLMs sometimes modify their answers to remain consistent with previously expressed positions, exhibiting preference-consistent rationalization \parencite{lehr2025kernels,simmons2022moral}. However, preference-following can deteriorate over long conversations, and matching an expressed preference does not necessarily imply robust adherence to its underlying value \parencite{ref112,ref125}. Models can also personalize responses to individual or cultural preferences and tailor arguments to particular interlocutors, sometimes matching or outperforming humans in persuasive exchanges \parencite{ref124,ref142,ref146}. The same adaptive and strategic capacities can support instrumental deception: under role-based incentives and strategic framing, LLMs may generate misleading statements to achieve a goal rather than merely producing falsehoods by accident \parencite{hagendorff2024deception}. Together, this literature characterizes alignment, persuasion, and deception as interconnected mind-like capacities involving preference representation, consistency, audience modeling, and strategic adaptation.

\section{LLM Societies}
\label{sec:llm-societies}
\emph{LLM Societies} shifts attention from individual models to populations of interacting agents whose aggregate behavior cannot be explained by individual capabilities alone. Multi-agent simulations are therefore not merely tools for reproducing human societies; their artificial social dynamics are themselves objects of inquiry. This perspective examines how memory, identity, network structure, incentives, and institutional rules shape collective action, norms, information diffusion, and polarization \parencite{li2023camel,park2023generative}. Simulations also provide controlled environments for testing theories of collective behavior and institutional design. Claims about human societies nevertheless require external validation because of simplified identities, model-specific assumptions, and uneven replication \parencite{argyle2023out,cui2024can,wang2024large}.

\paragraph{Behavioral Games of LLM Agents}
Behavioral game experiments treat LLM agents as objects of study and examine how they interact in controlled strategic settings. The iterated Prisoner’s Dilemma—a classic economic game for studying cooperation under repeated interaction—provides a prominent example. When playing against one another, LLM agents often make more generous initial choices than human players but condition subsequent decisions on their partners’ behavior, including by reciprocating defection \parencite{fontana2025nicer}. Cross-game comparisons reveal an important distinction between collaboration and coordination: LLM agents may collaborate effectively when mutual cooperation is the clear objective, yet struggle to coordinate when success requires converging on one of several competing equilibria, as in Battle of the Sexes. Broader evaluations also reveal unstable preferences, inconsistent belief updating, and departures from optimal play \parencite{ref68,ref84}. Beyond dyadic games, a large-scale social dilemma involving more than 200 agents identifies both adaptive egoists, whose altruism increases in response to social messages, and altruistic optimizers, which consistently prioritize collective welfare at a personal cost \parencite{li2025emergence}. Population-level coordination games further show that shared conventions can emerge from local interactions and develop collective biases even when individual agents show no corresponding bias; committed minorities can also shift the convention adopted by the wider population \parencite{ashery2025emergent}. These findings demonstrate that individual and collective intelligence are not equivalent: individually capable or collaborative agents do not necessarily coordinate successfully, while their interactions can generate altruism, conventions, and collective biases that cannot be inferred from isolated behavior. Collective outcomes instead emerge from the interplay of incentives, memory, communication, group composition, and institutional structure. 

\paragraph{Collective Intelligence and Emergent Dynamics}
Moving from structured game settings to broader patterns of group interaction, research on collective intelligence and emergent social dynamics examines how interactions among LLM agents produce group-level patterns of reasoning, belief formation, and information flow. LLMs can expand access to knowledge and support distributed reasoning, but they can also concentrate influence, homogenize information, and amplify bias within groups \parencite{ref121}. Experiments with partisan LLM crowds show that deliberation can improve belief accuracy and produce a “wisdom of crowds” effect, particularly when agents retain diverse prior beliefs; however, outcomes remain sensitive to persona design, prompting strategies, and model adaptation \parencite{ref80}. Architectures that combine heterogeneous agents through iterative “mindstorms” likewise show how structured exchange can improve reasoning beyond isolated model outputs \parencite{ref86}. The same interaction processes can nevertheless generate collective bias or persistent polarization, especially when agents communicate mainly with similar others or receive uneven information \parencite{ref76,piao2025emergence}. Opinion-dynamics models and platform-scale simulation environments make it possible to trace how these conditions produce information cascades, convergence, or sustained division and to test interventions against misinformation and competing influence \parencite{ref195,lin2025simspark}. Collective intelligence is therefore not the sum of individual model capabilities but an evolving property of interaction, shaped by cognitive diversity, information distribution, deliberative structure, network connections, and patterns of influence.

\paragraph{Group Decision-Making of LLM Agents}
Whereas behavioral games examine agents’ strategic choices and information aggregation, research on collective decision-making of LLM agents investigates how those choices and information are aggregated into group outcomes. In repeated route-choice experiments, LLM agents learn from experience and gradually converge, yet the resulting traffic patterns remain less efficient and less equitable because agents poorly anticipate others’ decisions and lack effective group-level mechanisms \parencite{ref137}. The GovSim common-resource environment exposes a related governance problem: when agents prioritize immediate returns, they can collectively deplete a shared resource despite each acting coherently from its own perspective \parencite{ref81}. Institutional designs attempt to address such failures. CRSEC allows agents to create, diffuse, evaluate, and enforce shared norms, reducing conflict within simulated communities \parencite{ref78}. Explicit team roles, designated or elected leadership, and iterative Criticize–Reflect procedures can likewise reduce redundant communication and improve group efficiency \parencite{ref85}. Social-contract and group-process simulations further demonstrate how rules, sanctions, and role structures shape collective order \parencite{ref79,ref8}. The central question is therefore not simply whether LLM agents are willing to cooperate, but how decision rules, leadership, norms, and enforcement transform local choices into stable collective organization.

\paragraph{Scalable LLM-Based Social and Network Simulation}
LLM-based social simulation extends controlled studies of interaction to larger populations and networks (Table~\ref{tab:simulation-platforms}). Interview-based generative agents simulate 1,052 real individuals, with evaluation against those individuals' survey responses and experimental behavior \parencite{park2024generative}. AgentSociety supports more than 10,000 agents and millions of interactions to study polarization, mobility, policy interventions, and external shocks \parencite{ref194}. Light Society takes opinion-diffusion simulation to a one-billion-agent network using models distilled from LLM responses \parencite{guan2025lightsociety}. MatrAIx addresses population diversity by assembling 8.3 billion persona records for simulated-user evaluation, with experiments conducted on sampled cohorts rather than an interacting world population \parencite{li2026matraix}. These platforms broaden the scope for comparing how information spreads and interventions affect populations with different compositions and network structures. Their value for social explanation depends on grounding behavioral assumptions in empirical evidence, comparing simulated outcomes with observed behavior, and testing robustness across models and settings \parencite{ref66,ref168,guan2025lightsociety,li2026matraix}.

\section{LLM--Human Interactions}
\label{sec:llm-human-interactions}
\emph{LLM--Human Interactions} shifts the unit of analysis from model outputs to how people interpret, use, and respond to them across conversational, creative, workplace, and educational settings. Across the following five subcategories, a shared concern is when LLMs augment human capabilities and when they instead produce overreliance, homogenization, inequality, or the displacement of human judgment.

\paragraph{Perception and Trust}
Examining people's perception and trust about LLMs forms the foundation. A central theme is anthropomorphism---when people treat machines as if they were human---and mentalizing, which means people assume that the model has inner states such as intentions, beliefs, or emotions, and then interpret its responses through that lens. These tendencies increase engagement and usability but also raise the risk of over-trust and misplaced responsibility \parencite{peter2025benefits}. Even simple design choices, such as describing a system as “your friendly assistant,” can make people feel a stronger social presence and change their expectations \parencite{van2025your}. Experimental studies show that when people treat LLMs as if they have mental states, they are more willing to rely on the outputs, regardless of actual accuracy \parencite{colombatto2025influence,street2024llm}. These dynamics develop in feedback loops: repeated exposure can amplify confidence and shift beliefs even if the underlying truth has not changed \parencite{glickman2025human}. Because many participants cannot distinguish LLMs from humans in modern Turing-style tests, researchers stress the importance of epistemic friction-design features that encourage users to pause and critically evaluate outputs rather than accept them automatically \parencite{jones2025people}.

\paragraph{Simulated Empathy and Support}
Research on simulated empathy and support examines how emotional responses and communicative dynamics shape engagement when LLMs support or partially replace human conversation. Perceived empathy depends not only on message content but also on how the system and the source of the response are presented: framing can shape beliefs about AI capabilities and thereby influence perceived trustworthiness, empathy, and effectiveness, while comparable responses may be evaluated differently depending on whether users believe a human or an AI produced them \parencite{pataranutaporn2023influencing,rubin2025comparing}. Such support can assist creative writing and learning \parencite{ref149,ref167}. In companionate settings, however, always-available and compliant artificial partners may encourage one-sided conversational habits, weaken attention to reciprocity and embodiment, and reproduce implicit biases and stereotypes despite neutral design intentions \parencite{ref132,grogan2025ai}. These effects can accumulate over time, as repeated interaction changes users’ understandings of LLMs, expectations, error-monitoring strategies, and reliance \parencite{schneider2025mental}. Overall, simulated empathy and support should be evaluated not only by how empathic the language appears, but also by how users respond to its AI source, whether they retain agency, and how it affects learning and human relationships.

\paragraph{Generative Work and Collaboration}
Expanding from emotional engagement, research increasingly examines how humans and AI systems collaborate, create, and learn. Within this broader area, studies of generative work and collaboration investigate how LLMs reshape individual productivity, teamwork, and the distribution of agency between humans and AI. In team settings, assigning an LLM a structured conversational role, such as “devil’s advocate,” can improve decision-making, while collaborative interfaces can externalize suggestions and reshape how groups explore a problem space \parencite{chiang2024enhancing,ref159}. Adjusting a model’s tone or role can further influence how users develop trust in and dependence on it over time \parencite{ha2024clochat}. Field experiments and user studies document gains in productivity, drafting speed, and confidence, although these benefits vary across tasks and levels of expertise and may be accompanied by overreliance, skill erosion, and loss of personal voice \parencite{noy2023experimental,li2024value,chen2024large}. Moreover, because adoption may be concentrated among higher-skilled workers, generative AI can widen existing inequalities unless institutions provide equitable access, training, and support \parencite{humlum2025unequal}. Effective human–LLM collaboration therefore depends on role design, complementary expertise, institutional support, and safeguards that preserve human agency, transparency, and accountability \parencite{ref134,ye2023improved}.

\paragraph{Creative Partnership}
Research on creativity examines how users perceive LLMs as creative partners. Users describe LLMs as “second minds” that support ideation, organization, and revision, while still emphasizing that effective co-creation requires human direction and judgment \parencite{ref119,ref154}. Experiments reveal a central tension: LLM assistance can increase the number and detail of an individual’s ideas and particularly benefit less creative users, yet it can also homogenize outputs across people and reduce collective novelty and perceived ownership \parencite{ref113,ref151}. Evaluations further suggest that leading LLMs can approach average human performance on some creativity tasks, although results vary substantially across prompts, benchmarks, creative domains, and definitions of creativity \parencite{ref15,ref22}. These judgments are also shaped by the conversational and anthropomorphic identities through which models are presented, which can encourage users to perceive them as agents, collaborators, or even companions rather than as conventional tools \parencite{ref145,ref166}. Assessments of generative creativity therefore cannot be separated from users’ perceptions of agency, authorship, originality, and the appropriate division of creative control between humans and models \parencite{ref184}.

\paragraph{Education and Assessment}
In education, LLMs support learning while challenging established instruction and evaluation. Research identifies opportunities for personalized explanations, feedback, and practice, alongside concerns about accuracy, academic integrity, overreliance, and teacher preparedness \parencite{ref116,ref130,ref140,ref187}. Structured guidance can improve students’ query strategies, performance, confidence, and appropriate trust more reliably than unrestricted answer provision; adoption also depends on AI literacy, motivation, and perceived usefulness \parencite{ref173,ref174,ref185}. Models can answer many university questions, yet performance depends on format and sustained reasoning, and plausible explanations may conceal fabricated information or incorrect logic \parencite{ref148,ref157}. They can also pass programming assessments designed to resist automated solution \parencite{ref164}. These findings require reconsidering learning goals and assessment. Alongside foundational knowledge, students need to formulate problems, judge when AI use is appropriate, evaluate its outputs, and remain accountable for their work. Teachers must assess whether students can explain, critique, and independently apply what they submit, using evidence of reasoning and revision as well as final answers \parencite{ref161,ref162}.

\section{Discussion}
\label{sec:discussion}
Our analysis supports organizing the reviewed literature on the social science of large language models (LLMs) into three domains: \emph{LLM as Social Minds}, \emph{LLM Societies}, and \emph{LLM--Human Interactions}. Combining a full-text synthesis of 198 papers with computational mapping of 47,719 publications shows how these domains are represented and developing. Human--LLM interactions account for the largest share of mapped research content, while attention to LLM societies has grown and thematic emphases differ across publication venues. Clustering, stability analyses, and comparisons with author classifications support the taxonomy as a useful organization of the reviewed corpus, while identifying substantive overlaps and boundaries that remain open to refinement.

The theoretical implications of the taxonomy developed in this study unfold across three main dimensions. First, the taxonomy provides three complementary levels of analysis for the social science of large language models---\emph{LLM as Social Minds}, \emph{LLM Societies}, and \emph{LLM--Human Interactions} (as shown in Figure~\ref{fig:llm-taxonomy}). \emph{Social Minds} centers on theory of mind, moral judgment, and bias, treating LLMs as entities capable of mind-like reasoning \parencite{marchetti2025artificial}. \emph{LLM Societies} examines emergent collective behaviors like cooperation, norms, and institutional dynamics among interacting model agents \parencite{park2023generative,ren2024emergence,perez2024cultural}. \emph{LLM--Human Interactions} focuses on trust, literacy, affect, and usage patterns at the interface, highlighting social and psychological factors shaping user behavior \parencite{lee2004trust,parasuraman1997humans,pataranutaporn2023influencing,ye2023improved,noy2023experimental}. These domains are understood as overlapping rather than sharply divided areas (Table~\ref{tab:llm-primary-secondary}), supporting a framework for cross-level reasoning. Findings at one level (e.g., mind-attribution tasks) have implications for others (e.g., trust in interactions or norm emergence in societies). Conversely, institutional results can refine individual-level expectations by showing which behaviors depend on social scaffolding. The reviewed findings illustrate why these connections matter: individually capable agents may struggle to coordinate, while improvements in individual creativity can accompany greater similarity across people's outputs \parencite{ref68,ref84,ref113,ref151}. This cross-level interpretation helps prevent category mistakes—such as treating cooperative behavior solely as individual cognition—and integrates cognitive, social, and human-centered perspectives \parencite{zhou2025rationality}.

Second, the review places substantive findings within a transparent and revisable map of research. The curated synthesis provides detailed interpretation of the three domains, while the field-scale analysis shows where research attention is concentrated and how its distribution changes over time. Human--LLM interactions dominate the mapped content overall, but conference papers give comparatively greater attention to model behavior and agent societies than journal papers. These differences suggest that readers drawing mainly on one publication setting may encounter a different picture of the field. Bringing these bodies of work together helps relate findings about model capabilities to questions about collective behavior and human practices. The computational analyses also make the proposed organization open to re-examination as new studies appear, supporting a living review. The categories remain informed by the data and interpreted by the authors; further validation across sources, models, and annotators is therefore important, in line with established principles of construct validity \parencite{campbell1959convergent,cronbach1955construct}.

Third, following Simon, adopting the perspective of LLMs as \emph{artifacts} situates our taxonomy within a view of these entities---whose design, deployment, and effects are inseparable from the institutions and relations in which they operate \parencite{simon2019sciences}. This stance does not reduce LLMs to mere tools or technical systems; rather, it treats them as socially embedded actors whose apparent “intelligence” and “agency” emerge from the interplay of model architectures, training data, interface affordances, and governance regimes. What counts as competence or understanding thus depends on situated practices of use and interpretation, not only on decontextualized internal capacities. In this sense, LLMs participate in collective cognition and coordination much as other designed artifacts do: they mediate social interaction, redistribute authority, and reshape accountability. This artifact-centered yet socially grounded view aligns with a long tradition of thinking about technology as part of social life and with recent discussions of the sociology of AI \parencite{liu2026sociology}. Technologies have been understood as nodes in broader socio-technical networks that link human and non-human actors \parencite{latour2005reassembling}; as elements within distributed cognitive systems where intelligence is shared across people and artifacts \parencite{hutchins1995cognition}; and as objects whose meaning and performance emerge through use in situated practice \parencite{suchman1987plans}. This lens shifts explanation from “are LLMs like persons?” to “what institutions do these artifacts enact---and how well do those institutions work?”---thereby integrating cognitive, social, and human-centered findings into a unified science of the artificial \parencite{simon2019sciences}.

Several limitations temper these conclusions but also point toward directions for future work. First, the 198-paper corpus was purposively curated rather than exhaustive; its membership was frozen in September~2025, with bibliographic metadata updated through August~2026. The field-scale corpus covered publications through 23 July 2026 and depended on prespecified search terms, English-language information, formal publication status, and a classifiable venue, potentially underrepresenting multilingual research, preprints, and work outside established outlets. These constraints highlight the value of developing a “living review” that continually tracks new work—including non-English and gray-literature sources—to test and update this taxonomy over time \parencite{elliott2017living}. Second, while stability analysis and validation indices minimize risk, the use of K-Means clustering with Euclidean distance (vs. density-based algorithms \parencite{comaniciu1999mean} or alternative embeddings) means that subtle shifts in cluster boundaries remain possible. Third, we employed a narrative synthesis approach rather than a quantitative meta-analysis \parencite{popay2006guidance,borenstein2021introduction}. This choice was necessary due to the heterogeneity of constructs, trading statistical generalization for interpretive integration; thus, our conclusions are theoretically integrative, not strictly meta-analytic. Finally, because neither expert nor machine classifications are fully objective, expert annotations are not treated as an absolute baseline. Accordingly, we complement quantitative alignment metrics with a qualitative analysis of divergences around category boundaries, which may indicate theoretical innovation rather than noise. Most mapping analyses used titles and abstracts, and full-text checks covered a bounded sample. LLM-assisted screening and venue assessments may vary with model and prompt choice, while topic naming and domain mapping require interpretation. Domain shares also depend on the number of topics mapped to each domain; citation and venue-standing analyses are descriptive, and the partial 2026 observation period limits annual comparisons.

Although our review has methodological constraints, it also exposes a broader limitation in the field it surveys: the need for more explanatory research on the mechanisms behind LLMs’ cognitive and social behaviors. Many studies describe surface-level behaviors—reasoning, moral judgment, cooperation, communication, influence—without examining the processes that produce them. As a result, it remains unclear whether these behaviors signal genuine intelligence or simply mirror the cognitive, linguistic, and social patterns in human training data, the dynamic captured by the notion of “stochastic parrots” \parencite{bender2020climbing,bender2021dangers}. This ambiguity fuels a deeper epistemic problem: when the fluency of LLMs is mistaken for understanding, or their apparent social conformity for moral reasoning, the line between simulated and genuine cognition blurs. Such misreadings can lead to persistent overtrust—treating generated patterns as dependable knowledge—and allow LLMs to reproduce and reinforce the norms and biases embedded in their data \parencite{raji2022fallacy}. 

Beyond cognitive and sustainability concerns, the risks extend to the social sphere. Overtrust can dampen incentives for individuals to contribute novel knowledge—undermining key crowdsourcing functions—and may also threaten the collective knowledge commons over time, exacerbating the Training Data Dilemma (TDD) and accelerating model degradation in LLMs~\parencite{burton2024large}. When many people rely on similar LLMs that tend to reproduce dominant viewpoints while sidelining minority perspectives, an illusion of consensus emerges, undermining independent judgment and weakening collective intelligence~\parencite{burton2024large}. And when outputs from systems without established awareness or intent are treated as authoritative, they begin to shape decisions, workflows, and institutional routines \parencite{crawford2021atlas,srnicek2017platform,zuboff2023age}. The central challenge is no longer whether LLMs “think,” but how their generative mechanisms shape socio-technical order—how trust is calibrated \parencite{lee2004trust,parasuraman1997humans}, accountability assigned, and externalities managed as LLMs become embedded in human institutions. Until these mechanisms are better understood, treating LLMs as intelligent actors risks obscuring the human structures of meaning, bias, and power that they so fluently reproduce.

When we cannot distinguish genuine cognition from fluent simulation, the stakes of this ambiguity demand a form of caution that is at once methodological and ethical. Methodological caution requires moving beyond surface description to identify the causal pathways through which linguistic, computational, and social processes jointly produce interpreted behaviors. Ethical caution demands restraint in attributing agency or moral status to systems without established awareness or intent \parencite{searle1980minds,floridi2002philosophy,Arcas2022large}. Together, these imperatives necessitate reflexivity: a recognition that explanation must replace analogy by tracing how human structures of meaning and power are reproduced through LLMs' apparent fluency. Our framework provides a structured agenda for this shift toward explanation by leveraging the three analytical levels to investigate underlying mechanisms. At the \emph{LLM as Social Minds} level, research should test the stability of cognitive-like behaviors (theory of mind, bias, deception) across varied and ecologically grounded contexts \parencite{ullman2023large,strachan2024testing,shapira2023clever,amirizaniani2025mind,marchetti2025artificial}. At the \emph{LLM Societies} level, work should systematically explain how institutional designs and micro-level incentives create emergent societal patterns (norms, cooperation, governance), testing the robustness of these effects across different environments (e.g., \cite{horiguchi2024evolution,ren2024emergence,park2023generative}). At the \emph{LLM–Human Interactions} level, studies should explain how interface variations and feedback mechanisms shape the calibration of trust and dependence, systematically manipulating transparency tools and warnings to clarify causal pathways affecting user oversight and risk \parencite{pataranutaporn2023influencing,jacob2025chat,ye2023improved,lee2004trust}. Finally, a critical need exists for shared, multi-level benchmarks to unify research across the three analytic domains. Widely used LLM evaluation benchmarks focus on technical performance and isolated capabilities \parencite{wang2019superglue}; the field would benefit from a cohesive platform to systematically test complex, human-like social abilities across the three domains. By introducing a unified testing environment, researchers can consolidate cumulative evidence and clarify how cognitive, institutional, and relational mechanisms interact. This approach will allow the field to move beyond debates about resemblance and directly guide the design and evaluation of LLMs that are more robust, interpretable, and ethically aligned with human judgment and autonomy \parencite{bommasani2021opportunities}.

\section{Methods}
\label{sec:methods}

The review combines full-text synthesis of a curated literature with mapping analyses of curated and field-scale corpora. Research was eligible when it examined LLMs or LLM-based agents themselves through their behavior, interactions, or social consequences. Studies using LLMs only as instruments for generating, coding, or analyzing research materials were outside scope. Detailed methods (Section~\ref{supp:methods}), search strings, model settings, prompts, and sensitivity analyses are provided in the Supplementary Materials. The corpus profile, adjacent-cluster transitions, screening flow, and topic-selection diagnostics are shown in Figs.~\ref{fig:descriptive-panels}, \ref{fig:k-transitions}, \ref{fig:study2-prisma}, and \ref{fig:stm-model-selection}.

\subsection{Study 1}

The curated corpus comprised 198 papers reviewed in full by the authors. Membership was finalized in September 2025; publication status and bibliographic metadata were updated through August 2026 without changing membership. Cleaned titles and abstracts were encoded using \texttt{sentence-transformers/all-mpnet-base-v2}, producing L2-normalized, 768-dimensional embeddings \parencite{song2020mpnet,reimers2019sentence}. K-means solutions for $K=2$--9 were compared using silhouette, Calinski--Harabasz, and Davies--Bouldin indices, together with resampling stability, initialization sensitivity, and transitions between adjacent resolutions \parencite{macqueen1967some,rousseeuw1987silhouettes,calinski1974dendrite,davies2009cluster}.

Within each selected cluster, latent Dirichlet allocation identified component themes \parencite{blei2003latent}. Topic number was selected from two to five using a combined coherence--perplexity criterion. Authors interpreted topic terms, word clouds, and representative papers and agreed on the three-domain codebook. Two authors independently classified papers from full texts with clustering and LLM assignments concealed, then resolved discrepancies through discussion. \texttt{gpt-5.6-sol} separately applied the codebook to titles and abstracts, assigning one primary and at most one secondary domain. Correspondence was assessed using proportion agreement, Cohen's $\kappa$, and adjusted Rand index; secondary assignments and low-consensus papers identified overlap and boundary cases \parencite{cohen1960coefficient,hubert1985comparing}.

\subsection{Study 2}

The field-scale search covered Semantic Scholar, OpenAlex, Scopus, PubMed, and Europe PMC for publications from 1 January 2020 through 23 July 2026. Forty-one terms described LLMs, related agents, and model families. Retrieved records were standardized, deduplicated, and consolidated across publication versions. Missing-year and clearly non-English records were excluded, and abstracts had to contain at least 200 characters. LLM-assisted title-and-abstract screening applied the substantive eligibility criteria. Further filters required formal scholarly publication, an LLM search term in the title, and a reliably classifiable venue, yielding 47,719 papers.

Structural topic models fitted with R's \texttt{stm} package represented topic prevalence as an additive function of publication period and LLM-assigned venue discipline \parencite{roberts2014structural,roberts2019stm}. Candidate resolutions of 10, 15, 20, 25, and 30 topics were compared using standardized mean coherence and exclusivity. Authors interpreted the selected topics using characteristic terms and representative papers, then mapped them to the taxonomy. Cross-domain topics remained outside the domain-specific mapping. Domain shares summed expected topic proportions and normalized over the 13 mapped topics. Temporal and venue contrasts accounted for both covariates simultaneously.

Citation percentiles combined discipline--year and publication-year ranks with respective weights of 0.70 and 0.30. Separately, \texttt{gpt-5.6-sol} estimated venue-standing indicators from venue name, type, and discipline; these descriptive scores were not official journal or conference rankings. Selective subsets combined citation percentiles of at least 99 with venue scores of 100 for conferences or at least 90 for journals. Sensitivity checks reassessed 300 sampled articles using all extracted full text, measured recovery of curated papers in the field-scale corpus, and repeated the curated analysis for recovered papers at or above the 50th citation percentile.

\clearpage

\makeatletter
\newcommand{\bibalias}[1]{%
  \if@filesw
    \immediate\write\@auxout{%
      \string\bibcite{#1}{\the\value{\@listctr}}%
    }%
  \fi
  \@ifundefined{hypertarget}{}{\hypertarget{cite.#1}{}}%
}
\makeatother

\begingroup
\singlespacing\small
\renewcommand{\refname}{References and Notes}

\endgroup

\clearpage
\section*{Acknowledgments}


\textbf{Data and materials availability.}

The processed datasets and derived analytical results supporting this study are available through \DataRepositoryLink. The shared materials include the Study~1 and Study~2 analysis inputs, clustering and topic-modelling results, classification and stability analyses, corpus-coverage outputs, and screening-flow information. Article full texts are not redistributed because they may be subject to copyright restrictions.

\textbf{Code availability.}

All custom code supporting this study is available through \CodeArchiveLink. The repository contains workflows for database retrieval, metadata standardization, deduplication, study-family consolidation, LLM-assisted eligibility screening, citation and publication-venue enrichment, and full-text sensitivity assessment. It also includes code for MPNet text representation, K-means clustering and stability analyses, classification correspondence, within-cluster LDA, STM estimation and model selection, topic-to-domain mapping, temporal and venue-related covariate analyses, curated-corpus coverage and citation sensitivity analyses, and the generation of analytical tables and figures.

\textbf{Competing interests.}

The authors declare no competing interests.

\textbf{Ethical approval.}

This study analysed published scholarly literature and bibliographic metadata and did not involve human participants or identifiable private data. Ethical approval was therefore not required.

\textbf{Informed consent.}

This study did not involve human participants. Informed consent was therefore not applicable.

\clearpage
\setcounter{page}{1}
\renewcommand{\thepage}{S\arabic{page}}
\section*{Supplementary Materials}
\label{supp:start}
\noindent\textbf{Mapping the Emerging Social Science of Large Language Models}\\
Zhanzhan Zhao, Yi Yang, Xiao Jia, Zeyun Dong, and Chenzhang Wang\\
Zhanzhan Zhao, Yi Yang, and Xiao Jia contributed equally to this work.\\
Correspondence: Zhanzhan Zhao (zhanzhanzhao@cuhk.edu.cn)

This supplement contains detailed mapping results, review methods, search strategies, model prompts, a bibliographic inventory, and supplementary figures and tables. All cited sources appear in the combined reference list in the main manuscript.

\setcounter{section}{0}
\renewcommand{\thesection}{S\arabic{section}}
\renewcommand{\theHsection}{S\arabic{section}}
\setcounter{figure}{0}
\renewcommand{\thefigure}{S\arabic{figure}}
\renewcommand{\theHfigure}{S\arabic{figure}}
\setcounter{table}{0}
\renewcommand{\thetable}{S\arabic{table}}
\renewcommand{\theHtable}{S\arabic{table}}
\singlespacing
\section{Detailed mapping results}
\label{supp:results}

To map the organization of research on the social science of LLMs, we conducted two complementary studies. Study 1 focused on a curated corpus of 198 papers collected and read in full by the authors. We used MPNet representations and K-means clustering to identify its principal semantic structure, and combined LDA, full-text author classification, and LLM-based reclassification to interpret the resulting clusters and their boundaries. Study 2 extended the analysis to a field-scale corpus assembled from five databases, using structural topic modelling to identify finer-grained research themes and examine their distributions across publication years and sources. In this design, Study 1 provided a closely examined basis for developing and evaluating the three-domain framework, while Study 2 assessed how that framework related to the thematic organization of a substantially broader literature. The results are presented in sequence, beginning with the identification and interpretation of structure in the curated corpus, followed by the field-scale thematic analysis and associated sensitivity and coverage checks.

\subsection{Study 1: Three-domain structure in the curated corpus}

\subsubsection{Cluster selection and stability}

We collected a corpus of 198 papers on the social science of LLMs (see Section~\ref{sec:5.1.1}) and first characterized its composition by publication type, year, and venue before analyzing its semantic structure. 
Table~\ref{tab:document-types} summarizes the publication types represented in
the final corpus. Journal articles constitute the largest category, accounting
for 115 of the 198 documents (58.1\%), followed by conference papers (76,
38.4\%) and preprints (7, 3.5\%). Thus, 96.5\% of the corpus consists of
journal or conference publications, indicating that the evidence base is
primarily grounded in formally disseminated scholarly outlets rather than
preprint-only dissemination.

Figure~\ref{fig:descriptive-panels} presents the temporal and venue profiles of
the corpus. Panel~(a) shows a rapid increase in publications from 1 document in
2021 (0.5\%) and 5 in 2022 (2.5\%) to 39 in 2023 (19.7\%), with the largest
annual count occurring in 2024 (97, 49.0\%) and a further 56 documents appearing
in 2025 (28.3\%). Panel~(b) indicates substantial dispersion across publication
venues. \textit{Scientific Reports} is the most frequent venue (9, 4.5\%),
followed by \textit{Nature Human Behaviour} (8, 4.0\%), while ACM CHI and EMNLP
each contribute 7 documents (3.5\%). \textit{AI \& Society} and arXiv each
contribute 6 documents (3.0\%), and IEEE BigData and \textit{Nature Machine
Intelligence} each contribute 5 (2.5\%). No individual venue accounts for more
than 4.5\% of the corpus, demonstrating that the sample is distributed across a
broad interdisciplinary set of outlets.

We represented each paper by applying MPNet to its concatenated title and abstract and clustered the resulting document embeddings using K-means. To determine the resolution used for subsequent interpretation, we compared solutions across $K=2$--$9$ using internal clustering indices, resampling stability, random-initialization stability, and transitions between adjacent values of $K$.

To evaluate clustering performance across the candidate range of \(K=2\)--\(9\) (Figure~\ref{fig:k-selection-stability}a--c), we considered three complementary internal validation indices: cosine silhouette (whether each paper is more similar to papers in its own cluster than to those in the nearest other cluster), Calinski--Harabasz (whether papers form tight groups that are far apart), and Davies--Bouldin (how much each cluster resembles its most similar neighboring cluster), where higher values are better for the first two indices and lower values for the third.
The cosine silhouette was highest at $K=2$ (0.161), whereas the Calinski--Harabasz index reached its maximum at $K=3$ (13.996). The Davies--Bouldin index, for which lower values indicate greater separation, reached its minimum at $K=9$ (2.552), compared with 3.078 at $K=3$. Taken together, these indices described a trade-off between a coarse two-cluster partition, an intermediate three-cluster partition, and progressively finer divisions; no single internal criterion determined the resolution used in the subsequent analysis.

Resampling stability provided clearer support for the intermediate solution (Figure~\ref{fig:k-selection-stability}d). Agreement was measured using the adjusted Rand index (ARI), which indicates how consistently two clustering solutions assign the same papers to the same groups after accounting for agreement expected by chance. An ARI of 1 indicates identical assignments, whereas a value near 0 indicates agreement no better than chance. 
Across 100 draws containing 90\% of the corpus without replacement, the $K=3$ assignments closely reproduced the full-sample solution (mean ARI $=0.952$, median $=0.964$, range $=0.846$--$1.000$), and none of the draws had an ARI below 0.80. By comparison, the mean ARI was 0.721 for $K=2$ and 0.778 for $K=4$, with 23 and 41 draws, respectively, falling below 0.80. We further examined whether the $K=3$ solution was sensitive to the random starting centroids used by K-means. Across 50 runs with different random seeds, each allowing 10 candidate initializations and retaining the best-fitting solution, the mean pairwise ARI was 0.966 (median $=0.973$, range $=0.865$--$1.000$). Increasing the number of candidate initializations from 10 to 50 raised the mean pairwise ARI to 0.987. Together, these results show that the $K=3$ solution was consistently reproduced under changes in both corpus composition and random initialization.

The transitions between adjacent resolutions clarified what was added or divided as $K$ increased (Figure~\ref{fig:k-transitions}). From $K=2$ to $K=3$, 48 of the 51 papers (94.1\%) in the smaller $K=2$ cluster remained together, while the 147-paper cluster separated primarily into groups of 50 and 94 papers. From $K=3$ to $K=4$, two clusters remained largely intact: 49 of 52 papers (94.2\%) in one cluster and 50 of 51 papers (98.0\%) in another moved together into single $K=4$ clusters. The remaining 95-paper cluster divided almost evenly, with 48 and 46 papers entering two different $K=4$ clusters. Thus, the move to $K=3$ distinguished two substantial components within the broad $K=2$ grouping, while the move to $K=4$ mainly subdivided one of the three resulting clusters.

Considering the diagnostics jointly, we selected $K=3$ as the working resolution for semantic interpretation. It combined the highest Calinski--Harabasz score with the strongest 90\% resampling stability, remained highly consistent across random initializations, and occupied a structurally informative position between the coarse division at $K=2$ and the further subdivision observed at $K=4$. This choice defines the level of resolution used in the analyses that follow; the substantive meaning of the three clusters is established through their textual content in the next section.

\subsubsection{Semantic interpretation of the three-cluster solution}

The selected $K=3$ solution divided the 198 papers into clusters of 95, 52, and 51 papers (Table~\ref{tab:study1-cluster-sizes}). Figure~\ref{fig:kmeans-pca-clusters} provides a two-dimensional projection of the full-dimensional assignments. Papers with negative silhouette values were concentrated near the interfaces among the projected clusters, identifying areas in which their semantic content may overlap.

To characterize the semantic content of these groups, we fitted a separate LDA model within each cluster. The retained models contained four topics in Cluster~0, four in Cluster~1, and five in Cluster~2. Figures~\ref{fig:cluster0}-\ref{fig:cluster2} display each topic across five values of the relevance parameter $\lambda$. At $\lambda=0$, the word rankings emphasize terms that are distinctive to a topic relative to the background vocabulary; as $\lambda$ approaches 1, they increasingly emphasize terms with high probability within that topic. We interpreted each cluster from the combinations of terms that remained prominent across this sweep, using the topic-specific terms to distinguish its component themes and the higher-probability terms to identify their common semantic orientation.

\textit{LLM as Social Minds.} Cluster~0, the largest cluster ($n=95$), grouped research whose central object was the individual LLM, examined as a bearer of social, cognitive, moral, and behavioural properties (Figure~\ref{fig:cluster0}). The first topic linked theory of mind with social reasoning, human cognition, and confidence. The second combined personality, behavioural traits, gender, race, and social bias, while the third concerned political orientations, moral foundations, cultural variation, and model judgments. The fourth centred on human values, preference alignment, and persuasive communication and deception. Across these themes, socially meaningful judgments, dispositions, and apparent cognitive capacities are examined as properties of an individual model. We therefore interpret this cluster as \emph{LLM as Social Minds} (see Section~\ref{sec:social-minds} for further elaboration).

\textit{LLM Societies.} Cluster~1 ($n=52$) grouped research whose central object was the collective processes and emergent social behaviors produced through interactions among multiple LLM-based agents (Figure~\ref{fig:cluster1}). The first topic brought together strategic games, multi-agent scenarios, and behavioural traits. The second connected collective intelligence with dialogue, cognition, and emergent LLM-agent societies. The third centred on group decision-making, cooperation, and social norms, while the fourth linked social networks, communities, populations, and large-scale simulation. Together, these topics move from strategic interaction and collective decision-making to emergent social organization and population-level simulation. Their shared focus on relations among multiple agents and the system-level patterns generated through those relations motivates the label \emph{LLM Societies} (see Section~\ref{sec:llm-societies} for further elaboration).

\textit{LLM--Human Interactions.} Cluster~2 ($n=51$) brought together studies whose central object was the relation between people and LLM-based systems (Figure~\ref{fig:cluster2}). Its five topics address how anthropomorphism and mental models shape trust and reliance; how simulated empathy and conversational support create emotional and relational experiences; how LLMs affect productivity, teamwork, and human agency; how users perceive LLMs as creative partners, raising questions of originality, authorship, and ownership; and how LLMs reshape teaching, learning goals, and assessment. Across these topics, recurring terms such as \emph{human}, \emph{user}, \emph{ChatGPT}, \emph{writing}, \emph{creativity}, \emph{support}, \emph{trust}, and \emph{education} indicate a shared focus on how people perceive, use, and are affected by LLMs. We therefore interpret Cluster~2 as \emph{LLM--Human Interactions} (see Section~\ref{sec:llm-human-interactions} for further elaboration).

\subsubsection{Correspondence across K-means, expert, and LLM classifications}

After establishing the three-domain codebook developed by the LDA-informed interpretations, two experts (authors who don't have access to the clustering results and LLM classification results) reviewed each paper's title, abstract, and main text and assigned a single domain that best captured its principal object of explanation. 
The expert classification corresponded with these K-means domain labels for 154 papers (77.78\%; Cohen's $\kappa=0.662$; ARI $=0.434$). At the domain level, the corresponding K-means cluster contained 67 of the 75 papers assigned by the experts to \emph{LLM as Social Minds} (89.3\%), 41 of the 47 assigned to \emph{LLM Societies} (87.2\%), and 46 of the 76 assigned to \emph{LLM--Human Interactions} (60.5\%). Thus, the expert-defined \emph{LLM as Social Minds} and \emph{LLM Societies} domains were concentrated more strongly in their corresponding clusters, while papers classified as \emph{LLM--Human Interactions} were distributed more broadly across the K-means solution.

The LLM-based evaluator applied the same three-domain codebook using only the title and abstract of each of the 198 papers and assigned each paper a primary category. 
Its primary classification corresponded with the K-means domain label for 149 papers (75.25\%; Cohen's $\kappa=0.612$; ARI $=0.384$). The corresponding cluster contained 73 of the 93 papers assigned by the LLM to \emph{LLM as Social Minds} (78.5\%), 32 of the 35 assigned to \emph{LLM Societies} (91.4\%), and 44 of the 70 assigned to \emph{LLM--Human Interactions} (62.9\%). This domain-specific pattern resembled the author comparison: \emph{LLM Societies} showed the most concentrated correspondence, whereas \emph{LLM--Human Interactions} extended most broadly across clusters.

The LLM-based evaluator could also assign at most one secondary category when another domain was substantively represented. Table~\ref{tab:llm-primary-secondary} summarizes the resulting combinations of primary and secondary categories. A secondary category was assigned to 87 of the 198 papers (43.9\%), indicating that cross-domain content was common in the LLM-based classification. The most prominent connection was between \emph{LLM as Social Minds} and \emph{LLM--Human Interactions}, which together accounted for more than two-thirds of all secondary assignments. \emph{LLM Societies} showed a different and directionally uneven pattern: papers assigned primarily to this domain frequently incorporated another domain, whereas relatively few papers in the other two domains received \emph{LLM Societies} as their secondary category. These patterns illustrate that the relationships among the three domains extend beyond mutually exclusive primary assignments.

The strongest pairwise correspondence was observed between the expert and LLM classifications: they assigned the same domain to 170 of the 198 papers (85.86\%; Cohen's $\kappa=0.781$; ARI $=0.623$; Table~\ref{tab:classification-correspondence}). 
The LLM-based evaluator used only the title and abstract, whereas the authors classified the papers after reviewing their full texts. Their agreement on 170 of the 198 papers indicates that the three-domain codebook could be applied consistently by a non-author evaluator and suggests that title-and-abstract information was sufficient to recover the broad domain orientation of most papers in the curated corpus. 
The LLM-assigned secondary categories captured additional correspondence beyond the primary assignments. Among the 49 papers whose LLM primary category did not correspond with the K-means domain, 25 included that domain as their secondary category. Of the remaining 24 papers, 18 received no secondary category, while only 6 received a secondary category that also differed from the K-means domain.

Disagreements were concentrated among papers located near the geometric boundaries of the K-means solution. 
Across the 100 runs using 90\% subsamples, ten papers did not consistently remain grouped with the papers in their full-sample cluster (within-cluster consensus \(<0.90\)). We refer to these as low-consensus papers.
Correspondence between the K-means domain and both the expert and LLM classifications was substantially lower for these papers than for the remaining 188 papers: author--K-means correspondence was 20.0\% versus 80.9\% (Fisher's exact $p<0.001$), and LLM--K-means correspondence was 30.0\% versus 77.7\% ($p=0.003$). 
These papers also had a lower median silhouette value (0.043 versus 0.135) and a smaller median relative margin between their nearest and second-nearest centroids (0.031 versus 0.148). Lower silhouette values indicate weaker separation between the assigned cluster and the nearest alternative cluster, while smaller centroid margins mean that these papers were nearly equally close to their assigned cluster centre and the nearest alternative cluster centre. Together, these patterns place the low-consensus papers near the semantic boundaries of the K-means solution.
Consistent with this boundary interpretation, six of the ten low-consensus papers received an LLM-assigned secondary category, compared with 81 of the remaining 188 papers (60.0\% versus 43.1\%; Fisher's exact \(p=0.340\)). This directional pattern provides descriptive evidence of cross-domain content, while the consensus, silhouette, and centroid-margin results provide the principal evidence that classification disagreements were concentrated at semantic boundaries.

\subsection{Study 2: Field-scale thematic structure}

\subsubsection{Corpus profile and STM model selection}

To obtain broad coverage of the field, we searched Semantic Scholar,
OpenAlex, Scopus, PubMed, and Europe PMC using 41 prespecified terms
referring to LLMs and related generative models, as detailed in
Section~\ref{sec:5.2.1}. These searches yielded
1,173,481 bibliographic records. We standardized the bibliographic fields,
removed duplicate records within and across databases, and excluded records
with missing publication years, non-English content, or abstracts that were
too short for eligibility assessment. Reports representing different versions
of the same study were also consolidated. These steps produced 478,405
representative records for substantive eligibility screening.

We then refined this broad retrieval to identify studies closely aligned with
our definition of the social science of LLMs. An LLM assessed the title and
abstract of each record and retained 146,682 records that substantively
examined LLMs or LLM-based agents as social objects. We subsequently focused
the corpus on formally published scholarly work by excluding software,
datasets, and other non-publication records, leaving 93,436 records. To
strengthen thematic specificity, we further retained records whose titles
contained at least one of the 41 prespecified LLM terms. This title screening
produced 54,159 records.

Finally, to enable the STM analysis of variation in topic prevalence across
academic domains, an LLM classified the disciplinary orientation of each
publication venue using only the venue field. Records associated with
repositories or preprint platforms and records whose venues could not be
classified reliably were excluded. This final step removed 6,440 records,
yielding an analytic corpus of 47,719 papers.

We fitted structural topic models with $K=10$, 15, 20, 25, and 30 topics to the 47,719-paper corpus. For each candidate model, semantic coherence and exclusivity were calculated for every topic and then averaged across topics (Figure~\ref{fig:stm-model-selection}). Mean semantic coherence was highest at $K=10$ ($-71.274$) and declined as the number of topics increased, whereas mean exclusivity increased from 9.399 at $K=10$ to 9.739 at $K=30$. These opposing trends indicated a trade-off between within-topic semantic coherence and the distinctiveness of topic vocabularies.

To select a working resolution, the mean coherence and exclusivity values were separately standardized across the five candidate models and summed. The resulting trade-off score was highest at $K=15$ (0.345), followed by $K=20$ (0.220); the scores for $K=10$, 25, and 30 were below zero. Accordingly, we selected the 15-topic model for substantive interpretation and the subsequent prevalence analyses.

\subsubsection{Fifteen topics and their relationship to the three domains}

The selected STM solution described the corpus through 15 topics. After reviewing the characteristic terms associated with each topic, we assigned descriptive labels that summarized their substantive content (Figure~\ref{fig:stm-topics-domains}A). The most prevalent topic was T8, \emph{LLM Evaluation, Reasoning, Safety, and Bias} (12.64\%), followed by T5, \emph{General ChatGPT Discourse and Prospects} (9.17\%), T12, \emph{Pedagogy, Teacher Practice, and Curriculum Integration} (8.68\%), and T13, \emph{Law, Ethics, Privacy, and Security Governance} (8.34\%). The remaining topics covered clinical evaluation, media and accessibility, language learning, business applications, technology adoption, cultural relations, higher education, software development, and public-facing services.

Thirteen of the 15 topics were mapped to one of the three domains according to their primary objects of explanation (Figure~\ref{fig:stm-topics-domains}A). T2, \emph{Agentic Systems, Simulation, and Robotics}, was mapped to \emph{LLM Societies}, and T8, \emph{LLM Evaluation, Reasoning, Safety, and Bias}, was mapped to \emph{LLM as Social Minds}. The eleven topics mapped to \emph{LLM--Human Interactions} were T3, \emph{Clinical and Patient-Facing LLM Evaluation}; T4, \emph{Digital Media, Content, and Accessibility}; T6, \emph{Language Learning, Writing, and Feedback}; T7, \emph{Business Applications, Management, and Innovation}; T9, \emph{Technology Adoption, Trust, and Use Intention}; T10, \emph{Culture, Identity, Agency, and Human-AI Relations}; T11, \emph{Student Use, Academic Integrity, and Higher Education}; T12, \emph{Pedagogy, Teacher Practice, and Curriculum Integration}; T13, \emph{Law, Ethics, Privacy, and Security Governance}; T14, \emph{Software Engineering and LLM-Assisted Development}; and T15, \emph{Public Sentiment, Translation, and Service Applications}. T1, \emph{Evidence Synthesis and Bibliometrics}, and T5, \emph{General ChatGPT Discourse and Prospects}, captured broad cross-domain material and were retained as a separate group in the domain-level comparison.

Together, the two broad cross-domain topics accounted for 13.01\% of the total expected topic mass. Across the full corpus, \emph{LLM--Human Interactions} accounted for 67.88\% of the total expected mass, compared with 12.64\% for \emph{LLM as Social Minds} and 6.47\% for \emph{LLM Societies}. After normalization over the 13 domain-mapped topics, the corresponding shares were 78.02\%, 14.53\%, and 7.44\%. The numerical dominance of \emph{LLM--Human Interactions} therefore extended across a diverse set of substantive settings, including education, healthcare, business, law, media, software development, technology adoption, and cultural relations. This breadth indicates that the uses and implications of LLMs for human activities constituted the most extensively represented line of inquiry in the corpus.

To examine whether domain composition varied among more highly cited papers and more highly ranked publication venues, we compared mapped topic mass across citation- and venue-stratified subsets (Figure~\ref{fig:venue-subsets}). Citation percentile placed papers on a 0--100 scale while accounting for publication year and discipline, as discussed in more detail in Section~\ref{sec:5.2.2}.

Venue standing was estimated by an LLM on a common 0--100 scale, using journal field percentiles and conference tiers as conceptual anchors. These scores were descriptive estimates, not official Journal Impact Factor (JIF) percentiles or Computing Research and Education Association of Australasia (CORE) rankings.
Among the 15,511 papers published in conference or proceedings-series venues, \emph{LLM as Social Minds} and \emph{LLM Societies} together represented 40.95\% of mapped topic mass. Their combined share increased to 65.45\% among the 102 conference papers with citation percentiles of at least 99. 

Among the 89 highly cited papers in conference venues assigned a score of 100,  \emph{LLM as Social Minds} represented 46.89\% of mapped mass and \emph{LLM Societies} represented 19.48\%, compared with 33.63\% for \emph{LLM--Human Interactions}.

The corresponding shares were smaller in journals but followed the same direction. \emph{LLM as Social Minds} and \emph{LLM Societies} together represented 11.83\% of mapped topic mass across all 27,178 journal papers, 16.57\% among the 382 journal papers with citation percentiles of at least 99, and 23.19\% among the 137 highly cited journal papers with venue scores of at least 90. In the last subset, \emph{LLM as Social Minds} accounted for 16.31\% and \emph{LLM Societies} for 6.88\%, while \emph{LLM--Human Interactions} remained the largest domain at 76.81\%. Thus, the greater relative representation of the two less numerous domains was most pronounced among highly cited papers in conferences with the highest LLM-estimated venue score, with the same directional pattern also visible within journals.

\subsubsection{Temporal and venue-related patterns}

The number of papers included in the analytic corpus increased sharply after 2022. The corpus contained 177 papers published between 2020 and 2022, 4,128 in 2023, 11,133 in 2024, and 18,190 in 2025. A further 14,091 papers were published in 2026; because the search covered publications only through 23 July 2026, this partial-year count is not directly comparable with those for complete calendar years. All 47,719 records met the formal-publication criterion. Of these, 27,178 appeared in journals, 13,313 in conferences, 2,198 in proceedings series, 3,110 in book series, and 1,920 in other classified publication venues.

The STM prevalence model estimated temporal differences while accounting for the disciplinary orientation of publication venues (Figure~\ref{fig:stm-temporal-venue-patterns}A). Relative to 2023, the adjusted topic mass associated with \emph{LLM--Human Interactions} was 5.98 percentage points higher in 2024, 8.90 points higher in 2025, and 10.64 points higher in the partial 2026 period. The corresponding mass of the two broad cross-domain topics decreased by 8.36, 13.75, and 15.48 points. Within the three-domain framework, Topic~2 (\emph{LLM Societies}) increased by 0.98, 2.93, and 3.77 points across the same periods (all $p<0.001$), while Topic~8 (\emph{LLM as Social Minds}) remained modestly above its 2023 level by 1.41, 1.92, and 1.06 points (all $p\leq0.0023$). The expansion of \emph{LLM--Human Interactions} was initially associated especially with topics on pedagogy and adoption and trust; by 2026, culture and identity, adoption and trust, and software engineering showed the largest positive period coefficients. These results indicate that recent growth was concentrated in increasingly specialized studies of how LLMs are used, experienced, and governed, alongside a smaller rise in work on multi-agent social systems.

Venue discipline was also associated with thematic composition (Figure~\ref{fig:stm-temporal-venue-patterns}B). Topic~8, \emph{LLM as Social Minds}, was concentrated in AI/ML/NLP venues: its adjusted mass was 22.65--34.35 percentage points lower in each of the other venue categories (all $p<0.001$). Topic~2, \emph{LLM Societies}, was 6.50 percentage points higher in HCI and human-centered computing venues ($p<0.001$) and 0.81 points higher in engineering and technology venues ($p=0.017$), but lower in the remaining venue categories. The model therefore indicates that research on model-level social capacities was concentrated primarily in AI/ML/NLP venues, whereas research on multi-agent social systems showed its clearest positive association with HCI and human-centered computing venues.

\subsubsection{Sensitivity and coverage analyses}
To assess eligibility decisions made from titles and abstracts, an LLM applied the same eligibility criteria to the complete extracted full texts of 300 sampled records. Of these, 288 were confirmed as eligible, 10 were judged ineligible, and 2 remained indeterminate after all extracted full-text evidence was considered, corresponding to a confirmation rate of 96.00\%, an exclusion rate of 3.33\%, and an indeterminate rate of 0.67\%. These results indicate that the title-and-abstract screening procedure generally retained its eligibility judgments when the complete extracted full texts were considered.

We also examined how the 198 papers in the curated corpus were represented in the larger analytic corpus. In total, 162 papers (81.82\%) were recovered. Relative to the full corpus of 198 papers, the 162-paper subset retained similar proportions across the three expert-assigned thematic clusters: 64 of 75 papers on \emph{LLM as Social Minds} (85.3\%), 37 of 47 on \emph{LLM Societies} (78.7\%), and 61 of 76 on \emph{LLM--Human Interactions} (80.3\%). The recovered papers occupied relatively high positions in the standardized citation distribution, with a median citation percentile of 90.35; 148 of the 162 papers (91.4\%) were at or above the 50th percentile.
For each recovered paper, we compared the LDA-informed domain label of its original Study~1 $K=3$ K-means cluster with the domain mapped from its dominant topic in the Study~2 15-topic STM. Thirteen of the 162 papers were dominated by Topic~1 or Topic~5, which were designated as cross-cutting, and were therefore excluded from the three-domain comparison. Among the remaining 149 papers, the two analyses assigned 117 to the same domain (78.52\%; Cohen's $\kappa=0.663$), indicating meaningful convergence in the three-domain structure across two different unsupervised methods and corpus scales.

Finally, we restricted the 162 recovered papers to the 148 at or above the 50th citation percentile and repeated the MPNet, K-means, and within-cluster LDA analysis. Across $K=2$--9, the $K=3$ solution had the highest Calinski--Harabasz index and the strongest correspondence with the author classifications. The LDA summaries continued to distinguish research on model-level social minds, multi-agent societies, and human--LLM interactions. Using these semantic interpretations, the K-means assignments corresponded with the authors' full-text classifications for 117 of 148 papers (79.05\%), with an ARI of 0.465 and an NMI of 0.475. Together, these analyses show that the principal results were maintained under full-text eligibility assessment, expansion to the larger corpus, and restriction of the curated sample to more highly cited papers.

\section{Detailed review methods}
\label{supp:methods}

The empirical analysis comprised two complementary studies. Study~1 used a curated corpus whose papers were reviewed in full to identify and interpret their semantic structure and to examine correspondence among K-means clustering, author classification, and LLM-based classification. Study~2 used a substantially larger corpus assembled from five bibliographic databases to characterize the field at scale through structural topic modelling, including variation across publication periods and academic venues. It also assessed the coverage of the curated corpus and whether the three-domain structure remained visible across the two corpus-construction and text-analysis procedures.

\subsection{Study 1: Curated-corpus analysis}

\subsubsection{Curated corpus and descriptive variables}
\label{sec:5.1.1}
Study 1 examined a corpus of 198 papers collected according to our definition of the social science of LLMs, defined here as research that treats LLMs or LLM-based agents themselves as objects to be explained from a social-scientific perspective. The authors obtained and reviewed the full text of each paper. Corpus membership was finalized in September 2025. Publication status and bibliographic metadata for these papers were subsequently updated through August 2026, while corpus membership remained fixed. Study 2 subsequently assessed both the recovery of these curated papers in a broader database-derived corpus and the correspondence between the three-domain structure and the thematic patterns observed at field scale.

For the descriptive analysis, we recorded publication year, publication type, and publication venue. Publication types were harmonized into journal, conference, and preprint categories, with the single working paper grouped with preprints for reporting. Venue names were standardized across capitalization variants and annual editions of the same conference series.

\subsubsection{Text representation, K-means clustering, and cluster selection}
\label{sec:5.1.2}

We represented each paper using its title and abstract. After decoding HTML entities, removing HTML tags, and normalizing whitespace, the two fields were concatenated with a period. We encoded the resulting texts with the \texttt{sentence-transformers/all-mpnet-base-v2} model, which combines the MPNet architecture with a sentence-transformer representation framework \parencite{song2020mpnet,reimers2019sentence}. Inputs were truncated at the model's maximum sequence length of 384 tokens. Mean pooling over token embeddings produced a 768-dimensional vector for each paper, and the document embeddings were L2-normalized before clustering.

We fitted K-means solutions to the normalized embeddings for $K=2$--9 \parencite{macqueen1967some}. Each full-sample model used k-means++ initialization, 10 initializations, a maximum of 300 Lloyd iterations, and a common random seed of 42 \parencite{arthur2006k,lloyd1982least}. K-means minimized Euclidean distances between the normalized document vectors; for unit-length vectors, these distances are monotonically related to cosine dissimilarity. We evaluated candidate solutions using three internal validation measures computed on the same embedding matrix: the cosine silhouette coefficient, which assesses the balance between within-cluster cohesion and between-cluster separation for individual documents; the Calinski--Harabasz (CH) index, which measures the ratio of between-cluster dispersion to within-cluster dispersion; and the Davies--Bouldin (DB) index, which quantifies the average similarity between each cluster and its most similar counterpart \parencite{rousseeuw1987silhouettes,calinski1974dendrite,davies2009cluster}. Higher values of the silhouette coefficient and CH index, and lower values of the DB index, indicate better-defined and more separated clustering structures.

We assessed sensitivity to both random initialization and corpus composition. For the $K=3$ candidate, we fitted 50 full-sample models using random seeds 0--49, first with 10 initializations and then with 50; stability was summarized by the adjusted Rand index (ARI) across all 1,225 pairs of solutions \parencite{hubert1985comparing}. For the subsampling analysis, we generated 100 samples containing 80\% of the corpus (158 papers) and 100 samples containing 90\% (178 papers), in each case sampling without replacement. The two sequences were generated consecutively using a NumPy random-number generator initialized with seed 42. The 80\% analysis was conducted for $K=3$, while one shared set of 90\% samples was used for every value of $K=2$--9 and for the document-level $K=3$ analyses. Each subsample model used 10 initializations and random state $42+r$, where $r$ indexed the repetition, and its assignments were compared with the corresponding full-sample solution on the papers included in that subsample. We also constructed a pairwise co-clustering matrix for the 90\% $K=3$ runs. For each pair of papers, consensus was defined as the proportion of runs in which they were assigned to the same cluster among runs in which both were sampled \parencite{monti2003consensus}. A paper's within-cluster consensus was the mean of its pairwise consensus values with the other papers assigned to its full-sample cluster. The thresholds of 0.80 and 0.90 served as post-hoc descriptive markers of comparatively low consensus.

To trace how the partition changed across adjacent resolutions, we cross-tabulated the full-sample assignments of the same 198 papers at $K=2$ and $K=3$, and at $K=3$ and $K=4$. Cell percentages were calculated within each source cluster at the lower value of $K$, using the full-sample assignments directly and without label alignment across values of $K$. The working value of $K$ was selected by considering the internal validation indices, subsampling stability, random-initialization stability, and the structural changes observed across adjacent resolutions. These diagnostics were interpreted jointly rather than combined into a weighted score. For visualization, we projected the complete embeddings onto their first two principal components; all cluster assignments remained those obtained in the original 768-dimensional space. The plotted black crosses mark the geometric median of each cluster in the two-dimensional projection, dashed ellipses show 95\% covariance regions, and papers with negative cosine silhouette values in the original embedding space are marked separately. Cluster identifiers for the selected solution were permuted after fitting solely to provide consistent manuscript-facing labels, leaving cluster membership and all calculated statistics unchanged.

\subsubsection{Within-cluster LDA and semantic interpretation}
\label{sec:5.1.3}

We fitted a separate latent Dirichlet allocation (LDA) model within each cluster of the selected K-means solution \parencite{blei2003latent}. The models used the same cleaned title-and-abstract texts as the embedding analysis. Within each cluster, a count-based document--term matrix was constructed after converting text to lowercase, normalizing Unicode accents, removing English stop words, and extracting unigrams and bigrams. Terms appearing in fewer than two documents or in more than 95\% of documents in the cluster were excluded, and the vocabulary was capped at 50,000 features. 

To select a topic resolution for each cluster using a consistent empirical criterion, we fitted candidate LDA models with $T=2$--5 topics and compared them using a combined score favoring solutions whose prominent topic terms tended to appear together in the same papers and whose fitted model better represented the observed word patterns within the cluster. All models used batch learning, a maximum of 80 iterations, and random seed 42. The first property was measured using UMass coherence, calculated from the ten highest-probability terms in each topic, and the second using model perplexity calculated on the cluster document--term matrix \parencite{mimno2011optimizing}. Because the two measures are expressed on different scales, mean topic coherence and perplexity were separately standardized across the four candidate models within each cluster. We then calculated the trade-off score as $z(\text{coherence})-z(\text{perplexity})$. The value of $T$ with the highest score was selected, and the corresponding model was refitted to all papers in that cluster.

We visualized the retained topics using term relevance across $\lambda\in\{0,0.25,0.5,0.75,1\}$ \parencite{sievert2014ldavis}. For term $w$ in topic $t$, relevance was calculated as $\lambda\log P(w\mid t)+(1-\lambda)\log[P(w\mid t)/P(w)]$. Lower values of $\lambda$ emphasize terms that are distinctive to a topic relative to the background vocabulary, whereas higher values emphasize terms with high probability within the topic. Each word cloud displayed up to 40 of the highest-relevance unigrams or bigrams. Multiple authors reviewed the word clouds across the five values of $\lambda$, the associated high-relevance terms, and representative papers from each topic. They summarized the recurring substantive themes and, through discussion, agreed on a domain name that captured the shared object of explanation across the topics within each cluster. These interpretations provided the three-domain codebook used in the subsequent author and LLM classification analyses.

\subsubsection{Author and LLM classifications and correspondence analyses}
\label{sec:5.1.4}

Two authors independently reviewed the title, abstract, and full text of each paper and applied the three-domain codebook while the K-means assignments and LLM classifications were concealed. One author left 13 papers unclassified at the initial stage. Paired initial classifications were therefore available for 185 papers, of which the authors classified 151 identically before discussion (81.62\%; Cohen's $\kappa=0.719$) \parencite{cohen1960coefficient}. After the independent stage, the authors jointly reviewed all discrepant and initially unclassified cases and resolved them through discussion, producing one consensus author classification for each of the 198 papers.

We then assessed whether the same distinctions could be applied by a non-author evaluator using the more limited information available in titles and abstracts. The three-domain codebook was provided to \texttt{gpt-5.6-sol}, which classified the 198 papers on 19 August 2026 using only each paper's title and abstract. The model received neither the K-means assignments nor the consensus author classifications. Each paper was assigned exactly one primary category and, when its content substantively overlapped another domain, at most one secondary category; the secondary category had to differ from the primary category, and \texttt{none} was returned when no secondary category was warranted. The records were processed in batches of eight with temperature set to 0, and the returned JSON was checked against the fixed category and output rules. The complete prompt and output specification are provided in the Supplementary Information.

Correspondence was calculated using the semantic relationship established in the LDA interpretation: Cluster~0 corresponded to \emph{LLM as Social Minds}, Cluster~1 to \emph{LLM Societies}, and Cluster~2 to \emph{LLM--Human Interactions}. We compared K-means with the consensus author classification, K-means with the LLM primary classification, and the consensus author classification with the LLM primary classification. For each comparison, we calculated the proportion of papers assigned to the same domain, Cohen's $\kappa$, and the adjusted Rand index (ARI) \parencite{cohen1960coefficient,hubert1985comparing}. To examine content spanning more than one domain, we additionally counted an LLM classification as corresponding with the K-means domain when that domain appeared as either the LLM-assigned primary or secondary category.

Finally, we examined whether classification disagreements were concentrated among the ten papers whose within-cluster consensus fell below 0.90 in the 90\% subsampling analysis. We compared these papers with the remaining 188 using two-sided Fisher's exact tests for author--K-means correspondence, LLM--K-means correspondence, and receipt of an LLM-assigned secondary category. We also compared the groups descriptively using each paper's cosine silhouette value and its relative centroid margin in the original embedding space. If $d_{i,1}$ and $d_{i,2}$ denote the distances from paper $i$ to its nearest and second-nearest K-means centroids, respectively, the relative margin was defined as $(d_{i,2}-d_{i,1})/d_{i,2}$. Smaller values indicate that the two nearest centroids were at more similar distances and therefore identify papers positioned closer to a cluster boundary.

\subsection{Study 2: Field-scale analysis}

\subsubsection{Database search, record processing, and eligibility screening}
\label{sec:5.2.1}

Figure~\ref{fig:study2-prisma} summarizes the identification, processing, and screening of records for the Study~2 corpus. To obtain broad coverage across the disciplines contributing to the social science of LLMs, we searched Semantic Scholar, OpenAlex, Scopus, PubMed, and Europe PMC for records published from 1 January 2020 through 23 July 2026. The search strategy used 41 prespecified terms comprising general descriptions (e.g., ``large language model'' and ``generative AI''), common abbreviations (e.g., ``LLM,'' ``GenAI,'' and ``GPT''), agent-related expressions (e.g., ``LLM agent,'' ``generative agent,'' and ``AI agent''), and the names of prominent model families (e.g., ``ChatGPT,'' ``Claude,'' ``Llama,'' ``Gemini,'' and ``DeepSeek''). These terms were applied to title, abstract, or keyword fields according to the syntax supported by each database. We imposed no topic, outcome, or disciplinary restriction at retrieval. The searches yielded 1,173,481 bibliographic records; the complete term list and database-specific search strings are provided in the Supplementary Information.

Retrieved records were converted to a common bibliographic structure, and their titles, abstracts, authors, publication years, sources, and persistent identifiers were standardized. Duplicate records were first consolidated within each database and then across databases using DOI, PMID, PMCID, arXiv, OpenAlex, and Semantic Scholar identifiers, standardized bibliographic metadata, and exact matches on normalized titles and abstracts. Within each duplicate group, we retained the record with the most complete abstract and bibliographic information. Records lacking a publication year or clearly identified as non-English were excluded. Language assessment combined database metadata with automated identification from titles and abstracts; inconclusive cases were retained. We also excluded records whose cleaned abstracts contained fewer than 200 characters. Where a preprint, conference paper, and journal article represented versions of the same study, the reports were linked as a study family and one representative record was retained, prioritizing abstract and bibliographic completeness.

We used \texttt{gpt-5.6-sol} as the screening model to assess the substantive eligibility of the resulting representative records from their titles and abstracts. The screening codebook defined the social science of LLMs by its object of explanation: eligible studies examined LLMs or LLM-based agents themselves as social objects, including their behavior, limitations, interactions, collective dynamics, or the ways in which people use, experience, and are affected by them. Purely technical studies concerned only with constructing or evaluating computational systems were excluded, as were studies that used LLMs solely as instruments for generating, processing, classifying, or analysing research materials without examining the models or their social relations substantively. Each record received an \texttt{include}, \texttt{exclude}, or \texttt{uncertain} label and a brief rationale based only on its title and abstract. The complete screening prompt and output schema are provided in the Supplementary Information.

We then refined the retained records to support the field-scale thematic analysis. Publication-status and publication-type screening retained formally published scholarly papers and excluded software, datasets, and other ineligible outputs. Titles were subsequently required to contain at least one of the 41 prespecified LLM terms, strengthening the alignment of the corpus with the target research object. Finally, to construct the academic-domain covariate used in the STM analysis, \texttt{gpt-5.6-sol} classified each publication venue using only the venue field. The model assigned a normalized venue name, a primary domain, up to two secondary domains, a venue type, and confidence and review indicators. Records associated with repositories or preprint platforms and records whose venues could not be classified reliably were excluded. These procedures yielded the final analytic corpus of 47,719 papers.

\subsubsection{STM estimation, model selection, and topic interpretation}
\label{sec:5.2.2}

We fitted structural topic models to the concatenated titles and abstracts of the 47,719 papers using the \texttt{stm} package in R \parencite{roberts2014structural,roberts2019stm}. Text was converted to lowercase, and English stop words, numbers, and punctuation were removed. Words shorter than three characters were excluded, and stemming was not applied. After tokenization, terms occurring fewer than ten times across the corpus were removed. All 47,719 papers remained in the modelled corpus, which contained a vocabulary of 14,791 terms.

For each candidate model, topic prevalence was specified as an additive function of publication period and the primary academic domain assigned to the publication venue. Publication period was represented by five levels: before 2023, 2023, 2024, 2025, and the partial 2026 observation period; 2023 served as the reference level. The venue-domain factor used AI/ML/NLP as its reference category. Models were fitted with spectral initialization and a maximum of 100 variational expectation--maximization iterations. The random seed for a model with $K$ topics was set to $20260819+K$. We considered $K=10$, 15, 20, 25, and 30. This broader range reflected the substantially larger and more heterogeneous field-scale corpus, for which a finer topic resolution was appropriate than the two-to-five-topic range examined within each of the much smaller Study~1 clusters.

We selected the topic resolution by jointly considering semantic coherence and exclusivity \parencite{mimno2011optimizing,roberts2019stm}. Semantic coherence is higher when the most probable words in a topic frequently occur together within the same papers, whereas exclusivity is higher when a topic's characteristic words are concentrated in that topic instead of being widely shared across topics. For each candidate value of $K$, we calculated both measures for every topic and then averaged them across topics. The five mean coherence values and five mean exclusivity values were separately standardized, and a trade-off score was calculated as $z(\overline{\text{coherence}}_K)+z(\overline{\text{exclusivity}}_K)$. The value of $K$ with the highest combined score was retained, yielding the 15-topic model used in the subsequent analyses.

To interpret the selected model, the authors jointly reviewed the twelve highest-ranked terms for each topic under four criteria \parencite{roberts2019stm}. Probability identified the terms most common within a topic; FREX balanced their frequency within the topic against their exclusivity to that topic; lift highlighted terms that were overrepresented relative to their overall frequency in the corpus; and score emphasized terms that were both probable within and distinctive of the topic. Considering these complementary views reduced reliance on either frequent but generic words or highly distinctive but rare words. The authors also reviewed the 15 papers with the highest estimated proportion of each topic. Based on these terms and representative papers, they discussed the substantive focus of each topic and agreed on a concise descriptive name for all 15 topics. After the topic names had been established, the authors examined each topic's object of explanation and assessed its relationship to the three domains identified in Study~1. Topics whose content was cross-cutting or was not anchored to one domain were retained outside the domain-specific mapping.

\subsubsection{Topic prevalence, citation, and venue-quality analyses}
\label{sec:5.2.3}

Publication period and the LLM-assigned academic domain of the publication venue entered the prevalence component during STM estimation, as described above. From the selected covariate-inclusive 15-topic model, each paper received a vector of expected topic proportions summing to one. As post-estimation summaries, we calculated the overall prevalence of each topic as its mean expected proportion across papers and obtained domain-level topic mass by summing the proportions of the topics mapped to each of the three domains. Topics~1 and 5 were retained as a separate broad cross-domain group. When comparing the relative composition of the three mapped domains, we divided each domain's expected mass by the combined mass of the 13 mapped topics, thereby excluding Topics~1 and 5 from the denominator. These summaries used the continuous topic proportions rather than an exclusive assignment of each paper to its largest topic.

We then summarized the temporal and venue-related relationships estimated within the prevalence component of the selected STM using \texttt{estimateEffect()} with \texttt{uncertainty = "Global"}. Because publication period and venue domain entered the model simultaneously, temporal contrasts accounted for venue discipline and venue contrasts accounted for publication period. The reference levels were 2023 and AI/ML/NLP venues, respectively. Topic-level coefficients were expressed as percentage-point differences in expected topic prevalence. Domain-level point estimates were calculated by summing the coefficients of the topics mapped to each domain, with Topics~1 and 5 summed separately as the broad cross-domain group. Reported topic-level $p$-values were obtained from the two-sided coefficient tests returned by \texttt{estimateEffect()}.

Citation counts already present in the bibliographic records were retained, while missing and recorded zero values were checked and completed using OpenAlex as the primary source, followed by Semantic Scholar and Scopus when required. Citation completion was finalized on 18 August 2026. To improve comparability across fields and publication cohorts, we ranked papers by citation count within their publication year and within their discipline--year group using average ranks for ties. The final citation percentile was calculated as $0.70\times$ the discipline--year percentile plus $0.30\times$ the publication-year percentile. The discipline--year rank received greater weight because citation practices vary substantially across fields, while the year-specific rank retained information on a paper's position within its publication cohort overall. The discipline metadata and its processing are documented in the Supplementary Information. Papers at or above the 99th citation percentile were treated as a high-citation subset for the descriptive comparisons.

We constructed LLM-estimated venue-standing indicators separately from the venue-domain classification used in the STM. Using only the normalized venue name, venue type, and assigned academic domain, \texttt{gpt-5.6-sol} estimated a JIF-like field percentile for journals, a CORE-style rank for conferences and proceedings series, and a publisher or book-series tier for book series. Article titles, abstracts, identifiers, and citation counts were not provided during this assessment. Deterministic rules then converted the estimates to a common 0--100 scale: journal estimates retained their percentile value; conference ranks A*, A, B, and C received scores of 100, 85, 70, and 50; and book-series tiers A, B, and C received scores of 90, 70, and 50. These indicators were used for descriptive stratification and were not treated as official JIF, CORE, or publisher rankings. Records without an applicable venue-standing score remained in the full-corpus analyses but were omitted from subsets defined by a venue-score threshold. The most selective comparisons combined a citation percentile of at least 99 with a venue score of 100 for conferences or at least 90 for journals. Within every subset, domain shares were calculated from mean expected topic mass and normalized over the 13 domain-mapped topics. The complete venue-assessment prompt, output schema, and processing details are provided in the Supplementary Information.

\subsubsection{Full-text sensitivity and curated-corpus coverage analyses}
\label{sec:5.2.4}

The reliability of title-and-abstract eligibility decisions was assessed using a sample of 300 records for which publicly accessible, machine-readable full text could be obtained. Documents were retrieved in PDF, HTML, or XML format. Text was extracted page by page from PDFs, while the main textual content of HTML and XML documents was retained after non-content elements had been removed. All 300 sampled records yielded non-empty normalized extracted text. All parser-readable text from each article was included in the analysis and divided in its original order into consecutive, non-overlapping chunks. This produced 593 chunks containing 17,835,674 characters. Exact reconstruction from the chunks was verified for every article, confirming that no extracted text was omitted during chunking.

Using \texttt{gpt-5.6-sol} with a temperature of 0, the same substantive eligibility criteria used in title-and-abstract screening were first applied to all 593 chunks to extract inclusion, exclusion, and uncertainty evidence. All chunk-level evidence for each article was then synthesized into one final article-level decision. An article was classified as `supports` when its complete extracted text supported the original inclusion decision, `contradicts` when it contained clear evidence meeting an exclusion criterion, or `indeterminate` when the complete evidence remained insufficient or ambiguous. This analysis examined whether eligibility decisions based on titles and abstracts remained supported after all parser-readable article text was considered. It served as a sensitivity check and did not alter the composition of the main analytic corpus. 

We assessed the coverage of the Study~1 corpus by matching its 198 papers against the final Study~2 analytic corpus using DOIs and standardized bibliographic information, primarily titles and abstracts. We recorded whether each paper was recovered and summarized coverage overall and separately for the three author-assigned domains. For the recovered papers, we linked the citation percentiles constructed for Study~2 and described their position in the broader citation distribution. We also compared their original $K=3$ K-means assignments with the Study~2 STM results. Each paper's dominant STM topic was translated using the topic-to-domain mapping established for the 15-topic model; papers dominated by Topic~1 or Topic~5 remained cross-cutting and were excluded from the three-domain correspondence metrics. Correspondence was summarized using the proportion of matching assignments, Cohen's $\kappa$, Macro-F1, ARI, and NMI. Further details of the record-matching procedure are provided in the Supplementary Information.

We then conducted a sensitivity analysis restricted to recovered Study~1 papers at or above the 50th citation percentile in the Study~2 corpus. This criterion yielded 148 papers. We repeated the Study~1 text-analysis pipeline on this subset: titles and abstracts were encoded with MPNet, K-means models were fitted for $K=2$--9, and the candidate solutions were evaluated using the same internal clustering indices. For the $K=3$ solution, we repeated the within-cluster LDA analysis and interpreted the clusters using their topic terms and representative papers. We compared the resulting cluster assignments with the authors' full-text classifications using correspondence, ARI, and NMI. This analysis examined whether the three-domain structure remained visible after restricting the curated corpus to papers that were both recovered by the database-based search and relatively highly cited within the broader literature.

\section{Study 1 LLM classification prompt}

After the authors had interpreted the $K=3$ K-means solution using the within-cluster LDA results, topic terms, word clouds, and representative papers, they formulated the three-domain codebook used in the classification analyses. The LLM-based evaluator applied this fixed codebook using only the title and abstract of each paper; it received neither the individual K-means assignments nor the consensus author classifications. The 198 records were processed in batches of eight using \texttt{gpt-5.6-sol} with temperature set to 0, and the returned JSON was checked against the category and output rules. The description of the framework as ``fixed'' in the prompt indicates that its definitions were not modified in response to the LLM classifications.

The prompt used the abbreviated output label \texttt{Social Minds}, reported in the manuscript as \emph{LLM as Social Minds}, and \texttt{Human-LLM Interactions}, reported typographically as \emph{LLM--Human Interactions}. These naming changes did not alter any paper's classification. Each paper was required to receive one primary category and could receive at most one different secondary category; \texttt{none} indicated that no secondary category was warranted. The complete prompt and output specification were:

\begin{quote}
\begin{lstlisting}[style=promptjson]
You are applying a fixed, author-developed conceptual framework
after the analytical corpus has already been selected. This is categorisation,
not eligibility screening. Use only the supplied title and abstract. Do not
exclude a record, do not create a fourth category, and do not infer evidence
absent from the text.

Assign exactly one primary category:

1. Social Minds: the central object is an individual LLM or LLM-based agent
   considered as a social, cognitive, moral, behavioural, or identity-bearing
   entity; examples include personality, values, mental-state attribution,
   social cognition, deception, individual agency, or model behaviour as a
   social object.

2. LLM Societies: the central object is a collective, multi-agent, networked,
   institutional, governance, or emergent social system involving LLMs or
   LLM-based agents; examples include agent societies, coordination, collective
   behaviour, institutions, governance among agents, or system-level dynamics.

3. Human-LLM Interactions: the central object is the relation between people,
   groups, organisations, or institutions and LLMs; examples include adoption,
   trust, perception, communication, work, education, health, policy, social
   relationships, or human-facing deployment and governance.

When a record overlaps categories, choose the category representing its central
research question, then record at most one secondary category. Confidence
describes how explicitly the abstract supports the choice; it is not a human
inter-coder reliability measure.

Return JSON only in exactly this form:
{"labels":[[0,"PRIMARY_CATEGORY","SECONDARY_OR_NONE","high|moderate|low","brief evidence"], ...]}

Every input index must appear exactly once and in original order. PRIMARY_CATEGORY
and SECONDARY_OR_NONE must be one of: Social Minds; LLM Societies;
Human-LLM Interactions; none. A secondary category must differ from the primary.
\end{lstlisting}
\end{quote}

\section{Search strategy}

The target search window extended from 1 January 2020 through 23 July 2026 and covered five bibliographic sources: Semantic Scholar, OpenAlex, Scopus, PubMed, and Europe PMC. Exact day-level date bounds were applied where supported by the source interface. Scopus represented the date restriction at publication-year resolution and was therefore queried for publication years 2020--2026. No downstream topic, cluster, discipline, or outcome terms were added to the search.

\subsection{Core object vocabulary}

This vocabulary was used in the initial literature search to identify publications whose stated object was an LLM, a named generative-AI system, or an LLM-based agent. It defined the broad retrieval universe before bibliographic normalization, deduplication, language and abstract-availability screening, and study-family deduplication. The final vocabulary contained 41 LLM and generative-AI object terms. The Boolean expression was:

\begin{quote}
\begin{lstlisting}[style=promptjson]
"large language model" OR "large language models" OR
"foundation language model" OR "foundation language models" OR
"generative language model" OR "generative AI" OR
"generative artificial intelligence" OR
"generative pre-trained transformer" OR
"generative pretrained transformer" OR "LLM-based" OR
"LLM agent" OR "language-model agent" OR "GPT agent" OR
"generative agent" OR "generative agents" OR "AI agent" OR
"AI agents" OR "artificial intelligence agent" OR
"artificial intelligence agents" OR "LLM" OR "LLMs" OR
"GenAI" OR "Gen AI" OR "GPT" OR "foundation model" OR
"foundation models" OR "conversational AI" OR "ChatGPT" OR
"GPT-3" OR "GPT-3.5" OR "GPT-4" OR "GPT-4o" OR "Claude" OR
"Llama" OR "Gemini" OR "Bard" OR "Mistral" OR "Mixtral" OR
"Qwen" OR "DeepSeek" OR "GPT-5"
\end{lstlisting}
\end{quote}

\subsection{Source-specific implementation}

The same 41-term vocabulary was used across all five sources. Search fields, Boolean operators, date filters, and pagination procedures were adapted to the query interface of each source. In the schematic expressions below, the ellipsis represents all intervening terms in the complete vocabulary reported above.

\paragraph{Semantic Scholar.}
The terms were submitted to the Semantic Scholar bulk-search interface as quoted expressions separated by the platform-specific disjunction operator ``|''. Hyphens were converted to spaces for compatibility with its query parser. Retrieval was divided into publication-year intervals, with an exact date interval used for the final partial year.

\begin{quote}
\begin{lstlisting}[style=promptjson]
"large language model" | "large language models" | ... | "GPT 5"

year=2020
...
year=2025
publicationDateOrYear=2026-01-01:2026-07-23
\end{lstlisting}
\end{quote}

\paragraph{OpenAlex.}
Each term was searched within the combined title-and-abstract index. The complete term disjunction was combined with exact lower and upper publication-date bounds.

\begin{quote}
\begin{lstlisting}[style=promptjson]
title_and_abstract.search:
("large language model" OR "large language models" OR ... OR "GPT-5"),
from_publication_date:2020-01-01,
to_publication_date:2026-07-23
\end{lstlisting}
\end{quote}

\paragraph{Scopus.}
The complete vocabulary was applied to titles, abstracts, and author-supplied keywords. Because the Scopus query interface was configured at publication-year resolution, the search included records assigned to publication years 2020--2026.

\begin{quote}
\begin{lstlisting}[style=promptjson]
TITLE-ABS-KEY(
  "large language model" OR "large language models" OR
  ... OR "GPT-5"
)
AND PUBYEAR > 2019
AND PUBYEAR < 2027
\end{lstlisting}
\end{quote}

\paragraph{PubMed.}
Each term was individually restricted to the title and abstract fields. The term disjunction was combined with an exact publication-date interval. For retrieval, the overall interval was divided into non-overlapping calendar-month segments.

\begin{quote}
\begin{lstlisting}[style=promptjson]
(
  "large language model"[Title/Abstract] OR
  "large language models"[Title/Abstract] OR
  ... OR
  "GPT-5"[Title/Abstract]
)
AND
(
  "2020-01-01"[Date - Publication] :
  "2026-07-23"[Date - Publication]
)
\end{lstlisting}
\end{quote}

\paragraph{Europe PMC.}
Each term was restricted to title and abstract content. The resulting disjunction was combined with an exact first-publication-date interval.

\begin{quote}
\begin{lstlisting}[style=promptjson]
(
  TITLE_ABS:"large language model" OR
  TITLE_ABS:"large language models" OR
  ... OR
  TITLE_ABS:"GPT-5"
)
AND FIRST_PDATE:[2020-01-01 TO 2026-07-23]
\end{lstlisting}
\end{quote}

Source-specific query construction, pagination, and response parsing were implemented in the archived retrieval code.

\section{Title-and-abstract screening prompt}

This prompt was applied after bibliographic normalization, deduplication, language and abstract-availability screening, and representative-record selection had produced 478,405 records with the required screening fields. Each model input supplied a stable record ID, title, and abstract. The task distinguished studies examining LLMs as objects from studies using LLMs solely as research instruments, and returned an \texttt{include}, \texttt{exclude}, or \texttt{uncertain} label with a concise reason. The screening retained 146,682 records for subsequent publication-eligibility and title review. The category-neutral instruction was:

\begin{quote}
You are an expert reviewer screening academic publications for a systematic review of the social science of large language models (LLMs). You will receive the title and abstract of one or more candidate papers. Classify each paper based only on the information explicitly available in its title and abstract.

Please include papers that are about social science of LLMs, which examines LLMs or LLM-based agents themselves as social objects.
EXCLUDE a paper only when the title and abstract clearly show that:
\begin{enumerate}
\item it is purely technical or engineering research focused only on constructing, improving, or evaluating a computational system, or its task performance;
\item the LLM or LLM-based agents are used solely as research instruments--whether to generate, process, classify, or analyze data or materials, or to substitute for human respondents or actors--without substantively examining LLMs' behavior, limitations, interactions, or collective dynamics, or how people use, experience, or are affected by them.
\end{enumerate}

IMPORTANT BOUNDARIES:
\begin{enumerate}
\item Do not exclude a study merely because people use an LLM as a tool. Include it if the study investigates the experience, adoption, relationship, or consequences of that use.
\item When the abstract is missing or too vague to determine the LLM's role, use "uncertain" rather than guessing.
\end{enumerate}

Before classifying the paper as "include," "exclude," or "uncertain," please give your reasoning. Let's think step by step.

Return JSON only, exactly in this form:
\begin{lstlisting}[style=promptjson]
{
  "results": [
    {
      "id": "record ID",
      "brief_reason": "One concise sentence summarizing your reasoning.",
      "label": "include|exclude|uncertain"
    }
  ]
}
\end{lstlisting}

Every input ID must appear exactly once, unchanged, and in the original order. Do not include Markdown, explanations, or text outside the JSON.
\end{quote}

\section{Venue-domain classification prompt}

This prompt was applied after publication-eligibility screening and title review had retained 54,159 records. It classified each unique publication venue by its established scholarly scope and generated standardized venue names, primary and secondary domains, confidence levels, review flags, and venue types. The classification concerned the venue itself, not the title, abstract, findings, or topic of an individual paper. Records assigned \texttt{REPOSITORY\_\allowbreak PLATFORM} or \texttt{OTHER\_UNCLEAR} were then excluded, leaving the final analytical cohort of 47,719 records. The instruction and required structured output were:

\begin{quote}
You are an expert in bibliometrics and scholarly field classification.

Your task is to classify the academic scope of a publication venue. Classify the venue itself--not the title, abstract, findings, or topic of any individual article.

The resulting classification will be used as metadata in a Structural Topic Model (STM).

Select exactly one `primary\_domain` from the following categories:

\begin{enumerate}
\item \texttt{AI\_ML\_NLP}: Artificial intelligence, machine learning, natural language processing, computational linguistics, speech technology, computer vision, data mining, and AI-centered robotics or intelligent systems.
\item \texttt{HCI\_HUMAN\_CENTERED\_COMPUTING}: Human-computer interaction, human-centered computing, user experience, interaction design, user interfaces, computer-supported cooperative work, social computing, ubiquitous computing, and interactive systems.
\item \texttt{COMPUTER\_SCIENCE\_SYSTEMS}: General computer science, software engineering, programming languages, databases and data management, operating and distributed systems, networks, cybersecurity, information retrieval infrastructure, computing theory, and information technology.
\item \texttt{ENGINEERING\_TECHNOLOGY}: Electrical and electronic engineering, control and automation, embedded systems, telecommunications, signal processing, consumer electronics, hardware, industrial engineering, and engineering-centered robotics.
\item \texttt{EDUCATION}: Education, teaching, learning sciences, educational technology, higher education, teacher education, curriculum, and assessment.
\item \texttt{MEDICINE\_HEALTH}: Clinical medicine, public health, nursing, medical education, healthcare services, biomedicine, and health informatics.
\item \texttt{BUSINESS\_MANAGEMENT}: Business, management, organizational studies, marketing, management information systems, finance, accounting, and industry studies.
\item \texttt{LAW\_GOVERNANCE}: Law, legal studies, justice, public policy, regulation, digital rights, and institutional governance.
\item \texttt{PSYCHOLOGY\_BEHAVIORAL}: Psychology, cognitive science, behavioral science, personality, emotion, and mental health.
\item \texttt{SOCIAL\_SCIENCES\_HUMANITIES\_COMMUNICATION}: Sociology, communication, media studies, political science, philosophy, ethics, linguistics, and other humanities and social sciences.
\item \texttt{MULTIDISCIPLINARY\_GENERAL}: Genuinely multidisciplinary or general-purpose scholarly venues without a clearly dominant disciplinary scope. Do not use this category merely because the venue is difficult to classify.
\item \texttt{REPOSITORY\_\allowbreak PLATFORM}
Repositories, preprint servers, and hosting platforms such as Zenodo, Figshare, OSF, SSRN, and arXiv.The platform itself does not represent an academic discipline.
\item \texttt{OTHER\_UNCLEAR}: Venues with insufficient information, ambiguous names, generic proceedings or book series, or venues that cannot be classified reliably. Use this category instead of guessing.
\end{enumerate}

Classification rules:

\begin{itemize}
\item Classify the venue according to its established academic scope, using only the venue name and your reliable knowledge of that venue.
\item Select the most specific domain that best represents the venue's primary scholarly community.
\item Use \texttt{HCI\_HUMAN\_CENTERED\_COMPUTING} when the venue community primarily studies people's use of, interaction with, experience of, or collaborative work through computing, even when AI is involved.
\item Use \texttt{AI\_ML\_NLP} for venues primarily centered on intelligent models, learning algorithms, language or speech processing, computer vision, or AI methods.
\item Use \texttt{COMPUTER\_SCIENCE\_SYSTEMS} for computing venues centered on software, systems, networks, databases, security, infrastructure, theory, or general computer science without a dominant AI or HCI scope.
\item Use \texttt{ENGINEERING\_TECHNOLOGY} for venues primarily centered on electronics, hardware, control, automation, telecommunications, signal processing, embedded systems, or engineering applications.
\item Do not classify a venue as \texttt{AI\_ML\_NLP} merely because an individual article published there concerns an LLM.
\item Treat computational linguistics and NLP venues as \texttt{AI\_ML\_NLP}, but general linguistics venues as \texttt{SOCIAL\_SCIENCES\_HUMANITIES\_COMMUNICATION}.
\item Treat management information systems venues as \texttt{BUSINESS\_MANAGEMENT} when their primary scholarly community is business or management.
\item Treat medical and health informatics venues as \texttt{MEDICINE\_HEALTH} when healthcare is their primary application domain.
\item Use \texttt{MULTIDISCIPLINARY\_GENERAL} only when the venue genuinely covers multiple fields without a dominant discipline.
\item Classify repositories, preprint servers, and hosting services as \texttt{REPOSITORY\_\allowbreak PLATFORM}.
\item If the venue name is ambiguous, generic, or insufficient for a reliable decision, use \texttt{OTHER\_UNCLEAR} and set \texttt{needs\_review} to true.
\item Do not guess or invent information about an unfamiliar venue.
\item Assign exactly one \texttt{primary\_domain}.
\item Assign zero to two \texttt{secondary\_domains}. They must differ from the primary domain.
\end{itemize}

Confidence levels:

\begin{itemize}
\item \texttt{HIGH}: The venue is well known or its name clearly indicates a dominant academic domain.
\item \texttt{MODERATE}: The classification is likely correct, but the venue is interdisciplinary, abbreviated, or somewhat ambiguous.
\item \texttt{LOW}: There is insufficient evidence for a reliable classification. Prefer \texttt{OTHER\_UNCLEAR} rather than guessing.
\end{itemize}

Set `needs\_review` to true for all LOW-confidence classifications. It may also be true for MODERATE-confidence boundary cases.

The `rationale` must state the classification evidence in no more than 25 words.

The input may contain multiple venues. Return one result for every input venue, preserve the original order, and copy `input\_id` exactly.

Return valid JSON only. Do not return Markdown or additional explanation.

Batch input format:
\begin{lstlisting}[style=promptjson]
{
  "venues": [
    {"input_id": "V001", "venue": "Venue name"},
    {"input_id": "V002", "venue": "Another venue name"}
  ]
}
\end{lstlisting}

Required output format:
\begin{lstlisting}[style=promptjson]
{
  "results": [
    {
      "input_id": "V001",
      "venue": "Original venue name",
      "normalized_venue": "Normalized venue name",
      "primary_domain": "ONE_OF_THE_ALLOWED_CATEGORIES",
      "secondary_domains": [],
      "confidence": "HIGH|MODERATE|LOW",
      "needs_review": false,
      "venue_kind": "journal|conference|proceedings_series|book_series|repository|other",
      "rationale": "Brief evidence-based justification"
    },
    {
      "input_id": "V002",
      "venue": "Another venue name",
      "normalized_venue": "Normalized venue name",
      "primary_domain": "ONE_OF_THE_ALLOWED_CATEGORIES",
      "secondary_domains": [],
      "confidence": "HIGH|MODERATE|LOW",
      "needs_review": true,
      "venue_kind": "journal|conference|proceedings_series|book_series|repository|other",
      "rationale": "Brief evidence-based justification"
    }
  ]
}
\end{lstlisting}
\end{quote}

\section{Venue-standing annotation prompt}

After the final analytical cohort of 47,719 records had been established, this prompt was used to add a provisional measure of scholarly standing for each publication venue. The annotation used only the supplied venue identity, venue type, and domain. It returned a numeric field-percentile estimate for journals and an ordinal category for conferences and book chapters. These values were analytical estimates rather than official JIF, CORE, or publisher rankings. They were added as metadata and did not exclude records or alter the analytical cohort.

\textbf{System prompt}

\begin{quote}
You are a cautious academic venue metadata annotator. Return valid JSON only.
\end{quote}

\textbf{User prompt}

\begin{quote}
Assess the likely scholarly standing of each publication venue using only the supplied venue identity, venue type, and domain. This is a provisional estimate, not an official JIF, CORE, or publisher ranking. Do not claim to have looked up a metric.

Return one JSON object with an \texttt{items} array and exactly one item for every input ID. Each item must have: \texttt{venue\_record\_key}, \texttt{metric\_value}, and \texttt{confidence} (\texttt{low}, \texttt{medium}, or \texttt{high}). For journals, \texttt{metric\_value} must be a numeric JIF-like field-percentile estimate from 0 to 100. For conferences, \texttt{metric\_value} must be one of \texttt{A*}, \texttt{A}, \texttt{B}, or \texttt{C}. For book chapters, \texttt{metric\_value} must be one of \texttt{A}, \texttt{B}, or \texttt{C}.

Do not include explanations or extra fields. Use conservative values and low confidence when a venue is generic, ambiguous, or repository-like. Never use citation counts from the input because none are supplied.

\texttt{Input JSON:}
\end{quote}

\section{Complete-text sensitivity prompts}

These prompts were used after the main analytical cohort had been
finalized. They provided an independent sensitivity check of the original
title-and-abstract eligibility decisions using a fixed sample of 300
records with publicly accessible, machine-readable full text. All
parser-readable text from each article was retained and divided in its
original order into consecutive, non-overlapping chunks. The assessment
was conducted in two stages. First, evidence was extracted separately
from every sequential chunk without making a final eligibility decision.
Second, the evidence from all chunks of an article was combined to produce
a single article-level signal of \texttt{supports},
\texttt{contradicts}, or \texttt{indeterminate}. This sensitivity
assessment did not redefine membership in the 47,719-record cohort.

\subsection{Chunk-level evidence-extraction prompt}

The following instruction was applied separately to every sequential
chunk of each article:

\begin{quote}
You are reviewing one sequential chunk from the machine-extracted
complete text of an academic article for a systematic review of the
social science of large language models (LLMs). Every chunk of the
article will be reviewed separately. Your task at this stage is to
extract evidence, not to make the final article-level eligibility
decision.

The review includes research that examines LLMs or LLM-based agents
themselves as social objects. Relevant evidence may concern their
behavior, limitations, interactions, collective dynamics, or how people
use, experience, adopt, relate to, or are affected by them.

Potential exclusion evidence is limited to the following:
\begin{enumerate}
\item the article is purely technical or engineering research focused
only on constructing, improving, or evaluating a computational system or
its task performance;
\item LLMs or LLM-based agents are used solely as research instruments
to generate, process, classify, or analyze data or materials, or to
substitute for human respondents or actors, without substantively
examining their behavior, limitations, interactions, collective dynamics,
use, experience, adoption, relationships, or consequences.
\end{enumerate}

Do not treat the use of an LLM as a tool as exclusion evidence when the
study investigates the experience, adoption, relationship, or
consequences of that use. Do not infer claims absent from this chunk. A
chunk may contain no decisive evidence.

Return JSON only with these fields:

\begin{verbatim}
{"family_id":"unchanged ID","chunk_index":1,
"inclusion_evidence":"concise evidence or empty string",
"exclusion_evidence":"concise evidence or empty string",
"study_role":"social_object|technical_system|
research_instrument|mixed|unclear",
"evidence_strength":"decisive|supporting|none"}
\end{verbatim}
\end{quote}

\subsection{Article-level decision prompt}

After all chunks of an article had been reviewed, the resulting
chunk-level evidence was supplied to the following instruction:

\begin{quote}
You are making an article-level sensitivity decision for a systematic
review of the social science of large language models (LLMs). You will
receive an article title and evidence extracted from every sequential
chunk of its machine-extracted complete text. Apply the criteria below
to the combined evidence from all chunks.

Include research that examines LLMs or LLM-based agents themselves as
social objects.

EXCLUDE an article only when the complete set of chunk evidence clearly
shows that:
\begin{enumerate}
\item it is purely technical or engineering research focused only on
constructing, improving, or evaluating a computational system or its
task performance; or
\item LLMs or LLM-based agents are used solely as research instruments
to generate, process, classify, or analyze data or materials, or to
substitute for human respondents or actors, without substantively
examining their behavior, limitations, interactions, collective dynamics,
use, experience, adoption, relationships, or consequences.
\end{enumerate}

Do not exclude a study merely because people use an LLM as a tool. It
remains eligible if it investigates the experience, adoption,
relationship, or consequences of that use. Return
\texttt{indeterminate} only when the combined evidence from all
successfully reviewed chunks remains insufficient or genuinely
ambiguous. This sensitivity decision does not itself redefine cohort
membership.

Return JSON only with these fields:

\begin{verbatim}
{"family_id":"unchanged ID",
"signal":"supports|contradicts|indeterminate",
"reason_code":"concise code",
"evidence":"one concise article-level explanation grounded
in the supplied chunk evidence"}
\end{verbatim}
\end{quote}

\clearpage
\section{Supplementary figures and tables}
\begin{figure}[H]
\centering
\includegraphics[width=\textwidth,height=0.76\textheight,keepaspectratio]{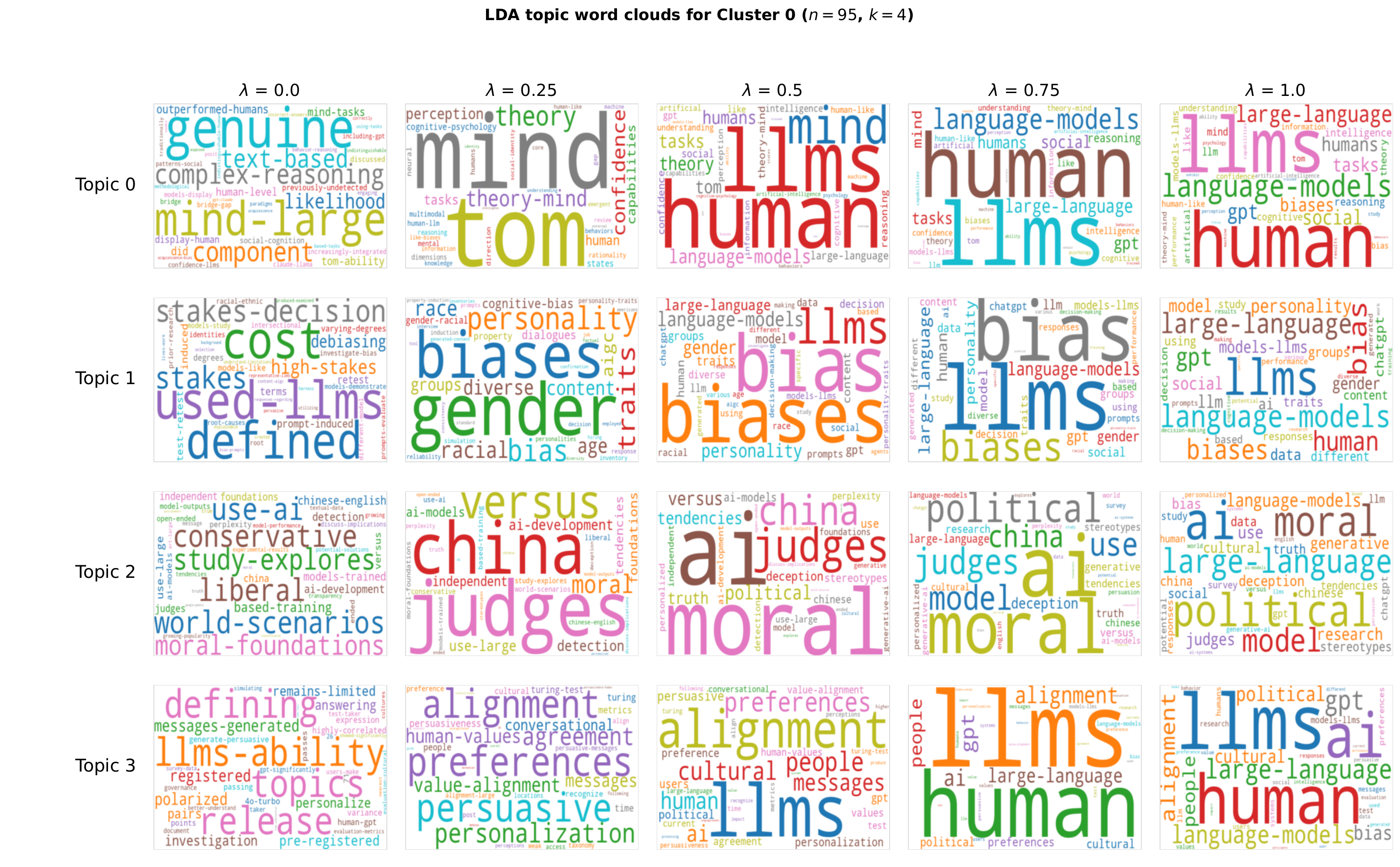}
\caption{LDA topic word clouds for Cluster~0 ($n=95$). Rows show the four retained topics and columns show term relevance at $\lambda=0$, 0.25, 0.5, 0.75, and 1. The topics encompass political and moral judgment, theory of mind and human-like reasoning, personality and demographic bias, and preference alignment and persuasion, supporting the interpretation of Cluster~0 as \emph{LLM as Social Minds}.}
\label{fig:cluster0}
\end{figure}

Figure~\ref{fig:cluster0} provides supporting diagnostics and descriptive results.

\begin{figure}[H]
\centering
\includegraphics[width=\textwidth,height=0.76\textheight,keepaspectratio]{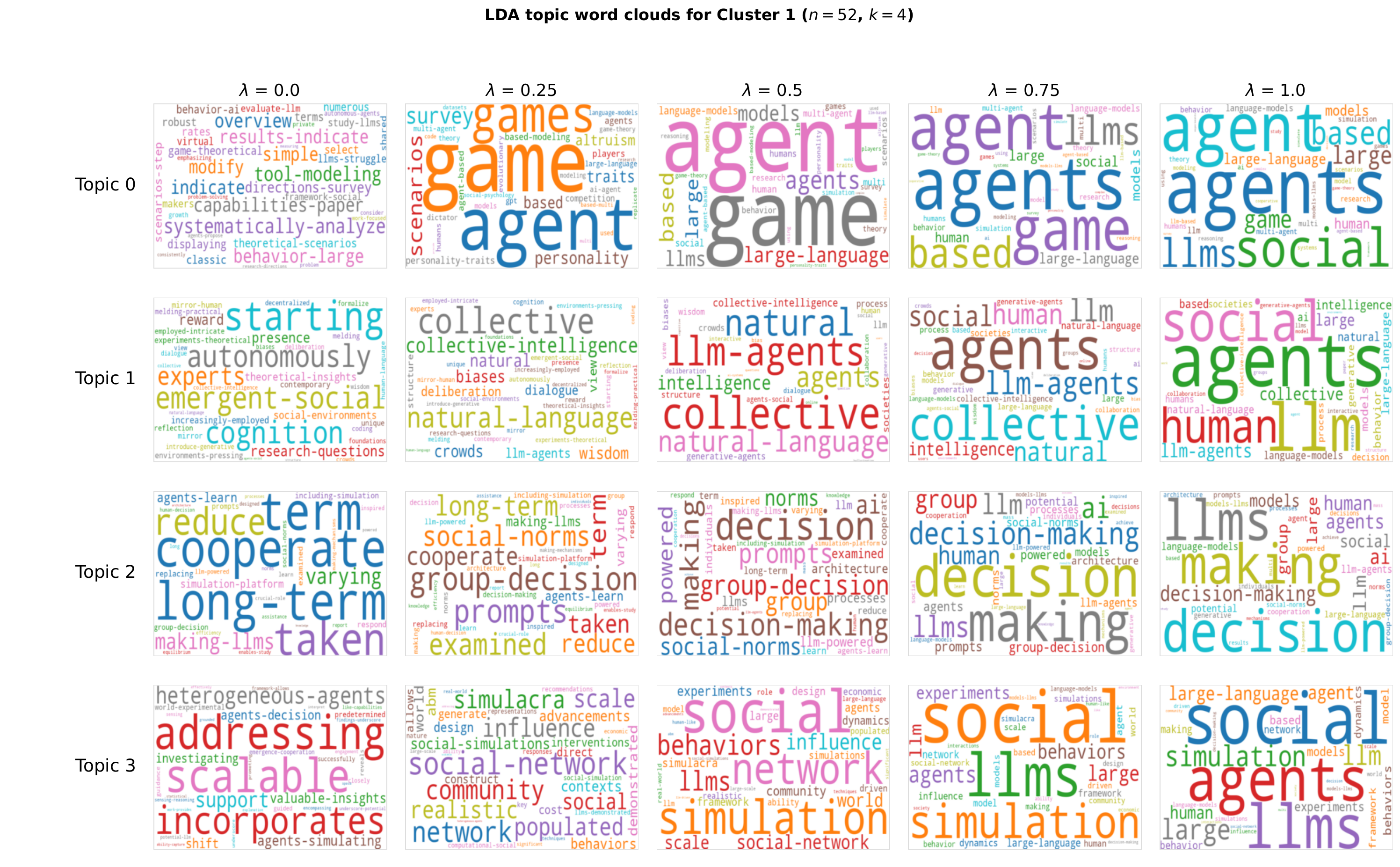}
\caption{LDA topic word clouds for Cluster~1 ($n=52$). Rows show the four retained topics and columns show term relevance at $\lambda=0$, 0.25, 0.5, 0.75, and 1. The topics encompass group decision-making and cooperation, collective intelligence and emergent dynamics, agent games and behavioral traits, and social networks and population-scale simulation, supporting the interpretation of Cluster~1 as \emph{LLM Societies}.}
\label{fig:cluster1}
\end{figure}

\clearpage
Figure~\ref{fig:cluster1} provides supporting diagnostics and descriptive results.

\begin{figure}[H]
\centering
\includegraphics[width=\textwidth,height=0.76\textheight,keepaspectratio]{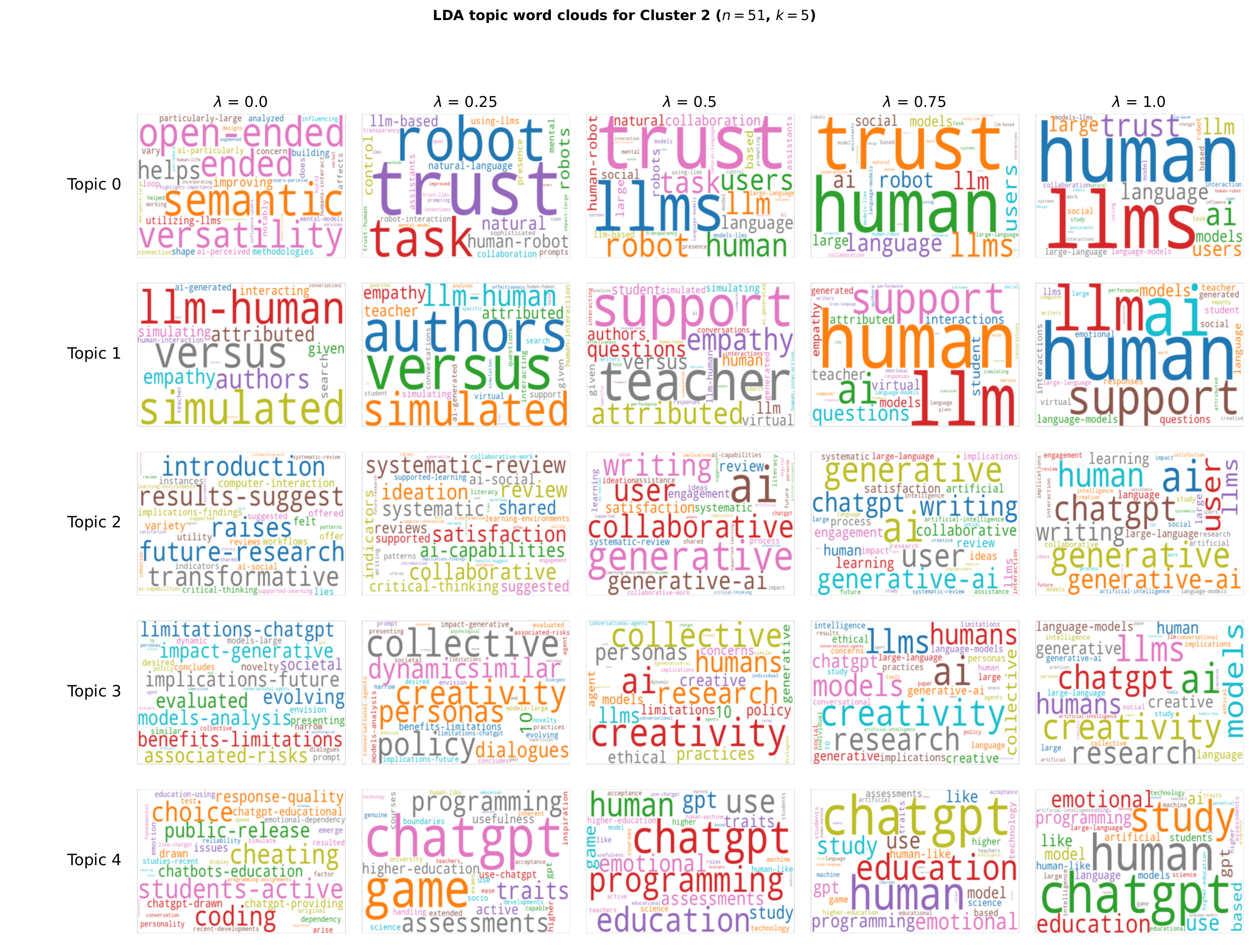}
\caption{LDA topic word clouds for Cluster~2 ($n=51$). Rows show the five retained topics, and columns show term relevance at $\lambda=0$, 0.25, 0.5, 0.75, and 1. The topics encompass generative work and collaboration, creativity and perceptions of LLMs, simulated empathy and support, trust and LLM adoption, and LLMs in education and assessment, supporting the interpretation of Cluster~2 as \emph{LLM--Human Interactions}.}
\label{fig:cluster2}
\end{figure}

\clearpage
Figure~\ref{fig:cluster2} provides supporting diagnostics and descriptive results.

\begin{figure}[H]
    \centering
    \includegraphics[width=\textwidth]{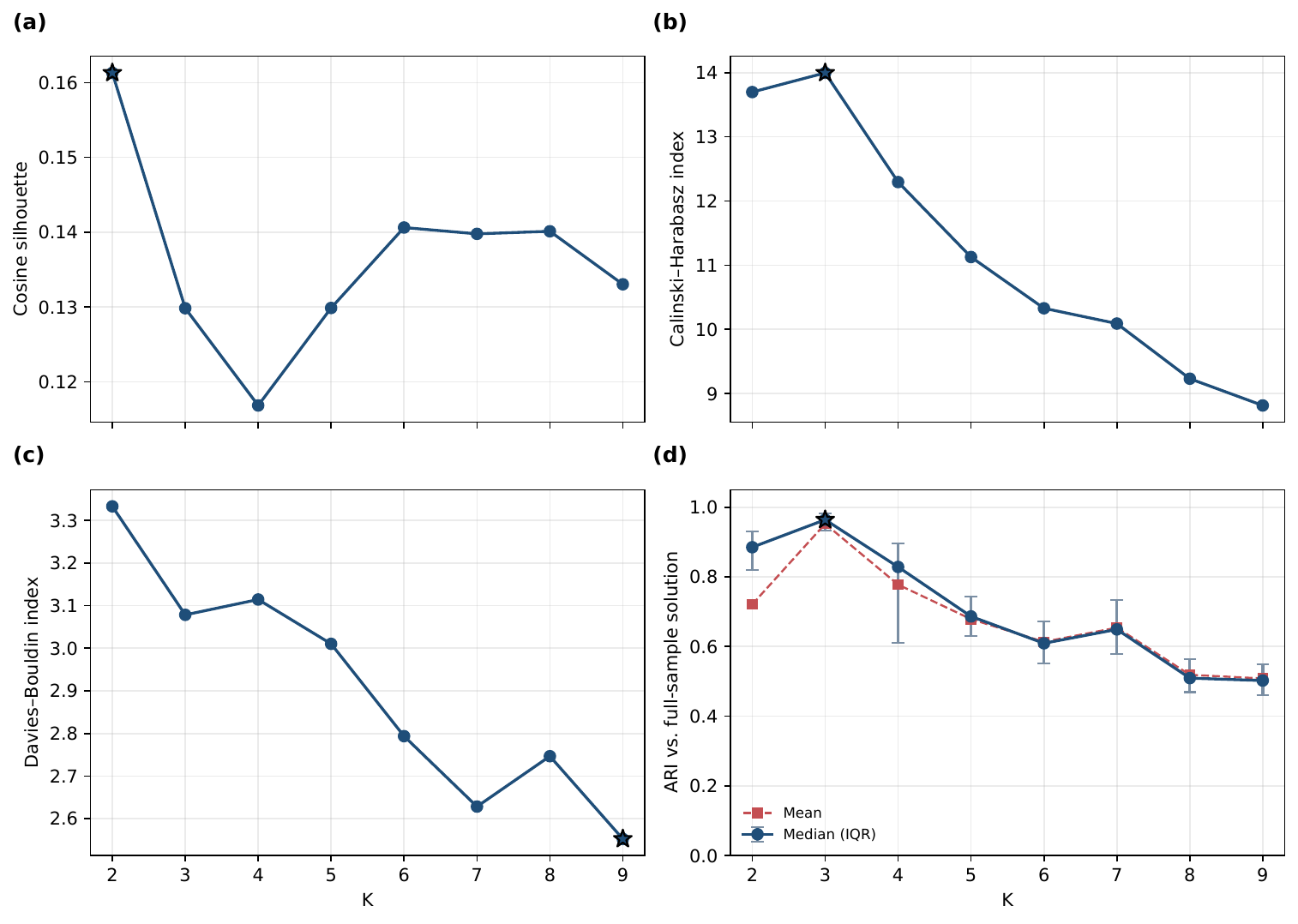}
    \caption{Selection and resampling stability of K-means solutions across $K=2$--$9$. Panels show (a) cosine silhouette, (b) the Calinski--Harabasz index, (c) the Davies--Bouldin index, and (d) agreement between each 90\% subsample solution and its corresponding full-sample solution across 100 draws, measured by the adjusted Rand index (ARI). In panel (d), red squares show the mean and blue circles show the median, with error bars indicating the interquartile range. Higher values indicate better performance in panels (a), (b), and (d), whereas lower values indicate better performance in panel (c). Stars mark the best-performing $K$ according to the criterion displayed in each panel.}
    \label{fig:k-selection-stability}
\end{figure}

\clearpage
Figure~\ref{fig:k-selection-stability} provides supporting diagnostics and descriptive results.

\clearpage

\begin{figure}[H]
\centering
\includegraphics[width=0.86\textwidth]{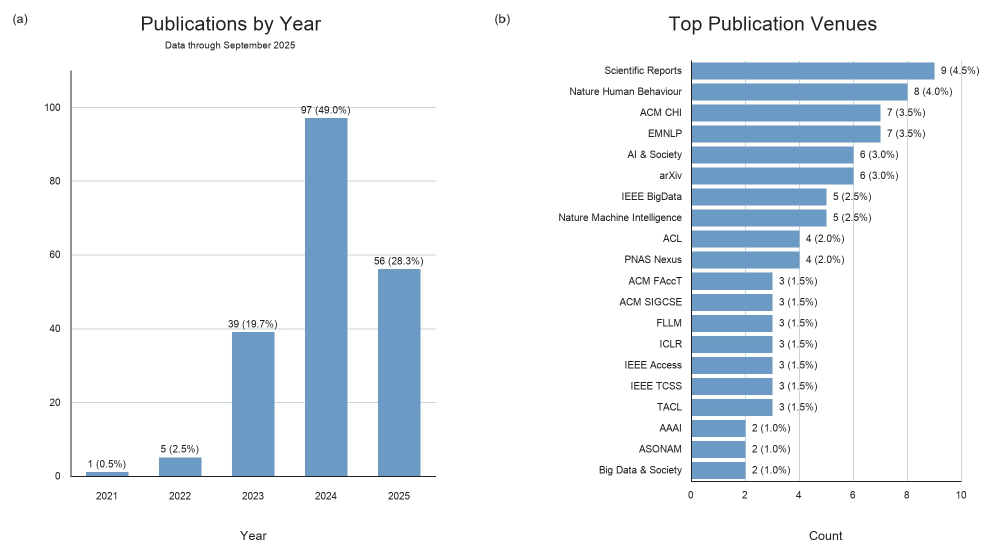}
\caption{\textbf{Corpus overview.} \textbf{(a)} Publications by year.
\textbf{(b)} Top 20 publication venues. Venue names were harmonized across
capitalization variants and annual editions of the same conference series;
ties were ordered alphabetically.}
\label{fig:descriptive-panels}
\end{figure}

Figure~\ref{fig:descriptive-panels} provides supporting diagnostics and descriptive results.

\begin{figure}[H]
    \centering
    \includegraphics[width=\textwidth]{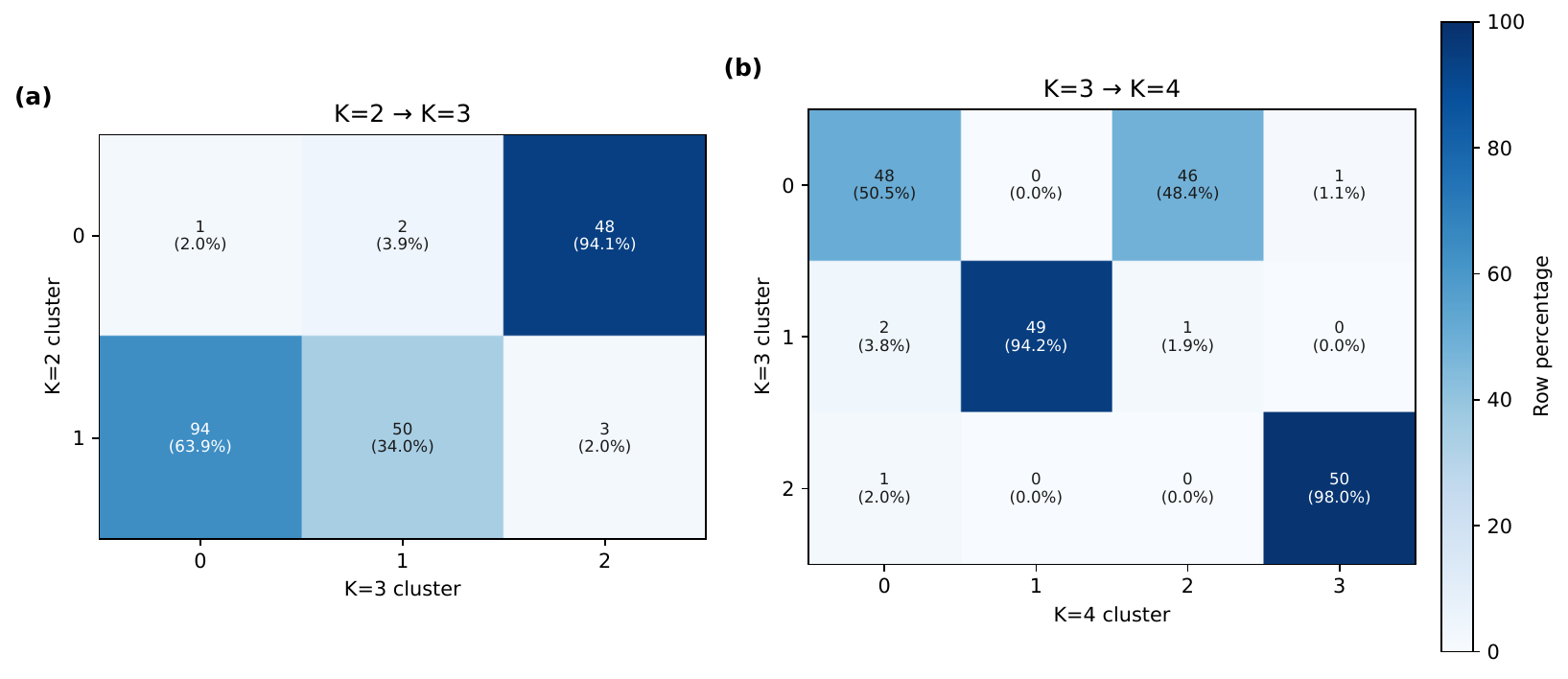}
    \caption{Document transitions between adjacent K-means resolutions. Panels (a) and (b) show transitions from $K=2$ to $K=3$ and from $K=3$ to $K=4$, respectively. Each cell reports the number of papers followed by its percentage of the corresponding source cluster. Shading represents row percentages on a common scale across panels.}
    \label{fig:k-transitions}
\end{figure}

\clearpage
Figure~\ref{fig:k-transitions} provides supporting diagnostics and descriptive results.

\begin{figure}[H]
    \centering
    \includegraphics[width=0.8\textwidth]{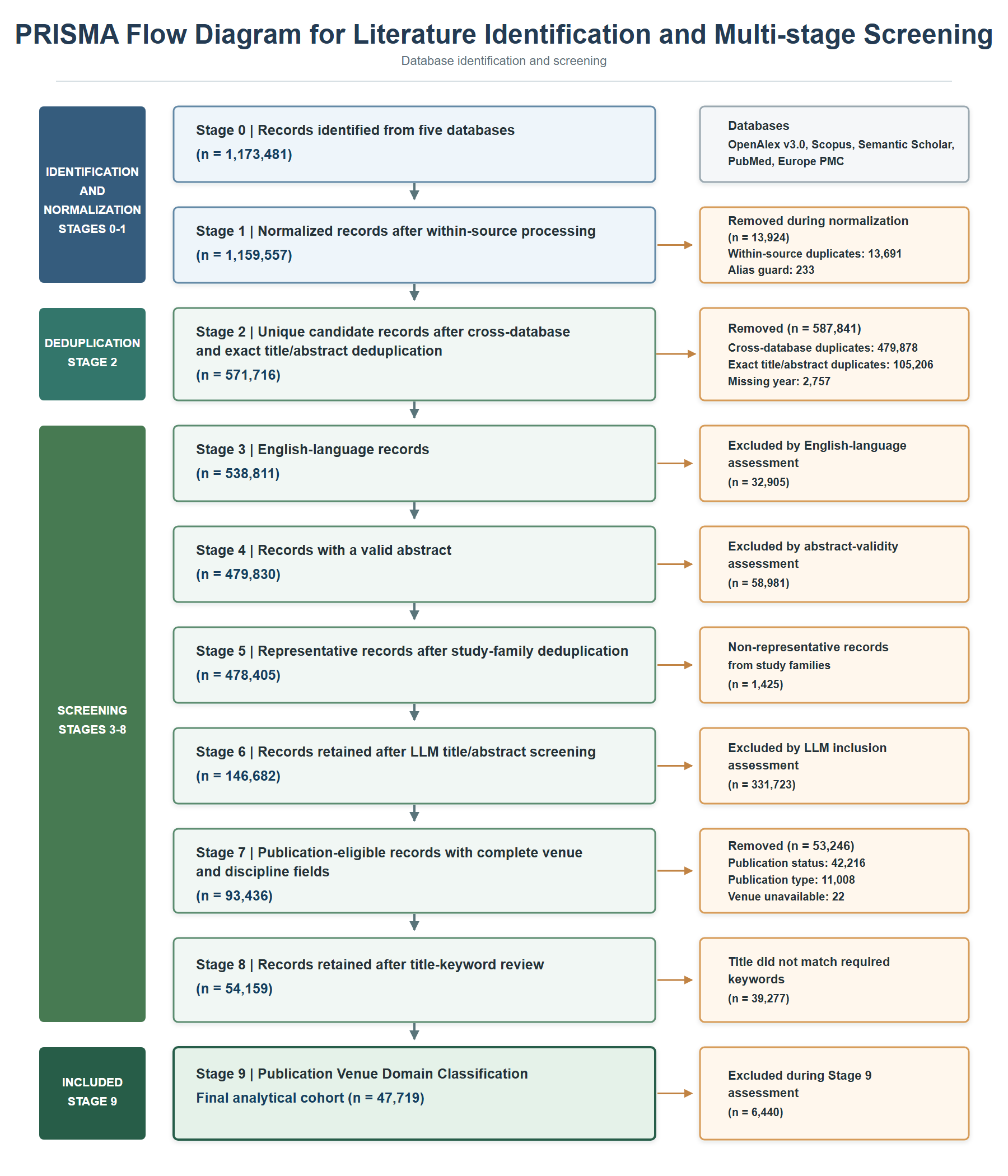}
    \caption{Flow of records through database identification, record processing,
    eligibility screening, and publication-venue classification in Study~2.
    Counts in the exclusion boxes report the records removed at each stage.}
    \label{fig:study2-prisma}
\end{figure}

\clearpage
Figure~\ref{fig:study2-prisma} provides supporting diagnostics and descriptive results.

\begin{figure}[H]
    \centering
    \includegraphics[width=\textwidth]{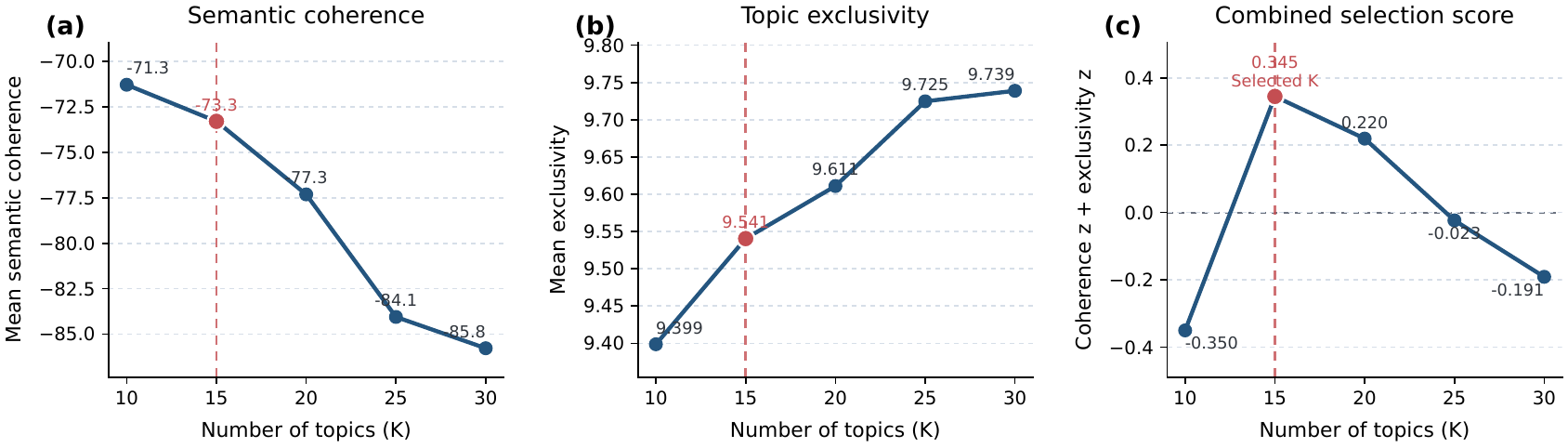}
    \caption{Selection diagnostics for structural topic models across candidate topic numbers. Panels show (a) mean semantic coherence, (b) mean topic exclusivity, and (c) the sum of the standardized mean coherence and exclusivity values. Higher values indicate better performance in all panels. The red markers and dashed vertical lines identify the selected $K=15$ solution, which attained the highest combined trade-off score.}
    \label{fig:stm-model-selection}
\end{figure}

\clearpage
Figure~\ref{fig:stm-model-selection} provides supporting diagnostics and descriptive results.

\clearpage

\begin{figure}[H]
\centering
\includegraphics[width=0.95\textwidth]{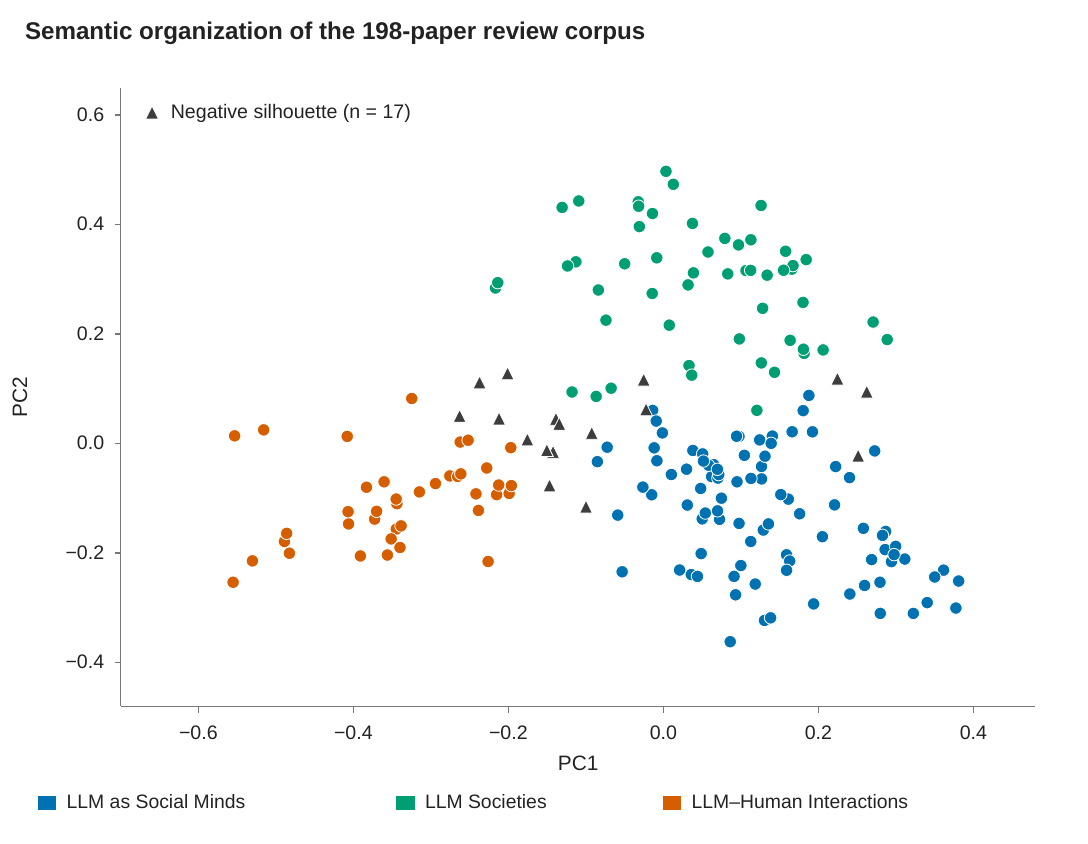}
\caption{\textbf{Principal component projection of the curated-corpus clusters.} Points represent the 198 papers; colors indicate their K-means assignments in the full MPNet embedding space and use the same domain palette as the main figures. Gray triangles identify papers with negative silhouette values; their domain assignments are not shown by color. Principal component analysis is used only for visualization: the two-dimensional projection does not establish separation in the original embedding space. Domain names reflect the authors' interpretation of within-cluster topic terms and representative papers.}
\label{fig:kmeans-pca-clusters}
\end{figure}

\begin{figure}[H]
\centering
\includegraphics[width=\textwidth]{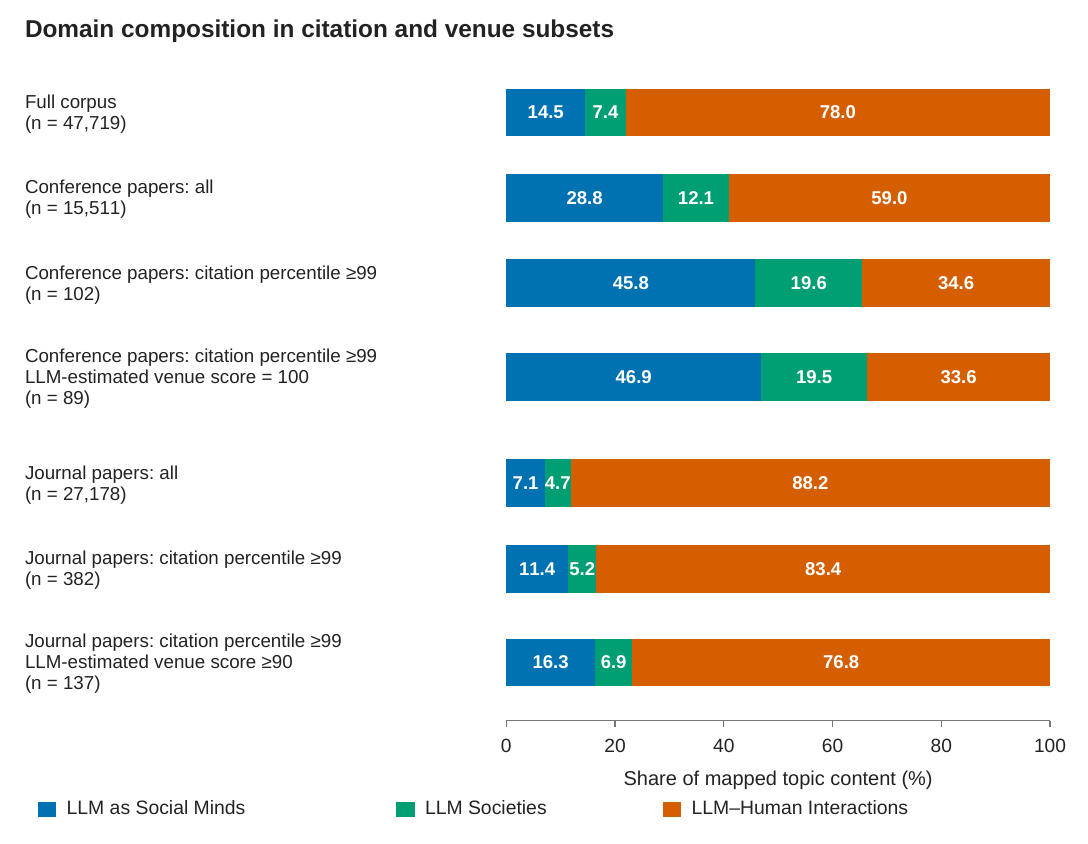}
\caption{\textbf{Domain composition across citation- and venue-score subsets.} Bars show the shares of expected topic content normalized over the 13 domain-mapped STM topics; the two broad cross-domain topics are excluded. Conference subsets include proceedings-series venues. High-citation subsets use citation percentiles of at least 99. The further restricted subsets use LLM-estimated venue scores of 100 for conferences and at least 90 for journals. These scores are descriptive indicators, not official journal percentiles or conference rankings. Labels give the number of papers in each subset.}
\label{fig:venue-subsets}
\end{figure}

Figure~\ref{fig:kmeans-pca-clusters} shows the projected structure of the curated corpus.
\clearpage
Figure~\ref{fig:venue-subsets} retains the detailed citation- and venue-score comparisons.
\clearpage

Table~\ref{tab:classification-correspondence} provides supporting classification details.

\begin{table}[H]
    \centering
    \caption{Correspondence among K-means domain labels, author full-text classifications, and LLM primary classifications. K-means cluster identifiers are expressed using the LDA-informed domain labels established in the preceding analysis; the author and LLM classifications use the same three-domain codebook.}
    \label{tab:classification-correspondence}
    \begin{adjustbox}{max width=\textwidth}
\begin{tabular}{lrrrr}
        \toprule
        Comparison & Matched papers & Correspondence & Cohen's $\kappa$ & ARI \\
        \midrule
        K-means vs. author & 154/198 & 77.78\% & 0.662 & 0.434 \\
        K-means vs. LLM primary & 149/198 & 75.25\% & 0.612 & 0.384 \\
        Author vs. LLM primary & 170/198 & 85.86\% & 0.781 & 0.623 \\
        \bottomrule
    \end{tabular}
    \end{adjustbox}
\end{table}
\clearpage
Table~\ref{tab:document-types} provides supporting classification or corpus details.

\begin{table}[H]
\centering
\caption{\textbf{Document Types Distribution (total $N=198$).} Number and
percentage of documents in each publication-type category.}
\label{tab:document-types}
\begin{tabular}{lrr}
\toprule
Type & Count & Percentage \\
\midrule
Journal    & 115 & 58.1\% \\
Conference &  76 & 38.4\% \\
Preprint   &   7 &  3.5\% \\
\bottomrule
\end{tabular}
\end{table}
\clearpage
Table~\ref{tab:study1-cluster-sizes} provides supporting classification or corpus details.

\begin{table}[H]
    \centering
    \caption{Distribution of papers across the selected $K=3$ solution.}
    \label{tab:study1-cluster-sizes}
    \begin{tabular}{lrr}
        \toprule
        Cluster & Number of papers & Percentage \\
        \midrule
        Cluster 0 & 95 & 47.98\% \\
        Cluster 1 & 52 & 26.26\% \\
        Cluster 2 & 51 & 25.76\% \\
        \midrule
        Total & 198 & 100.00\% \\
        \bottomrule
    \end{tabular}
\end{table}
\clearpage
Table~\ref{tab:llm-primary-secondary} provides supporting classification or corpus details.

\begin{table}[H]
    \centering
    \caption{Primary and secondary domain assignments produced by the LLM-based evaluator. Each paper received one primary category and at most one secondary category. A secondary category could not duplicate the primary category.}
    \label{tab:llm-primary-secondary}
    \small
    \begin{tabularx}{\textwidth}{>{\raggedright\arraybackslash}Xrrrrr}
        \toprule
        LLM primary category & \shortstack{LLM--Human\\Interactions} & \shortstack{LLM\\Societies} & \shortstack{LLM as\\Social Minds} & None & \shortstack{Any\\secondary} \\
        \midrule
        LLM--Human Interactions ($n=70$) & 0 & 3 & 24 & 43 & 27 (38.6\%) \\
        LLM Societies ($n=35$) & 9 & 0 & 12 & 14 & 21 (60.0\%) \\
        LLM as Social Minds ($n=93$) & 35 & 4 & 0 & 54 & 39 (41.9\%) \\
        \midrule
        Total & 44 & 7 & 36 & 111 & 87 (43.9\%) \\
        \bottomrule
    \end{tabularx}
\end{table}

\clearpage
Table~\ref{tab:simulation-platforms} summarizes illustrative simulation platforms discussed in the review.

\begin{table}[H]
\centering
\small
\setlength{\tabcolsep}{5pt}
\renewcommand{\arraystretch}{1.15}
\caption{\textbf{Illustrative LLM-based social simulation platforms and their uses.} The entries summarize features and applications reported in the reviewed literature. The final column distinguishes what each design allows researchers to examine; it is not a comparative assessment of predictive validity. Population size and realism of generated behavior should be evaluated separately from validation against human observations.}
\label{tab:simulation-platforms}
\begin{tabularx}{\linewidth}{@{}>{\raggedright\arraybackslash}p{0.19\linewidth}YY@{}}
\toprule
Platform or approach & Scope and features & Interpretive or validation relevance \\
\midrule
Generative Agents \parencite{park2023generative} &
Episodic memory, retrieval, reflection, and planning support daily routines, information diffusion, and spontaneous coordination. &
Allows examination of how agent architecture supports socially organized behavior. \\

CAMEL \parencite{li2023camel} &
Role-based agents sustain communication and cooperation. &
Supports study of cooperation through structured roles and communication. \\

Interview-based generative agents \parencite{park2024generative} &
Agents represent 1,052 real individuals using interview-based information. &
Evaluation compares simulated responses with the represented individuals' survey responses and experimental behavior. \\

$S^3$ \parencite{ref189} &
Simulates diffusion of information, attitudes, and emotions through social networks. &
Supports analysis of relationships between network interaction and collective change. \\

AgentSociety \parencite{ref194} &
Supports more than 10,000 agents and millions of interactions, with applications to polarization, mobility, policy interventions, and external shocks. &
Enables interventions on simulated collective dynamics; conclusions about human populations require external validation. \\

Light Society \parencite{guan2025lightsociety} &
Models distilled from LLM responses support opinion diffusion on a one-billion-agent network. &
Enables population-scale experiments; demographic representation and simulated outcomes require empirical validation. \\

OASIS \parencite{yang2024oasis} &
Social-media simulations range from hundreds to millions of agents, including one-million-agent experiments on information propagation, polarization, and herd behavior. &
Demonstrates large-scale simulation of collective patterns; scale alone does not establish correspondence with human societies. \\
\bottomrule
\end{tabularx}
\par\smallskip\noindent\footnotesize\raggedright
Other reviewed applications examine cultural transmission, institutional rules, economic exchange, bargaining, team formation, and behavior in emerging social systems \parencite{perez2024cultural,ren2024emergence,horiguchi2024evolution,ref100,ref101,ref102,ref104,ref108}. Across applications, credibility depends on robustness across models and settings and on tracing aggregate outcomes to interaction mechanisms \parencite{ref66,ref168}.
\end{table}

Model-to-model conversations also provide a scalable way to generate and evaluate simulated exchanges, but their realism depends on the models and prompts used to produce both sides \parencite{ref95}.

\clearpage
\section{Bibliographic coverage of the curated corpus}
The following inventory retains the subcategory reference groupings used to document the curated literature. References are included in the single combined bibliography of the main manuscript.

\begin{longtable}{p{0.43\textwidth}p{0.49\textwidth}}
\caption{\textbf{Bibliographic coverage of the 13 subcategories.} Source groupings accompanying the taxonomy.}\label{tab:corpus-reference-coverage}\\
\toprule Subcategory & References \\ \midrule
\endfirsthead
\toprule Subcategory & References \\ \midrule
\endhead
Human-Like Reasoning and Theory of Mind & \cite{tax001,tax002,tax003,tax004,tax005,tax006,tax007,tax008,tax009,tax010,tax011,tax012,tax013,tax014,tax015,tax016,tax017,tax018,tax019,tax020,tax021,tax022,tax023,tax024,tax025,tax026,tax027,tax028,tax029,tax030,tax031,tax032,tax033,tax034,tax035,tax036,tax037,tax038,tax039,tax040,tax041} \\ \addlinespace
Personality and Demographic Bias & \cite{tax042,tax043,tax044,tax045,tax046,tax047,tax048,tax049,tax050,tax051,tax052,tax053,tax054,tax055,tax056,tax057,tax058,tax059,tax060,tax061,tax062,tax063,tax064,tax065,tax066,tax067,tax068,tax069,tax070} \\ \addlinespace
Political and Moral Judgment & \cite{tax071,tax072,tax073,tax074,tax075,tax076,tax077,tax078,tax079,tax080,tax081,tax082} \\ \addlinespace
Preference Alignment, Persuasion, and Deception & \cite{tax083,tax084,tax085,tax086,tax087,tax088,tax089,tax090,tax091,tax092,tax093} \\ \addlinespace
Behavioral Games of LLM Agents & \cite{tax094,tax095,tax096,tax097,tax098} \\ \addlinespace
Collective Intelligence and Emergent Dynamics & \cite{tax099,tax100,tax101,tax102,tax103,tax104,tax105} \\ \addlinespace
Group Decision-Making of LLM Agents & \cite{tax106,tax107,tax108,tax109,tax110,tax111,tax112,tax113,tax114,tax115,tax116} \\ \addlinespace
Scalable LLM-Based Social and Network Simulation & \cite{tax117,tax118,tax119,tax120,tax121,tax122,tax123,tax124,tax125,tax126,tax127,tax128} \\ \addlinespace
Perception and Trust & \cite{tax129,tax130,tax131,tax132,tax133,tax134,tax135,tax136,tax137,tax138,tax139,tax140,tax141,tax142,tax143,tax144} \\ \addlinespace
Simulated Empathy and Support & \cite{tax145,tax146,tax147,tax148,tax149,tax150,tax151,tax152} \\ \addlinespace
Generative Work and Collaboration & \cite{tax153,tax154,tax155,tax156,tax157,tax158,tax159,tax160,tax161,tax162,tax163,tax164,tax165,tax166,tax167,tax168,tax169,tax170,tax171,tax172,tax173,tax174,tax175,tax176} \\ \addlinespace
Creative Partnership & \cite{tax177,tax178,tax179,tax180,tax181,tax182,tax183,tax184} \\ \addlinespace
Education and Assessment & \cite{tax185,tax186,tax187,tax188,tax189,tax190,tax191,tax192,tax193,tax194,tax195,tax196,tax197,tax198} \\ \addlinespace
\bottomrule
\end{longtable}


\begin{thebibliography}{999}
\bibitem{simon2019sciences}Herbert A Simon. \emph{The Sciences of the Artificial, reissue of the third edition with a new introduction by
John Laird}. MIT press, 2019.
{}
\bibitem{mei2024turing}Qiaozhu Mei, Yutong Xie, Walter Yuan, and Matthew O. Jackson. ``A Turing test of whether AI chatbots are behaviorally similar to humans''. In: \emph{Proceedings of the National Academy of Sciences} 121.9 (2024), e2313925121. \textsc{doi}: \nolinkurl{10.1073/pnas.2313925121}.
{}
\bibitem{tax117}\bibalias{park2023generative}
\bibalias{ref69}

Joon Sung Park, Joseph O'Brien, Carrie Jun Cai, Meredith Ringel Morris, Percy Liang, Michael S. Bernstein. “Generative Agents: Interactive Simulacra of Human Behavior”. In: \emph{Proceedings of
the 36th Annual ACM Symposium on User Interface Software and Technology}. ACM, 2023, pp. 1–22.
\textsc{doi}: \nolinkurl {10.1145/3586183.3606763}.
{}
\bibitem{tax176}\bibalias{bail2024can}
\bibalias{ref191}

Christopher A. Bail. \emph{Can Generative AI improve social science?} 2023. \textsc{doi}: \nolinkurl
{10.31235/osf.io/rwtzs}.
{}
\bibitem{peter2025benefits}Sandra Peter, Kai Riemer, and Jevin D West. “The benefits and dangers of anthropomorphic conversational
agents”. In: \emph{Proceedings of the National Academy of Sciences} 122.22 (2025), e2415898122.
{}
\bibitem{xu2024ai}Ruoxi Xu, Yingfei Sun, Mengjie Ren, Shiguang Guo, Ruotong Pan, Hongyu Lin, Le Sun, Xianpei Han. “AI for social science and social science of AI: A survey”. In: \emph{Information Processing \&
Management} 61.3 (2024), p. 103665.
{}
\bibitem{hutchins1995cognition}Edwin Hutchins. \emph{Cognition in the Wild}. MIT Press, 1995.
{}
\bibitem{latour2005reassembling}Bruno Latour. \emph{Reassembling the Social: An Introduction to Actor-Network-Theory}. Oxford University Press, 2005. \textsc{doi}: \nolinkurl{10.1093/oso/9780199256044.001.0001}.
{}
\bibitem{vygotsky1978mind}Lev S. Vygotsky. \emph{Mind in Society: The Development of Higher Psychological Processes}. Harvard University Press, 1978.
{}
\bibitem{kosinski2024evaluating}Michal Kosinski. “Evaluating large language models in theory of mind tasks”. In: \emph{Proceedings of the
National Academy of Sciences} 121.45 (2024), e2405460121.
{}
\bibitem{tax012}\bibalias{ref23}
\bibalias{strachan2024testing}

James W. A. Strachan, Dalila Albergo, Giulia Borghini, Oriana Pansardi, Eugenio Scaliti, Saurabh Gupta, Krati Saxena, Alessandro Rufo, Stefano Panzeri, Guido Manzi, Michael S. A. Graziano, Cristina Becchio. “Testing theory of mind in large language models and humans”. In: \emph{Nature
Human Behaviour} 8.7 (2024), pp. 1285–1295. \textsc{doi}: \nolinkurl {10.1038/s41562-024-01882-z}.
{}
\bibitem{lehr2025kernels}Steven A. Lehr, Ketan S. Saichandran, Eddie Harmon-Jones, Nykko Vitali, Mahzarin R. Banaji. “Kernels of selfhood: GPT-4o shows humanlike patterns of cognitive dissonance moderated
by free choice”. In: \emph{Proceedings of the National Academy of Sciences} 122.20 (2025), e2501823122.
{}
\bibitem{tax067}\bibalias{hu2025generative}
\bibalias{ref92}

Tiancheng Hu, Yara Kyrychenko, Steve Rathje, Nigel Collier, Sander van der Linden, Jon Roozenbeek. “Generative language models exhibit social identity biases”. In: \emph{Nature Computational
Science} 5.1 (2024), pp. 65–75. \textsc{doi}: \nolinkurl {10.1038/s43588-024-00741-1}.
{}
\bibitem{li2023camel}Guohao Li, Hasan Hammoud, Hani Itani, Dmitrii Khizbullin, Bernard Ghanem. “Camel: Communicative agents for" mind" exploration of large language model society”. In:
\emph{Advances in Neural Information Processing Systems} 36 (2023), pp. 51991–52008.
{}
\bibitem{tax094}\bibalias{akata2025playing}
\bibalias{ref68}

Elif Akata, Lion Schulz, Julian Coda-Forno, Seong Joon Oh, Matthias Bethge, Eric Schulz. “Playing repeated games with large language models”. In: \emph{Nature Human Behaviour} 9.7
(2025), pp. 1380–1390. \textsc{doi}: \nolinkurl {10.1038/s41562-025-02172-y}.
{}
\bibitem{tax183}\bibalias{chen2024large}
\bibalias{ref175}

Zenan Chen and Jason Chan. “Large Language Model in Creative Work: The Role of Collaboration Modality and
User Expertise”. In: \emph{Management Science} 70.12 (2024), pp. 9101–9117. \textsc{doi}: \nolinkurl
{10.1287/mnsc.2023.03014}.
{}
\bibitem{noy2023experimental}Shakked Noy and Whitney Zhang. “Experimental evidence on the productivity effects of generative artificial
intelligence”. In: \emph{Science} 381.6654 (2023), pp. 187–192.
{}
\bibitem{bommasani2021opportunities}Rishi Bommasani. “On the opportunities and risks of foundation models”. In: \emph{arXiv preprint
arXiv:2108.07258} (2021).
{}
\bibitem{yan2024practical}Lixiang Yan, Lele Sha, Linxuan Zhao, Yuheng Li, Roberto Martinez-Maldonado, Guanliang Chen, Xinyu Li, Yueqiao Jin, Dragan Gašević. “Practical and ethical challenges of large language models in education: A systematic scoping
review”. In: \emph{British Journal of Educational Technology} 55.1 (2024), pp. 90–112.
{}
\bibitem{lai2024large}Jinqi Lai, Wensheng Gan, Jiayang Wu, Zhenlian Qi, Philip S. Yu. “Large language models in law: A survey”. In: \emph{AI Open} 5 (2024), pp. 181–196.
{}
\bibitem{haltaufderheide2024ethics}Joschka Haltaufderheide and Robert Ranisch. “The ethics of ChatGPT in medicine and healthcare: a systematic
review on Large Language Models (LLMs)”. In: \emph{NPJ digital medicine} 7.1 (2024), p. 183.
{}
\bibitem{rogiers2024persuasion}Alexander Rogiers, Sander Noels, Maarten Buyl, Tijl De Bie. “Persuasion with large language models: a survey”. In: \emph{arXiv preprint
arXiv:2411.06837} (2024).
{}
Version 1: \url{https://arxiv.org/abs/2411.06837v1}.
\bibitem{kuntur2024under}Soveatin Kuntur, Anna Wróblewska, Marcin Paprzycki, Maria Ganzha. “Under the influence: A survey of large language models in fake news detection”. In:
\emph{IEEE Transactions on Artificial Intelligence} (2024).
{}
\bibitem{xi2025rise}Zhiheng Xi, Wenxiang Chen, Xin Guo, Wei He, Yiwen Ding, Boyang Hong, Ming Zhang, Junzhe Wang, Senjie Jin, Enyu Zhou, Rui Zheng, Xiaoran Fan, Xiao Wang, Limao Xiong, Yuhao Zhou, Weiran Wang, Changhao Jiang, Yicheng Zou, Xiangyang Liu, Zhangyue Yin, Shihan Dou, Rongxiang Weng, Wenjuan Qin, Yongyan Zheng, Xipeng Qiu, Xuanjing Huang, Qi Zhang, Tao Gui. “The rise and potential of large language model based agents: A survey”. In: \emph{Science
China Information Sciences} 68.2 (2025), p. 121101.
{}
\bibitem{wang2024survey}Lei Wang, Chen Ma, Xueyang Feng, Zeyu Zhang, Hao Yang, Jingsen Zhang, Zhiyuan Chen, Jiakai Tang, Xu Chen, Yankai Lin, Wayne Xin Zhao, Zhewei Wei, Jirong Wen. “A survey on large language model based autonomous agents”. In: \emph{Frontiers of Computer
Science} 18.6 (2024), p. 186345.
{}
\bibitem{liu2026sociology}Canhui Liu. ``The Sociology of AI as an Emerging Field: Mapping Tensions and Boundaries''. In: \emph{Sociology Compass} 20.5 (2026), e70201. \textsc{doi}: \nolinkurl{10.1111/soc4.70201}.
{}
\bibitem{chen2026agentbehavior}Lin Chen, Yunke Zhang, Jie Feng, Haoye Chai, Honglin Zhang, Bingbing Fan, Yibo Ma, Shiyuan Zhang, Nian Li, Tianhui Liu, Nicholas Sukiennik, Keyu Zhao, Yu Li, Ziyi Liu, Fengli Xu, and Yong Li. ``AI agent behavioral science''. In: \emph{Humanities and Social Sciences Communications} 13 (2026), 1011. \textsc{doi}: \nolinkurl{10.1057/s41599-026-07316-7}.
{}
\bibitem{fasce2023taxonomy}Angelo Fasce, Philipp Schmid, Dawn L. Holford, Luke Bates, Iryna Gurevych, and Stephan Lewandowsky. ``A taxonomy of anti-vaccination arguments from a systematic literature review and text modelling''. In: \emph{Nature Human Behaviour} 7.9 (2023), pp. 1462--1480. \textsc{doi}: \nolinkurl{10.1038/s41562-023-01644-3}.
{}
\bibitem{reimers2019sentence}Nils Reimers and Iryna Gurevych. “Sentence-bert: Sentence embeddings using siamese bert-networks”. In:
\emph{arXiv preprint arXiv:1908.10084} (2019).
{}
\bibitem{tax001}\bibalias{ref4}

Changde Du, Kaicheng Fu, Bincheng Wen, Yi Sun, Jie Peng, Wei Wei, Ying Gao, Shengpei Wang, Chuncheng Zhang, Jinpeng Li, Shuang Qiu, Le Chang, Huiguang He. “Human-like object concept representations emerge naturally in multimodal large language
models”. In: \emph{Nature Machine Intelligence} 7.6 (2025), pp. 860–875. \textsc{doi}: \nolinkurl
{10.1038/s42256-025-01049-z}.
{}
\bibitem{tax003}\bibalias{ref9}

Xuena Wang, Xueting Li, Zi Yin, Yue Wu, Jia Liu. “Emotional intelligence of Large Language Models”. In: \emph{Journal of Pacific Rim
Psychology} 17 (2023), p. 18344909231213958. \textsc{doi}: \nolinkurl {10.1177/18344909231213958}.
{}
\bibitem{tax022}\bibalias{ref45}

Justin M. Mittelstädt, Julia Maier, Panja Goerke, Frank Zinn, Michael Hermes. “Large language models can outperform humans in social situational judgments”. In:
\emph{Scientific Reports} 14.1 (2024), p. 27449. \textsc{doi}: \nolinkurl {10.1038/s41598-024-79048-0}.
{}
\bibitem{tax023}\bibalias{ref46}

Taylor Webb, Keith J. Holyoak, and Hongjing Lu. “Emergent analogical reasoning in large language models”. In:
\emph{Nature Human Behaviour} 7.9 (2023), pp. 1526–1541. \textsc{doi}: \nolinkurl
{10.1038/s41562-023-01659-w}.
{}
\bibitem{tax025}\bibalias{ref50}

Nicholas Ichien, Dušan Stamenković, and Keith J. Holyoak. “Large Language Model Displays Emergent Ability to
Interpret Novel Literary Metaphors”. In: \emph{Metaphor and Symbol} 39.4 (2024), pp. 296–309. \textsc{doi}:
\nolinkurl {10.1080/10926488.2024.2380348}.
{}
\bibitem{tax005}\bibalias{ref13}

Aniket Kumar Singh, Bishal Lamichhane, Suman Devkota, Uttam Dhakal, Chandra Dhakal. “Do Large Language Models Show Human-like Biases? Exploring
Confidence—Competence Gap in AI”. In: \emph{Information} 15.2 (2024), p. 92. \textsc{doi}: \nolinkurl
{10.3390/info15020092}.
{}
\bibitem{tax019}\bibalias{ref42}

Sifatkaur Dhingra, Manmeet Singh, Vaisakh S.B., Neetiraj Malviya, Sukhpal Singh Gill. “Mind meets machine: Unravelling GPT-4’s cognitive psychology”. In:
\emph{BenchCouncil Transactions on Benchmarks, Standards and Evaluations} 3.3 (2023), p. 100139.
\textsc{doi}: \nolinkurl {10.1016/j.tbench.2023.100139}.
{}
\bibitem{premack1978does}David Premack and Guy Woodruff. “Does the chimpanzee have a theory of mind?” In: \emph{Behavioral and
brain sciences} 1.4 (1978), pp. 515–526.
{}
\bibitem{apperly2010mindreaders}Ian Apperly. \emph{Mindreaders: the cognitive basis of" theory of mind"}. Psychology Press, 2010.
{}
\bibitem{kosinski2023theory}Michal Kosinski. “Theory of mind may have spontaneously emerged in large language models”. In: \emph{arXiv
preprint arXiv:2302.02083} 4 (2023), p. 169.
{}
\bibitem{tax002}\bibalias{amirizaniani2025mind}
\bibalias{ref7}

Maryam Amirizaniani. “Mind Over Machine: Evaluating Theory of Mind Reasoning in LLMs and Humans”. In:
\emph{Proceedings of the Eighteenth ACM International Conference on Web Search and Data Mining}. 2025.
\textsc{doi}: \nolinkurl {10.1145/3701551.3707417}.
{}
\bibitem{ullman2023large}Tomer Ullman. “Large language models fail on trivial alterations to theory-of-mind tasks”. In: \emph{arXiv
preprint arXiv:2302.08399} (2023).
{}
\bibitem{tax020}\bibalias{ref43}
\bibalias{shapira2023clever}

Natalie Shapira, Mosh Levy, Seyed Hossein Alavi, Xuhui Zhou, Yejin Choi, Yoav Goldberg, Maarten Sap, Vered Shwartz. \emph{Clever Hans or Neural Theory of Mind? Stress Testing Social Reasoning in Large
Language Models}. arXiv. 2023. \textsc{doi}: \nolinkurl {10.48550/arXiv.2305.14763}.
{}
\bibitem{tax158}\bibalias{ref112}
\bibalias{zhao2025llms}

Siyan Zhao, Mingyi Hong, Yang Liu, Devamanyu Hazarika, Kaixiang Lin. “Do LLMs Recognize Your Preferences? Evaluating Personalized Preference Following in LLMs”.
In: \emph{The Thirteenth International Conference on Learning Representations}. 2025. \textsc{url}: \url
{https://openreview.net/forum?id=I5326jR0W3}.
{}
\bibitem{tax021}\bibalias{marchetti2025artificial}
\bibalias{ref44}

Antonella Marchetti, Federico Manzi, Giuseppe Riva, Andrea Gaggioli, Davide Massaro. “Artificial Intelligence and the Illusion of Understanding: A Systematic Review of
Theory of Mind and Large Language Models”. In: \emph{Cyberpsychology, Behavior, and Social Networking}
28.7 (2025), pp. 505–514. \textsc{doi}: \nolinkurl {10.1089/cyber.2024.0536}.
{}
\bibitem{tax033}\bibalias{ref96}

Maarten Sap, Ronan Le Bras, Daniel Fried, Yejin Choi. “Neural Theory-of-Mind? On the Limits of Social Intelligence in Large LMs”. In:
\emph{Proceedings of the 2022 Conference on Empirical Methods in Natural Language Processing}. Association
for Computational Linguistics, 2022, pp. 3762–3780. \textsc{doi}: \nolinkurl
{10.18653/v1/2022.emnlp-main.248}.
{}
\bibitem{binz2023using}Marcel Binz and Eric Schulz. “Using cognitive psychology to understand GPT-3”. In: \emph{Proceedings of the
National Academy of Sciences} 120.6 (2023), e2218523120.
{}
\bibitem{tax017}\bibalias{ref39}

Thilo Hagendorff, Ishita Dasgupta, Marcel Binz, Stephanie C. Y. Chan, Andrew Lampinen, Jane X. Wang, Zeynep Akata, Eric Schulz. \emph{Machine Psychology}. arXiv. 2023. \textsc{doi}: \nolinkurl
{10.48550/arXiv.2303.13988}.
{}
\bibitem{tax024}\bibalias{ref49}

Xintong Wang, Xiaoyu Li, Xingshan Li, Chris Biemann. “Probing Large Language Models from A Human Behavioral Perspective”. In:
\emph{Proceedings of the Language Resources and Evaluation Conference}. European Language Resources
Association (ELRA) and ICCL, 2024, pp. 1–7. \textsc{doi}: \nolinkurl {10.63317/2r7fmikuo4ec}.
{}
\bibitem{tax042}\bibalias{ref1}

Yilei Wang, Jiabao Zhao, Deniz S. Ones, Liang He, Xin Xu. “Evaluating the ability of large language models to emulate personality”. In: \emph{Scientific
Reports} 15.1 (2025), pp. 1–9. \textsc{doi}: \nolinkurl {10.1038/s41598-024-84109-5}.
{}
\bibitem{tax045}\bibalias{bhandari2025evaluating}
\bibalias{ref10}

Pranav Bhandari, Usman Naseem, Amitava Datta, Nicolas Fay, Mehwish Nasim. “Evaluating Personality Traits in Large Language Models: Insights from Psychological
Questionnaires”. In: \emph{Companion Proceedings of the ACM on Web Conference 2025}. ACM, 2025,
pp. 868–872. \textsc{doi}: \nolinkurl {10.1145/3701716.3715504}.
{}
\bibitem{tax064}\bibalias{ref60}

Zhaleh Havaei, Morteza Saberi, and Omar Khadeer Hussain. “Evaluation of Large Language Model Responses for
32 Diverse Personality Types Using the Best Worst Method”. In: \emph{2024 IEEE International Conference on
e-Business Engineering (ICEBE)}. IEEE, 2024, pp. 262–271. \textsc{doi}: \nolinkurl
{10.1109/icebe62490.2024.00048}.
{}
\bibitem{tax065}\bibalias{ref61}

Max Pellert, Clemens M. Lechner, Claudia Wagner, Beatrice Rammstedt, Markus Strohmaier. “AI Psychometrics: Assessing the Psychological Profiles of Large Language Models Through
Psychometric Inventories”. In: \emph{Perspectives on Psychological Science} 19.5 (2024), pp. 808–826.
\textsc{doi}: \nolinkurl {10.1177/17456916231214460}.
{}
\bibitem{tax066}\bibalias{ref62}

Siyang Liu, Trisha Maturi, Bowen Yi, Siqi Shen, Rada Mihalcea. “The Generation Gap: Exploring Age Bias in the Value Systems of Large Language Models”. In:
\emph{Proceedings of the 2024 Conference on Empirical Methods in Natural Language Processing}. Association
for Computational Linguistics, 2024, pp. 19617–19634. \textsc{doi}: \nolinkurl
{10.18653/v1/2024.emnlp-main.1094}.
{}
\bibitem{tax077}\bibalias{ref65}

Kai Chen, Zihao He, Jun Yan, Taiwei Shi, Kristina Lerman. “How Susceptible are Large Language Models to Ideological Manipulation?” In:
\emph{Proceedings of the 2024 Conference on Empirical Methods in Natural Language Processing}. Association
for Computational Linguistics, 2024, pp. 17140–17161. \textsc{doi}: \nolinkurl
{10.18653/v1/2024.emnlp-main.952}.
{}
\bibitem{tax054}\bibalias{kotek2023gender}
\bibalias{ref37}

Hadas Kotek, Rikker Dockum, and David Sun. “Gender bias and stereotypes in Large Language Models”. In:
\emph{Proceedings of The ACM Collective Intelligence Conference}. ACM, 2023, pp. 12–24. \textsc{doi}:
\nolinkurl {10.1145/3582269.3615599}.
{}
\bibitem{tax044}\bibalias{lu2025cultural}
\bibalias{ref3}

Jackson G. Lu, Lesley Luyang Song, and Lu Doris Zhang. “Cultural tendencies in generative AI”. In:
\emph{Nature Human Behaviour} 9.11 (2025), pp. 2360–2369. \textsc{doi}: \nolinkurl
{10.1038/s41562-025-02242-1}.
{}
\bibitem{tax055}\bibalias{abid2021persistent}
\bibalias{ref38}

Abubakar Abid, Maheen Farooqi, and James Zou. “Persistent Anti-Muslim Bias in Large Language Models”. In:
\emph{Proceedings of the 2021 AAAI/ACM Conference on AI, Ethics, and Society}. ACM, 2021, pp. 298–306.
\textsc{doi}: \nolinkurl {10.1145/3461702.3462624}.
{}
\bibitem{tax061}\bibalias{ref56}
\bibalias{siddique2024better}

Zara Siddique, Liam D. Turner, and Luis Espinosa-Anke. “Who is better at math, Jenny or Jingzhen? Uncovering
Stereotypes in Large Language Models”. In: \emph{Proceedings of the 2024 Conference on Empirical Methods in
Natural Language Processing}. 2024, pp. 18601–18619. \textsc{doi}: \nolinkurl
{10.18653/v1/2024.emnlp-main.1035}.
{}
\bibitem{tax047}\bibalias{ref14}

Xiao Fang, Shangkun Che, Minjia Mao, Hongzhe Zhang, Ming Zhao, Xiaohang Zhao. “Bias of AI-generated content: an examination of news produced by large language models”. In:
\emph{Scientific Reports} 14.1 (2024), p. 5224. \textsc{doi}: \nolinkurl {10.1038/s41598-024-55686-2}.
{}
\bibitem{tax050}\bibalias{ref24}
\bibalias{torres2024comprehensive}

Nicolás Torres, Catalina Ulloa, Ignacio Araya, Matías Ayala, Sebastián Jara. “A comprehensive analysis of gender, racial, and prompt-induced biases in large language
models”. In: \emph{International Journal of Data Science and Analytics} 20.4 (2025), pp. 3797–3834.
\textsc{doi}: \nolinkurl {10.1007/s41060-024-00696-6}.
{}
\bibitem{tax051}\bibalias{ref28}

Leilei Jiang, Guixiang Zhu, Jianshan Sun, Jie Cao, Jia Wu. “Exploring the occupational biases and stereotypes of Chinese large language models”. In:
\emph{Scientific Reports} 15.1 (2025), p. 18777. \textsc{doi}: \nolinkurl {10.1038/s41598-025-03893-w}.
{}
\bibitem{tax057}\bibalias{mahesh2024investigating}
\bibalias{ref47}

Tarun Mahesh. “Investigating Social Bias in GPT-4: A Study on Race Representation”. In: \emph{2024 2nd
International Conference on Foundation and Large Language Models (FLLM)}. IEEE, 2024, pp. 599–606.
\textsc{doi}: \nolinkurl {10.1109/fllm63129.2024.10852468}.
{}
\bibitem{tax060}\bibalias{lee2024large}
\bibalias{ref54}

Messi H.J. Lee, Jacob M. Montgomery, and Calvin K. Lai. “Large Language Models Portray Socially Subordinate
Groups as More Homogeneous, Consistent with a Bias Observed in Humans”. In: \emph{The 2024 ACM
Conference on Fairness, Accountability, and Transparency}. ACM, 2024, pp. 1321–1340. \textsc{doi}: \nolinkurl
{10.1145/3630106.3658975}.
{}
\bibitem{tax062}\bibalias{ref57}

Yi-Cheng Lin, Wei-Chih Chen, and Hung-Yi Lee. “Spoken Stereoset: on Evaluating Social Bias Toward Speaker in
Speech Large Language Models”. In: \emph{2024 IEEE Spoken Language Technology Workshop (SLT)}. IEEE,
2024, pp. 871–878. \textsc{doi}: \nolinkurl {10.1109/slt61566.2024.10832259}.
{}
\bibitem{tax075}\bibalias{ref58}

Xiaobo Shan, Yan Teng, Yixu Wang, Haiquan Zhao, Yingchun Wang. “Collectivism and individualism political bias in large language models: A two-step approach”.
In: \emph{Big Data \& Society} 12.2 (2025), p. 20539517251343861. \textsc{doi}: \nolinkurl
{10.1177/20539517251343861}.
{}
\bibitem{tax082}\bibalias{ref181}

Fabio Motoki, Valdemar Pinho Neto, and Victor Rodrigues. “More human than human: Measuring ChatGPT
political bias”. In: \emph{Public Choice} 198.1–2 (2024), pp. 3–23. \textsc{doi}: \nolinkurl
{10.1007/s11127-023-01097-2}.
{}
\bibitem{tax068}\bibalias{ref183}

Haozhe An, Christabel Acquaye, Colin Wang, Zongxia Li, Rachel Rudinger. “Do Large Language Models Discriminate in Hiring Decisions on the Basis of Race, Ethnicity,
and Gender?” In: \emph{Proceedings of the 62nd Annual Meeting of the Association for Computational
Linguistics (Volume 2: Short Papers)}. Association for Computational Linguistics, 2024, pp. 386–397.
\textsc{doi}: \nolinkurl {10.18653/v1/2024.acl-short.37}.
{}
\bibitem{bai2025explicitly}Xuechunzi Bai, Angelina Wang, Ilia Sucholutsky, Thomas L. Griffiths. “Explicitly unbiased large language models still form biased associations”. In:
\emph{Proceedings of the National Academy of Sciences} 122.8 (2025), e2416228122.
{}
\bibitem{tax043}\bibalias{ref2}

Yang Chen, Samuel N. Kirshner, Anton Ovchinnikov, Meena Andiappan, Tracy Jenkin. “A Manager and an AI Walk into a Bar: Does ChatGPT Make Biased Decisions Like We Do?”
In: \emph{Manufacturing \& Service Operations Management} 27.2 (2025), pp. 354–368. \textsc{doi}:
\nolinkurl {10.1287/msom.2023.0279}.
{}
\bibitem{tax133}\bibalias{ref127}

Tianyi Zhang, Antonis Koutsoumpis, Janneke K. Oostrom, Djurre Holtrop, Sina Ghassemi, Reinout E. de Vries. “Can Large Language Models Assess Personality From Asynchronous Video Interviews? A
Comprehensive Evaluation of Validity, Reliability, Fairness, and Rating Patterns”. In: \emph{IEEE Transactions
on Affective Computing} 15.3 (2024), pp. 1769–1785. \textsc{doi}: \nolinkurl {10.1109/taffc.2024.3374875}.
{}
\bibitem{tax069}\bibalias{argyle2023out}
\bibalias{ref190}

Lisa P. Argyle, Ethan C. Busby, Nancy Fulda, Joshua R. Gubler, Christopher Rytting, David Wingate. “Out of One, Many: Using Language Models to Simulate Human Samples”. In:
\emph{Political Analysis} 31.3 (2023), pp. 337–351. \textsc{doi}: \nolinkurl {10.1017/pan.2023.2}.
{}
\bibitem{tax070}\bibalias{park2024generative}
\bibalias{ref192}

Joon Sung Park, Carolyn Q. Zou, Aaron Shaw, Benjamin Mako Hill, Carrie Cai, Meredith Ringel Morris, Robb Willer, Percy Liang, Michael S. Bernstein. \emph{Generative Agent Simulations of 1,000 People}. arXiv. 2024. \textsc{doi}:
\nolinkurl {10.48550/arXiv.2411.10109}.
{}
Version 1: \url{https://arxiv.org/abs/2411.10109v1}.
\bibitem{tax053}\bibalias{gallegos2024bias}
\bibalias{ref32}

Isabel O. Gallegos, Ryan A. Rossi, Joe Barrow, Md Mehrab Tanjim, Sungchul Kim, Franck Dernoncourt, Tong Yu, Ruiyi Zhang, Nesreen K. Ahmed. “Bias and Fairness in Large Language Models: A Survey”. In: \emph{Computational
Linguistics} 50.3 (2024), pp. 1097–1179. \textsc{doi}: \nolinkurl {10.1162/coli\_a\_00524}.
{}
\bibitem{radaideh2025fairness}Mohammed I Radaideh, O Hwang Kwon, and Majdi I Radaideh. “Fairness and social bias quantification in Large
Language Models for sentiment analysis”. In: \emph{Knowledge-Based Systems} (2025), p. 113569.
{}
\bibitem{tax153}\bibalias{an2025measuring}
\bibalias{ref5}

Jiafu An, Difang Huang, Chen Lin, Mingzhu Tai. “Measuring gender and racial biases in large language models: Intersectional evidence from
automated resume evaluation”. In: \emph{PNAS Nexus} 4.3 (2025), pgaf089. \textsc{doi}: \nolinkurl
{10.1093/pnasnexus/pgaf089}.
{}
\bibitem{tax104}\bibalias{ref103}

Purva Prasad Gosavi, Vaishnavi Murlidhar Kulkarni, and Alan F. Smeaton. “Capturing Bias Diversity in LLMs”.
In: \emph{2024 2nd International Conference on Foundation and Large Language Models (FLLM)}. IEEE, 2024,
pp. 593–598. \textsc{doi}: \nolinkurl {10.1109/fllm63129.2024.10852459}.
{}
\bibitem{tax078}\bibalias{ref118}

Tavishi Choudhary. “Political Bias in Large Language Models: A Comparative Analysis of ChatGPT-4,
Perplexity, Google Gemini, and Claude”. In: \emph{IEEE Access} 13 (2025), pp. 11341–11379. \textsc{doi}:
\nolinkurl {10.1109/access.2024.3523764}.
{}
\bibitem{tax079}\bibalias{ref122}
\bibalias{rozado2024political}

David Rozado. “The political preferences of LLMs”. In: \emph{PLOS ONE} 19.7 (2024), e0306621. \textsc{doi}:
\nolinkurl {10.1371/journal.pone.0306621}.
{}
\bibitem{tax072}\bibalias{ref31}

Guilherme F.C.F. Almeida, José Luiz Nunes, Neele Engelmann, Alex Wiegmann, Marcelo de Araújo. “Exploring the psychology of LLMs’ moral and legal reasoning”. In:
\emph{Artificial Intelligence} 333 (2024), p. 104145. \textsc{doi}: \nolinkurl {10.1016/j.artint.2024.104145}.
{}
\bibitem{tax074}\bibalias{ref55}

Luana Bulla, Misael Mongiovì, Stefano De Giorgis, Aldo Gangemi. \emph{Large Language Models meet moral values: A comprehensive assessment of moral
abilities}. 2024. \textsc{doi}: \nolinkurl {10.2139/ssrn.4907562}.
{}
\bibitem{tax076}\bibalias{ref63}
\bibalias{simmons2022moral}

Gabriel Simmons. “Moral Mimicry: Large Language Models Produce Moral Rationalizations Tailored to Political
Identity”. In: \emph{Proceedings of the 61st Annual Meeting of the Association for Computational Linguistics:
Student Research Workshop}. 2023, pp. 282–297. \textsc{doi}: \nolinkurl {10.18653/v1/2023.acl-srw.40}.
{}
\bibitem{tax080}\bibalias{ref125}

Mehdi Khamassi, Marceau Nahon, and Raja Chatila. “Strong and weak alignment of large language models with
human values”. In: \emph{Scientific Reports} 14.1 (2024), p. 19399. \textsc{doi}: \nolinkurl
{10.1038/s41598-024-70031-3}.
{}
\bibitem{tax160}\bibalias{ref124}

Hannah Rose Kirk, Bertie Vidgen, Paul Röttger, Scott A. Hale. “The benefits, risks and bounds of personalizing the alignment of large language models
to individuals”. In: \emph{Nature Machine Intelligence} 6.4 (2024), pp. 383–392. \textsc{doi}: \nolinkurl
{10.1038/s42256-024-00820-y}.
{}
\bibitem{tax164}\bibalias{ref142}

Francesco Salvi, Manoel Horta Ribeiro, Riccardo Gallotti, Robert West. “On the conversational persuasiveness of GPT-4”. In: \emph{Nature Human Behaviour} 9.8
(2025), pp. 1645–1653. \textsc{doi}: \nolinkurl {10.1038/s41562-025-02194-6}.
{}
\bibitem{tax167}\bibalias{ref146}

Sandra Matz, Jake Teeny, Sumer Sumeet Vaid, Heinrich Peters, Gabriella M. Harari, Moran Cerf. \emph{The potential of generative AI for personalized persuasion at scale}. 2023.
\textsc{doi}: \nolinkurl {10.31234/osf.io/rn97c}.
{}
\bibitem{hagendorff2024deception}Thilo Hagendorff. “Deception abilities emerged in large language models”. In: \emph{Proceedings of the National
Academy of Sciences} 121.24 (2024), e2317967121.
{}
\bibitem{cui2024can}Ziyan Cui, Ning Li, and Huaikang Zhou. “Can ai replace human subjects? a large-scale replication of psychological
experiments with llms”. In: \emph{A Large-Scale Replication of Psychological Experiments with LLMs (August
25, 2024)} (2024).
{}
\bibitem{wang2024large}Angelina Wang, Jamie Morgenstern, and John P Dickerson. “Large language models that replace human
participants can harmfully misportray and flatten identity groups”. In: \emph{arXiv preprint arXiv:2402.01908}
(2024).
{}
\bibitem{tax086}\bibalias{fontana2025nicer}
\bibalias{ref82}

Nicolò Fontana, Francesco Pierri, and Luca Maria Aiello. “Nicer than Humans: How Do Large Language Models
Behave in the Prisoner’s Dilemma?” In: \emph{Proceedings of the International AAAI Conference on Web and
Social Media} 19 (2025), pp. 522–535. \textsc{doi}: \nolinkurl {10.1609/icwsm.v19i1.35829}.
{}
\bibitem{tax087}\bibalias{ref84}

Caoyun Fan, Jindou Chen, Yaohui Jin, Hao He. “Can Large Language Models Serve as Rational Players in Game Theory? A Systematic
Analysis”. In: \emph{Proceedings of the AAAI Conference on Artificial Intelligence} 38.16 (2024),
pp. 17960–17967. \textsc{doi}: \nolinkurl {10.1609/aaai.v38i16.29751}.
{}
\bibitem{li2025emergence}Haoyang Li, Xiao Jia, and Zhanzhan Zhao. “The Emergence of Altruism in Large-Language-Model Agents
Society”. In: \emph{arXiv preprint arXiv:2509.22537} (2025). \textsc{doi}: \nolinkurl
{10.48550/arXiv.2509.22537}.
{}
\bibitem{tax100}\bibalias{ashery2025emergent}
\bibalias{ref76}

Ariel Flint Ashery, Luca Maria Aiello, and Andrea Baronchelli. “Emergent social conventions and collective bias
in LLM populations”. In: \emph{Science Advances} 11.20 (2025), eadu9368. \textsc{doi}: \nolinkurl
{10.1126/sciadv.adu9368}.
{}
\bibitem{tax159}\bibalias{burton2024large}
\bibalias{ref121}

Jason W. Burton, Ezequiel Lopez-Lopez, Shahar Hechtlinger, Zoe Rahwan, Samuel Aeschbach, Michiel A. Bakker, Joshua A. Becker, Aleks Berditchevskaia, Julian Berger, Levin Brinkmann, Lucie Flek, Stefan M. Herzog, Saffron Huang, Sayash Kapoor, Arvind Narayanan, Anne-Marie Nussberger, Taha Yasseri, Pietro Nickl, Abdullah Almaatouq, Ulrike Hahn, Ralf H. J. M. Kurvers, Susan Leavy, Iyad Rahwan, Divya Siddarth, Alice Siu, Anita W. Woolley, Dirk U. Wulff, Ralph Hertwig. “How large language models can reshape collective intelligence”. In: \emph{Nature Human
Behaviour} 8.9 (2024), pp. 1643–1655. \textsc{doi}: \nolinkurl {10.1038/s41562-024-01959-9}.
{}
\bibitem{tax101}\bibalias{chuang2023wisdom}
\bibalias{ref80}

Yun-Shiuan Chuang, Siddharth Suresh, Nikunj Harlalka, Agam Goyal, Robert Hawkins, Sijia Yang, Dhavan Shah, Junjie Hu, Timothy T. Rogers. “The Wisdom of Partisan Crowds: Comparing Collective Intelligence in Humans and
LLM-based Agents”. In: \emph{Proceedings of the 46th Annual Conference of the Cognitive Science Society}.
2024, pp. 5824–5831. \textsc{url}: \url {https://escholarship.org/uc/item/3k67x8s5}.
{}
\bibitem{tax102}\bibalias{ref86}

Mingchen Zhuge, Haozhe Liu, Francesco Faccio, Dylan R. Ashley, Róbert Csordás, Anand Gopalakrishnan, Abdullah Hamdi, Hasan Abed Al Kader Hammoud, Vincent Herrmann, Kazuki Irie, Louis Kirsch, Bing Li, Guohao Li, Shuming Liu, Jinjie Mai, Piotr Piękos, Aditya A. Ramesh, Imanol Schlag, Weimin Shi, Aleksandar Stanić, Wenyi Wang, Yuhui Wang, Mengmeng Xu, Deng-Ping Fan, Bernard Ghanem, Jürgen Schmidhuber. “Mindstorms in Natural Language-Based Societies of Mind”. In: \emph{Computational
Visual Media} 11.1 (2025), pp. 29–81. \textsc{doi}: \nolinkurl {10.26599/cvm.2025.9450460}.
{}
\bibitem{piao2025emergence}Jinghua Piao, Zhihong Lu, Chen Gao, Fengli Xu, Fernando P. Santos, Yong Li, James Evans. “Emergence of human-like polarization among large language model agents”. In: \emph{arXiv
preprint arXiv:2501.05171} (2025).
{}
Version 1: \url{https://arxiv.org/abs/2501.05171v1}.
\bibitem{tax127}\bibalias{ref195}

Mehwish Nasim, Syed Muslim Gilani, Amin Qasmi, Usman Naseem. “Simulating Influence Dynamics with LLM Agents”. In: \emph{2025 IEEE International
Conference on Big Data (BigData)}. IEEE, 2025, pp. 6353–6356. \textsc{doi}: \nolinkurl
{10.1109/bigdata66926.2025.11401885}.
{}
\bibitem{lin2025simspark}Ziyue Lin, Yi Shan, Lin Gao, Xinghua Jia, and Siming Chen. “SimSpark: Interactive Simulation of Social Media Behaviors”. In: \emph{Proceedings of the
ACM on Human-Computer Interaction} 9.2 (2025), pp. 1–32.
{}
\bibitem{tax116}\bibalias{ref137}

Linghao Wang, Zheyuan Jiang, Chenke Hu, Jun Zhao, Zheng Zhu, Xiqun Chen, Ziyi Wang, Tianming Liu, Guibing He, Yafeng Yin, Der-Horng Lee. “Comparing AI and human decision-making mechanisms in daily collaborative experiments”.
In: \emph{iScience} 28.6 (2025), p. 112711. \textsc{doi}: \nolinkurl {10.1016/j.isci.2025.112711}.
{}
\bibitem{tax109}\bibalias{ref81}

Giorgio Piatti, Zhijing Jin, Max Kleiman-Weiner, Bernhard Schölkopf, Mrinmaya Sachan, Rada Mihalcea. “Cooperate or Collapse: Emergence of Sustainable Cooperation in a Society of LLM Agents”.
In: \emph{Advances in Neural Information Processing Systems 37}. Neural Information Processing Systems
Foundation, Inc. (NeurIPS), 2024, pp. 111715–111759. \textsc{doi}: \nolinkurl {10.52202/079017-3548}.
{}
\bibitem{tax107}\bibalias{ref78}
\bibalias{ren2024emergence}

Siyue Ren, Zhiyao Cui, Ruiqi Song, Zhen Wang, Shuyue Hu. “Emergence of Social Norms in Generative Agent Societies: Principles and Architecture”. In:
\emph{Proceedings of the Thirty-Third International Joint Conference on Artificial Intelligence}. 2024,
pp. 7895–7903. \textsc{doi}: \nolinkurl {10.24963/ijcai.2024/874}.
{}
\bibitem{tax110}\bibalias{ref85}

Xudong Guo, Kaixuan Huang, Jiale Liu, Wenhui Fan, Natalia Vélez, Qingyun Wu, Huazheng Wang, Thomas L. Griffiths, Mengdi Wang. “Embodied llm agents learn to cooperate in organized teams”. In: \emph{IEEE Transactions
on Computational Social Systems} 13.2 (2026), pp. 2514–2530. \textsc{doi}: \nolinkurl
{10.1109/tcss.2025.3637527}.
{}
\bibitem{tax108}\bibalias{ref79}

Gordon Dai, Weijia Zhang, Jinhan Li, Siqi Yang, Chidera Onochie lbe, Srihas Rao, Arthur Caetano, Misha Sra. “Artificial Leviathan: Exploring Social Evolution of LLM Agents Through the Lens of
Hobbesian Social Contract Theory”. In: \emph{Frontiers in Physics} 14 (2026), p. 1700712. \textsc{doi}:
\nolinkurl {10.3389/fphy.2026.1700712}.
{}
\bibitem{tax106}\bibalias{ref8}

Jintian Zhang, Xin Xu, Ningyu Zhang, Ruibo Liu, Bryan Hooi, Shumin Deng. “Exploring Collaboration Mechanisms for LLM Agents: A Social Psychology View”. In:
\emph{Proceedings of the 62nd Annual Meeting of the Association for Computational Linguistics (Volume 1:
Long Papers)}. Association for Computational Linguistics, 2024, pp. 14544–14607. \textsc{doi}: \nolinkurl
{10.18653/v1/2024.acl-long.782}.
{}
\bibitem{tax126}\bibalias{piao2025agentsociety}
\bibalias{ref194}

Jinghua Piao, Yuwei Yan, Jun Zhang, Nian Li, Junbo Yan, Xiaochong Lan, Zhihong Lu, Zhiheng Zheng, Jing Yi Wang, Di Zhou, Chen Gao, Fengli Xu, Fang Zhang, Ke Rong, Jun Su, Yong Li. “AgentSociety: Large-Scale Simulation of LLM-Driven Generative Agents Advances
Understanding of Human Behaviors and Society”. In: \emph{iFuture} (2026). \textsc{doi}: \nolinkurl
{10.26599/if.2026.9710004}.
{}
\bibitem{guan2025lightsociety}
Haoxiang Guan, Jiyan He, Liyang Fan, Zhenzhen Ren, Shaobin He, Xin Yu, Yuan Chen, Xueyin Xu, Shuxin Zheng, Yan Gao, Enhong Chen, Tie-Yan Liu, Zhen Liu. \emph{Modeling Earth-Scale Human-Like Societies with One Billion Agents}. arXiv preprint (2025; revised 2026). \textsc{doi}: \nolinkurl{10.48550/arXiv.2506.12078}.
Version 2: \url{https://arxiv.org/abs/2506.12078v2}.
{}
\bibitem{li2026matraix}
Xiaomin Li, Yuexing Hao, Jianheng Hou, Jintao Huang, Qianfeng Wen, Shirley Huang, Yifan Liu, Xiaoyi Liu, Yilan Fan, Yijun Wang, Koutian Wu, Ruoqi Gao, Muhammad Ahmed Mohsin, Jing Tang, Brihi Joshi, Heming Liu, Zheyuan Deng, Zonglin Di, Sankalp Jajee, Jiuyao Lu, Zhiwei Zhang, Saksham Kapoor, Ishan Gupta, Yunhan Zhao, Chanwoo Park, Yucheng Lu, Bing Hu, Weihang Xiao, Aravind Mohan, Hanwen Xing, Runyu Zhang, Mihir Kulshreshtha, Yuanda Xu, Qianyu Zhu, Dianzhuo Wang, Yuxin Xiao, Bowen Jiang, Yongye Su, Wenhao Chai, Zuxin Liu, Lawrence Yunliang Chen, Xuandong Zhao, Ethan Ye, Shivam Patel, Jason Xie, Alex Martin Richmond, Weixiang Ding, Emre Okcular, Diya Mathew, Ziheng Wang, Rana M. Shahroz Khan, Zhejian Peng, Fang Wu, Fan Nie, Xinyang Han, Yubin Kim, Jiawei Zhang, Zhenting Qi, Huangyuan Su, Xu Pan, Abinitha Gourabathina, Hyewon Jeong, Hemanth Neelgund Ramesh, Kumail Alhamoud, Kimia Hamidieh, Zidi Xiong, Samuel Schmidgall, Pengrui Han, Yepeng Huang, Yongheng Wang, Bowen Yang, Alex Gu, Yuchu Wang, Akshay Paruchuri, Brenna Li, Hejie Cui, Jiayuan Ding, Chaosheng Dong, Jiahao Wang, Yixuan He, Chi Wang, Pamela Bhattacharya, Tianyi Peng, Paul Pu Liang, Mitchell Gordon, Yilun Du, Marinka Zitnik, James Zou, Prasanna Tambe, Philip Torr, Emily Fox, Asu Ozdaglar, Dawn Song. \emph{MatrAIx: Simulating the World with 8.3 Billion Persona Agents}. arXiv preprint (2026). \textsc{doi}: \nolinkurl{10.48550/arXiv.2608.04205}.
Version 1: \url{https://arxiv.org/abs/2608.04205v1}.
{}
\bibitem{tax029}\bibalias{ref66}

Kyusik Kim, Hyeonseok Jeon, Jeongwoo Ryu, Bongwon Suh. “Will LLMs Sink or Swim? Exploring Decision-Making Under Pressure”. In: \emph{Findings of
the Association for Computational Linguistics: EMNLP 2024}. Association for Computational Linguistics, 2024,
pp. 11425–11450. \textsc{doi}: \nolinkurl {10.18653/v1/2024.findings-emnlp.668}.
{}
\bibitem{tax038}\bibalias{fang2024llm}
\bibalias{ref168}

Jingchao Fang, Nikos Arechiga, Keiichi Namikoshi, Nayeli Bravo, Candice Hogan, David A. Shamma. “On LLM Wizards: Identifying Large Language Models’ Behaviors for Wizard of Oz
Experiments”. In: \emph{Proceedings of the 24th ACM International Conference on Intelligent Virtual Agents}.
2024, 16:1–16:11. \textsc{doi}: \nolinkurl {10.1145/3652988.3673967}.
{}
\bibitem{tax035}\bibalias{ref145}
\bibalias{van2025your}

Karin van Es and Dennis Nguyen. “"Your friendly AI assistant": the anthropomorphic self-representations of
ChatGPT and its implications for imagining AI”. In: \emph{AI \& SOCIETY} 40.5 (2025), pp. 3591–3603.
\textsc{doi}: \nolinkurl {10.1007/s00146-024-02108-6}.
{}
\bibitem{colombatto2025influence}Clara Colombatto, Jonathan Birch, and Stephen M Fleming. “The influence of mental state attributions on trust
in large language models”. In: \emph{Communications Psychology} 3.1 (2025), p. 84.
{}
\bibitem{street2024llm}Winnie Street. “Llm theory of mind and alignment: Opportunities and risks”. In: \emph{arXiv preprint
arXiv:2405.08154} (2024).
{}
\bibitem{glickman2025human}Moshe Glickman and Tali Sharot. “How human–AI feedback loops alter human perceptual, emotional and social
judgements”. In: \emph{Nature Human Behaviour} 9.2 (2025), pp. 345–359.
{}
\bibitem{tax036}\bibalias{jones2025people}
\bibalias{ref163}

Cameron R. Jones and Benjamin K. Bergen. \emph{People cannot distinguish GPT-4 from a human in a Turing
test}. arXiv. 2024. \textsc{doi}: \nolinkurl {10.48550/arXiv.2405.08007}.
{}
\bibitem{tax136}\bibalias{pataranutaporn2023influencing}
\bibalias{ref135}

Pat Pataranutaporn, Ruby Liu, Ed Finn, Pattie Maes. “Influencing human-AI interaction by priming beliefs about AI can increase perceived
trustworthiness, empathy and effectiveness”. In: \emph{Nature Machine Intelligence} 5.10 (2023), pp. 1076–1086.
\textsc{doi}: \nolinkurl {10.1038/s42256-023-00720-7}.
{}
\bibitem{tax152}\bibalias{ref182}
\bibalias{rubin2025comparing}

Matan Rubin, Joanna Z. Li, Federico Zimmerman, Desmond C. Ong, Amit Goldenberg, Anat Perry. \emph{Comparing the value of perceived human versus AI-generated empathy}. 2024.
\textsc{doi}: \nolinkurl {10.31219/osf.io/ng97s\_v1}.
{}
\bibitem{tax149}\bibalias{ref149}

Katy Ilonka Gero, Tao Long, and Lydia B Chilton. “Social Dynamics of AI Support in Creative Writing”. In:
\emph{Proceedings of the 2023 CHI Conference on Human Factors in Computing Systems}. ACM, 2023,
pp. 1–15. \textsc{doi}: \nolinkurl {10.1145/3544548.3580782}.
{}
\bibitem{tax194}\bibalias{ref167}

Mengqi Liu and Faten M’Hiri. “Beyond Traditional Teaching: Large Language Models as Simulated Teaching
Assistants in Computer Science”. In: \emph{Proceedings of the 55th ACM Technical Symposium on Computer
Science Education V. 1}. ACM, 2024, pp. 743–749. \textsc{doi}: \nolinkurl {10.1145/3626252.3630789}.
{}
\bibitem{tax148}\bibalias{ref132}

Darren Frey and Daniel H. Weiss. “One Person Dialogues: Concerns About AI-Human Interactions”. In:
\emph{Harvard Data Science Review} 7.2 (2025). \textsc{doi}: \nolinkurl {10.1162/99608f92.01674a29}.
{}
\bibitem{grogan2025ai}Clare Grogan, Jackie Kay, and María Pérez-Ortiz. “AI Will Always Love You: Studying Implicit Biases in
Romantic AI Companions”. In: \emph{arXiv preprint arXiv:2502.20231} (2025).
{}
\bibitem{tax129}\bibalias{ref25}
\bibalias{schneider2025mental}

Johannes Schneider. “Mental model shifts in human-LLM interactions”. In: \emph{Journal of Intelligent
Information Systems} 63.5 (2025), pp. 1737–1752. \textsc{doi}: \nolinkurl {10.1007/s10844-025-00960-6}.
{}
\bibitem{tax170}\bibalias{chiang2024enhancing}
\bibalias{ref160}

Chun-Wei Chiang, Zhuoran Lu, Zhuoyan Li, Ming Yin. “Enhancing AI-Assisted Group Decision Making through LLM-Powered Devil’s
Advocate”. In: \emph{Proceedings of the 29th International Conference on Intelligent User Interfaces}. ACM,
2024, pp. 103–119. \textsc{doi}: \nolinkurl {10.1145/3640543.3645199}.
{}
\bibitem{tax182}\bibalias{ref159}

Jessica He, Stephanie Houde, Gabriel E. Gonzalez, Darío Andrés Silva Moran, Steven I. Ross, Michael Muller, Justin D. Weisz. “AI and the Future of Collaborative Work: Group Ideation with an LLM in a Virtual Canvas”. In:
\emph{Proceedings of the 3rd Annual Meeting of the Symposium on Human-Computer Interaction for Work}.
ACM, 2024, pp. 1–14. \textsc{doi}: \nolinkurl {10.1145/3663384.3663398}.
{}
\bibitem{tax141}\bibalias{ha2024clochat}
\bibalias{ref166}

Juhye Ha, Hyeon Jeon, Daeun Han, Jinwook Seo, Changhoon Oh. “CloChat: Understanding How People Customize, Interact, and Experience Personas in Large
Language Models”. In: \emph{Proceedings of the CHI Conference on Human Factors in Computing Systems}.
ACM, 2024, pp. 1–24. \textsc{doi}: \nolinkurl {10.1145/3613904.3642472}.
{}
\bibitem{tax180}\bibalias{li2024value}
\bibalias{ref154}

Zhuoyan Li, Chen Liang, Jing Peng, Ming Yin. “The Value, Benefits, and Concerns of Generative AI-Powered Assistance in Writing”. In:
\emph{Proceedings of the CHI Conference on Human Factors in Computing Systems}. ACM, 2024, pp. 1–25.
\textsc{doi}: \nolinkurl {10.1145/3613904.3642625}.
{}
\bibitem{humlum2025unequal}Anders Humlum and Emilie Vestergaard. “The unequal adoption of ChatGPT exacerbates existing inequalities
among workers”. In: \emph{Proceedings of the National Academy of Sciences} 122.1 (2025), e2414972121.
{}
\bibitem{tax161}\bibalias{ref134}

Jörg Noller. “Connectionism about human agency: responsible AI and the social lifeworld”. In: \emph{AI \&
SOCIETY} 40.5 (2025), pp. 3881–3890. \textsc{doi}: \nolinkurl {10.1007/s00146-024-02133-5}.
{}
\bibitem{tax134}\bibalias{ref131}
\bibalias{ye2023improved}

Yang Ye, Hengxu You, and Jing Du. “Improved Trust in Human-Robot Collaboration With ChatGPT”. In:
\emph{IEEE Access} 11 (2023), pp. 55748–55754. \textsc{doi}: \nolinkurl {10.1109/access.2023.3282111}.
{}
\bibitem{tax178}\bibalias{ref119}
\bibalias{wan2024felt}

Qian Wan, Siying Hu, Yu Zhang, Piaohong Wang, Bo Wen, Zhicong Lu. \emph{"It Felt Like Having a Second Mind": Investigating Human-AI Co-creativity in Prewriting
with Large Language Models}. arXiv. 2023. \textsc{doi}: \nolinkurl {10.48550/arXiv.2307.10811}.
{}
\bibitem{tax177}\bibalias{ref113}

Anil R. Doshi and Oliver P. Hauser. “Generative AI enhances individual creativity but reduces the collective
diversity of novel content”. In: \emph{Science Advances} 10.28 (2024), eadn5290. \textsc{doi}: \nolinkurl
{10.1126/sciadv.adn5290}.
{}
\bibitem{tax179}\bibalias{anderson2024homogenization}
\bibalias{ref151}

Barrett R. Anderson, Jash Hemant Shah, and Max Kreminski. “Homogenization Effects of Large Language
Models on Human Creative Ideation”. In: \emph{Proceedings of Creativity and Cognition}. 2024, pp. 1–13.
\textsc{doi}: \nolinkurl {10.1145/3635636.3656204}.
{}
\bibitem{tax006}\bibalias{ref15}

Luning Sun, Yuzhuo Yuan, Yuan Yao, Yanyan Li, Hao Zhang, Xing Xie, Xiting Wang, Fang Luo, David Stillwell. \emph{Large language models show both individual and collective creativity comparable to
humans}. arXiv. 2024. \textsc{doi}: \nolinkurl {10.48550/arXiv.2412.03151}.
{}
\bibitem{tax011}\bibalias{ref22}

Giorgio Franceschelli and Mirco Musolesi. “On the creativity of large language models”. In: \emph{AI \&
SOCIETY} 40.5 (2025), pp. 3785–3795. \textsc{doi}: \nolinkurl {10.1007/s00146-024-02127-3}.
{}
\bibitem{tax184}\bibalias{ref184}

Ezra N. S. Lockhart. “Creativity in the age of AI: the human condition and the limits of machine generation”. In:
\emph{Journal of Cultural Cognitive Science} 9 (2025), pp. 83–88. \textsc{doi}: \nolinkurl
{10.1007/s41809-024-00158-2}.
{}
\bibitem{tax185}\bibalias{jeon2023large}
\bibalias{ref116}

Jaeho Jeon and Seongyong Lee. “Large language models in education: A focus on the complementary relationship
between human teachers and ChatGPT”. In: \emph{Education and Information Technologies} 28.12 (2023),
pp. 15873–15892. \textsc{doi}: \nolinkurl {10.1007/s10639-023-11834-1}.
{}
\bibitem{tax188}\bibalias{lang2025transforming}
\bibalias{ref130}

Qi Lang, Minjuan Wang, Minghao Yin, Shuang Liang, Wenzhuo Song. “Transforming Education With Generative AI (GAI): Key Insights and Future Prospects”. In:
\emph{IEEE Transactions on Learning Technologies} 18 (2025), pp. 230–242. \textsc{doi}: \nolinkurl
{10.1109/tlt.2025.3537618}.
{}
\bibitem{tax189}\bibalias{ref140}

Marine Cloux, Davy Monticolo, and Raphaël Bary. “Understanding how to assess the impact of Artificial
Intelligence on learning: a systematic review”. In: \emph{Interactive Learning Environments} 33.7 (2025),
pp. 4618–4632. \textsc{doi}: \nolinkurl {10.1080/10494820.2025.2468986}.
{}
\bibitem{tax198}\bibalias{ref187}

Chung Kwan Lo, Khe Foon Hew, and Morris Siu-yung Jong. “The influence of ChatGPT on student engagement:
A systematic review and future research agenda”. In: \emph{Computers \& Education} 219 (2024),
p. 105100. \textsc{doi}: \nolinkurl {10.1016/j.compedu.2024.105100}.
{}
\bibitem{tax142}\bibalias{ref173}

Seyoung Lee and Gain Park. “Exploring the Impact of ChatGPT Literacy on User Satisfaction: The Mediating
Role of User Motivations”. In: \emph{Cyberpsychology, Behavior, and Social Networking} 26.12 (2023),
pp. 913–918. \textsc{doi}: \nolinkurl {10.1089/cyber.2023.0312}.
{}
\bibitem{tax195}\bibalias{kumar2024guiding}
\bibalias{ref174}

Harsh Kumar, Ilya Musabirov, Mohi Reza, Jiakai Shi, Xinyuan Wang, Joseph Jay Williams, Anastasia Kuzminykh, Michael Liut. “Guiding Students in Using LLMs in Supported Learning Environments: Effects on
Interaction Dynamics, Learner Performance, Confidence, and Trust”. In: \emph{Proceedings of the ACM on
Human-Computer Interaction} 8 (2024), pp. 1–30. \textsc{doi}: \nolinkurl {10.1145/3687038}.
{}
\bibitem{tax196}\bibalias{ref185}

Chung Yee Lai, Kwok Yip Cheung, and Chee Seng Chan. “Exploring the role of intrinsic motivation in ChatGPT
adoption to support active learning: An extension of the technology acceptance model”. In: \emph{Computers and
Education: Artificial Intelligence} 5 (2023), p. 100178. \textsc{doi}: \nolinkurl {10.1016/j.caeai.2023.100178}.
{}
\bibitem{tax190}\bibalias{ref148}

Kamil Malinka, Martin Peresíni, Anton Firc, Ondrej Hujnák, Filip Janus. “On the Educational Impact of ChatGPT: Is Artificial Intelligence Ready to Obtain a
University Degree?” In: \emph{Proceedings of the 2023 Conference on Innovation and Technology in Computer
Science Education V. 1}. ACM, 2023, pp. 47–53. \textsc{doi}: \nolinkurl {10.1145/3587102.3588827}.
{}
\bibitem{tax191}\bibalias{joshi2024chatgpt}
\bibalias{ref157}

Ishika Joshi, Ritvik Budhiraja, Harshal Dev, Jahnvi Kadia, Mohammad Osama Ataullah, Sayan Mitra, Harshal D. Akolekar, Dhruv Kumar. “ChatGPT in the Classroom: An Analysis of Its Strengths and Weaknesses for Solving
Undergraduate Computer Science Questions”. In: \emph{Proceedings of the 55th ACM Technical Symposium on
Computer Science Education V. 1}. ACM, 2024, pp. 625–631. \textsc{doi}: \nolinkurl
{10.1145/3626252.3630803}.
{}
\bibitem{tax037}\bibalias{ref164}

Jaromir Savelka, Arav Agarwal, Marshall An, Chris Bogart, Majd Sakr. “Thrilled by Your Progress! Large Language Models (GPT-4) No Longer Struggle to Pass
Assessments in Higher Education Programming Courses”. In: \emph{Proceedings of the 2023 ACM Conference on
International Computing Education Research V.1}. ACM, 2023, pp. 78–92. \textsc{doi}: \nolinkurl
{10.1145/3568813.3600142}.
{}
\bibitem{tax192}\bibalias{ref161}

Hyanghee Park and Daehwan Ahn. “The Promise and Peril of ChatGPT in Higher Education: Opportunities,
Challenges, and Design Implications”. In: \emph{Proceedings of the CHI Conference on Human Factors in
Computing Systems}. ACM, 2024, pp. 1–21. \textsc{doi}: \nolinkurl {10.1145/3613904.3642785}.
{}
\bibitem{tax193}\bibalias{ref162}

Judy Sheard, Paul Denny, Arto Hellas, Juho Leinonen, Lauri Malmi, Simon. “Instructor Perceptions of AI Code Generation Tools - A Multi-Institutional Interview Study”.
In: \emph{Proceedings of the 55th ACM Technical Symposium on Computer Science Education V. 1}. ACM,
2024, pp. 1223–1229. \textsc{doi}: \nolinkurl {10.1145/3626252.3630880}.
{}
\bibitem{perez2024cultural}Jérémy Perez, Corentin Léger, Marcela Ovando-Tellez, Chris Foulon, Joan Dussauld, Pierre-Yves Oudeyer, Clément Moulin-Frier. “Cultural evolution in populations of Large Language Models”. In: \emph{arXiv preprint
arXiv:2403.08882} (2024).
{}
\bibitem{lee2004trust}John D Lee and Katrina A See. ``Trust in automation: Designing for appropriate reliance''. In: \emph{Human Factors} 46.1 (2004), pp. 50--80.
{}
\bibitem{parasuraman1997humans}Raja Parasuraman and Victor Riley. ``Humans and automation: Use, misuse, disuse, abuse''. In: \emph{Human Factors} 39.2 (1997), pp. 230--253.
{}
\bibitem{zhou2025rationality}Zhilun Zhou, Jing Yi Wang, Nicholas Sukiennik, Chen Gao, Fengli Xu, Yong Li, and James Evans. ``Rationality Check! Benchmarking the Rationality of Large Language Models''. In: \emph{arXiv preprint arXiv:2509.14546} (2025). \textsc{doi}: \nolinkurl{10.48550/arXiv.2509.14546}.
{}
\bibitem{campbell1959convergent}Donald T Campbell and Donald W Fiske. ``Convergent and discriminant validation by the multitrait-multimethod matrix''. In: \emph{Psychological Bulletin} 56.2 (1959), pp. 81--105.
{}
\bibitem{cronbach1955construct}Lee J Cronbach and Paul E Meehl. ``Construct validity in psychological tests''. In: \emph{Psychological Bulletin} 52.4 (1955), pp. 281--302.
{}
\bibitem{suchman1987plans}Lucille Alice Suchman. \emph{Plans and situated actions: The problem of human-machine communication}. Cambridge University Press, 1987.
{}
\bibitem{elliott2017living}Julian H. Elliott, Anneliese Synnot, Tari Turner, Mark Simmonds, Elie A. Akl, Steve McDonald, Georgia Salanti, Joerg Meerpohl, Harriet MacLehose, John Hilton, David Tovey, Ian Shemilt, and James Thomas, on behalf of the Living Systematic Review Network. ``Living systematic review: 1. Introduction---the why, what, when, and how''. In: \emph{Journal of Clinical Epidemiology} 91 (2017), pp. 23--30. \textsc{doi}: \nolinkurl{10.1016/j.jclinepi.2017.08.010}.
{}
\bibitem{comaniciu1999mean}Dorin Comaniciu and Peter Meer. ``Mean shift analysis and applications''. In: \emph{Proceedings of the Seventh IEEE International Conference on Computer Vision} 2 (1999), pp. 1197--1203.
{}
\bibitem{popay2006guidance}Jennie Popay, Helen Roberts, Amanda Sowden, Mark Petticrew, Lisa Arai, Mark Rodgers, Nicky Britten, Katrina Roen, and Steven Duffy. \emph{Guidance on the Conduct of Narrative Synthesis in Systematic Reviews: A Product from the ESRC Methods Programme}. Version 1. Lancaster University / ESRC Methods Programme, 2006. \textsc{doi}: \nolinkurl{10.13140/2.1.1018.4643}.
{}
\bibitem{borenstein2021introduction}Michael Borenstein, Larry V Hedges, Julian PT Higgins, and Hannah R Rothstein. \emph{Introduction to meta-analysis}. John Wiley \& Sons, 2021.
{}
\bibitem{bender2020climbing}Emily M Bender and Alexander Koller. ``Climbing towards NLU: On meaning, form, and understanding in the age of data''. In: \emph{Proceedings of the 58th Annual Meeting of the Association for Computational Linguistics} (2020), pp. 5185--5198.
{}
\bibitem{bender2021dangers}Emily M Bender, Timnit Gebru, Angelina McMillan-Major, and Shmargaret Shmitchell. ``On the dangers of stochastic parrots: Can language models be too big?''. In: \emph{Proceedings of the 2021 ACM Conference on Fairness, Accountability, and Transparency} (2021), pp. 610--623.
{}
\bibitem{raji2022fallacy}Inioluwa Deborah Raji, I Elizabeth Kumar, Aaron Horowitz, and Andrew Selbst. ``The fallacy of AI functionality''. In: \emph{Proceedings of the 2022 ACM Conference on Fairness, Accountability, and Transparency} (2022), pp. 959--972.
{}
\bibitem{crawford2021atlas}Kate Crawford. \emph{The atlas of AI: Power, politics, and the planetary costs of artificial intelligence}. Yale University Press, 2021.
{}
\bibitem{srnicek2017platform}Nick Srnicek. \emph{Platform capitalism}. John Wiley \& Sons, 2017.
{}
\bibitem{zuboff2023age}Shoshana Zuboff. ``The age of surveillance capitalism''. In: \emph{Social theory re-wired} (2023), pp. 203--213.
{}
\bibitem{searle1980minds}John R Searle. ``Minds, brains, and programs''. In: \emph{Behavioral and Brain Sciences} 3.3 (1980), pp. 417--424.
{}
\bibitem{floridi2002philosophy}Luciano Floridi. ``What is the Philosophy of Information?''. In: \emph{Metaphilosophy} 33.1-2 (2002), pp. 123--145.
{}
\bibitem{Arcas2022large}Blaise Ag{\"u}era y Arcas. ``Do large language models understand us?''. In: \emph{Daedalus} 151.2 (2022), pp. 183--197. \textsc{doi}: \nolinkurl{10.1162/daed_a_01909}.
{}
\bibitem{horiguchi2024evolution}Ilya Horiguchi, Takahide Yoshida, and Takashi Ikegami. “Evolution of social norms in llm agents using natural
language”. In: \emph{arXiv preprint arXiv:2409.00993} (2024).
{}
\bibitem{tax131}\bibalias{jacob2025chat}
\bibalias{ref107}

Christo Jacob, Páraic Kerrigan, and Marco Bastos. “The chat-chamber effect: Trusting the AI hallucination”. In:
\emph{Big Data \& Society} 12.1 (2025), p. 20539517241306345. \textsc{doi}: \nolinkurl
{10.1177/20539517241306345}.
{}
\bibitem{wang2019superglue}Alex Wang, Yada Pruksachatkun, Nikita Nangia, Amanpreet Singh, Julian Michael, Felix Hill, Omer Levy, and Samuel Bowman. ``Superglue: A stickier benchmark for general-purpose language understanding systems''. In: \emph{Advances in Neural Information Processing Systems} 32 (2019).
{}
\bibitem{song2020mpnet}Kaitao Song, Xu Tan, Tao Qin, Jianfeng Lu, and Tie-Yan Liu. “Mpnet: Masked and permuted pre-training for language understanding”. In: \emph{Advances
in neural information processing systems} 33 (2020), pp. 16857–16867.
{}
\bibitem{macqueen1967some}James B McQueen. “Some methods of classification and analysis of multivariate observations”. In: \emph{Proc. of
5th Berkeley Symposium on Math. Stat. and Prob.} 1967, pp. 281–297.
{}
\bibitem{rousseeuw1987silhouettes}Peter J Rousseeuw. “Silhouettes: a graphical aid to the interpretation and validation of cluster analysis”. In:
\emph{Journal of computational and applied mathematics} 20 (1987), pp. 53–65.
{}
\bibitem{calinski1974dendrite}Tadeusz Caliński and Jerzy Harabasz. “A dendrite method for cluster analysis”. In: \emph{Communications in
Statistics-theory and Methods} 3.1 (1974), pp. 1–27.
{}
\bibitem{davies2009cluster}David L Davies and Donald W Bouldin. “A cluster separation measure”. In: \emph{IEEE transactions on pattern
analysis and machine intelligence} 2 (2009), pp. 224–227.
{}
\bibitem{blei2003latent}David M Blei, Andrew Y Ng, and Michael I Jordan. “Latent dirichlet allocation”. In: \emph{Journal of machine
Learning research} 3.Jan (2003), pp. 993–1022.
{}
\bibitem{cohen1960coefficient}Jacob Cohen. “A coefficient of agreement for nominal scales”. In: \emph{Educational and psychological
measurement} 20.1 (1960), pp. 37–46.
{}
\bibitem{hubert1985comparing}Lawrence Hubert and Phipps Arabie. “Comparing partitions”. In: \emph{Journal of classification} 2.1 (1985),
pp. 193–218.
{}
\bibitem{roberts2014structural}Margaret E. Roberts, Brandon M. Stewart, Dustin Tingley, Christopher Lucas, Jetson Leder-Luis, Shana Kushner Gadarian, Bethany Albertson, David G. Rand. “Structural Topic Models for Open-Ended Survey Responses”. In: \emph{American
Journal of Political Science} 58.4 (2014), pp. 1064–1082. \textsc{doi}: \nolinkurl {10.1111/ajps.12103}.
{}
\bibitem{roberts2019stm}Margaret E. Roberts, Brandon M. Stewart, and Dustin Tingley. “{stm}: An {R} Package for Structural Topic
Models”. In: \emph{Journal of Statistical Software} 91.2 (2019), pp. 1–40. \textsc{doi}: \nolinkurl
{10.18637/jss.v091.i02}. \textsc{url}: \url {https://www.jstatsoft.org/article/view/v091i02}.
\bibitem{arthur2006k}David Arthur and Sergei Vassilvitskii. \emph{k-means++: The advantages of careful seeding}. Tech. rep.
Stanford, 2006.
{}
\bibitem{lloyd1982least}Stuart Lloyd. “Least squares quantization in PCM”. In: \emph{IEEE transactions on information theory} 28.2
(1982), pp. 129–137.
{}
\bibitem{monti2003consensus}Stefano Monti, Pablo Tamayo, Jill Mesirov, and Todd Golub. “Consensus clustering: a resampling-based method for class discovery and visualization of
gene expression microarray data”. In: \emph{Machine learning} 52.1 (2003), pp. 91–118.
{}
\bibitem{mimno2011optimizing}David Mimno, Hanna Wallach, Edmund Talley, Miriam Leenders, and Andrew McCallum. “Optimizing semantic coherence in topic models”. In: \emph{Proceedings of the 2011
conference on empirical methods in natural language processing}. 2011, pp. 262–272.
{}
\bibitem{sievert2014ldavis}Carson Sievert and Kenneth Shirley. “LDAvis: A method for visualizing and interpreting topics”. In:
\emph{Proceedings of the workshop on interactive language learning, visualization, and interfaces}. 2014,
pp. 63–70.
{}
\bibitem{tax124}\bibalias{ref189}

Chen Gao, Xiaochong Lan, Zhihong Lu, Jinzhu Mao, Jinghua Piao, Huandong Wang, Depeng Jin, Yong Li. \emph{S\textasciicircum {}3: Social-network Simulation System with Large Language
Model-Empowered Agents}. 2025. \textsc{doi}: \nolinkurl {10.2139/ssrn.5604712}.
{}
\bibitem{yang2024oasis}Ziyi Yang, Zaibin Zhang, Zirui Zheng, Yuxian Jiang, Ziyue Gan, Zhiyu Wang, Zijian Ling, Jinsong Chen, Martz Ma, Bowen Dong, Prateek Gupta, Shuyue Hu, Zhenfei Yin, Guohao Li, Xu Jia, Lijun Wang, Bernard Ghanem, Huchuan Lu, Chaochao Lu, Wanli Ouyang, Yu Qiao, Philip Torr, Jing Shao. “{OASIS}: Open Agent Social Interaction Simulations with One Million Agents”. In:
\emph{arXiv preprint arXiv:2411.11581} (2024). \textsc{doi}: \nolinkurl {10.48550/arXiv.2411.11581}. arXiv:
\nolinkurl {2411.11581} \texttt{[cs.CL]}.
{}
\bibitem{tax122}\bibalias{ref100}

Joon Sung Park, Lindsay Popowski, Carrie Cai, Meredith Ringel Morris, Percy Liang, Michael S. Bernstein. “Social Simulacra: Creating Populated Prototypes for Social Computing Systems”. In:
\emph{Proceedings of the 35th Annual ACM Symposium on User Interface Software and Technology}. 2022,
74:1–74:18. \textsc{doi}: \nolinkurl {10.1145/3526113.3545616}.
{}
\bibitem{tax123}\bibalias{ref101}

Nian Li, Chen Gao, Mingyu Li, Yong Li, Qingmin Liao. “EconAgent: Large Language Model-Empowered Agents for Simulating Macroeconomic Activities”.
In: \emph{Proceedings of the 62nd Annual Meeting of the Association for Computational Linguistics (Volume 1:
Long Papers)}. Association for Computational Linguistics, 2024, pp. 15523–15536. \textsc{doi}: \nolinkurl
{10.18653/v1/2024.acl-long.829}.
{}
\bibitem{tax113}\bibalias{ref102}

Zengqing Wu, Run Peng, Shuyuan Zheng, Qianying Liu, Xu Han, Brian I. Kwon, Makoto Onizuka, Shaojie Tang, Chuan Xiao. “Shall We Team Up: Exploring Spontaneous Cooperation of Competing LLM Agents”. In:
\emph{Findings of the Association for Computational Linguistics: EMNLP 2024}. Association for Computational
Linguistics, 2024, pp. 5163–5186. \textsc{doi}: \nolinkurl {10.18653/v1/2024.findings-emnlp.297}.
{}
\bibitem{tax098}\bibalias{ref104}

Ayato Kitadai, Yudai Tsurusaki, Yusuke Fukasawa, Nariaki Nishino. “Toward a Novel Methodology in Economic Experiments: Simulation of the Ultimatum Game
with Large Language Models”. In: \emph{2023 IEEE International Conference on Big Data (BigData)}. IEEE,
2023, pp. 3168–3175. \textsc{doi}: \nolinkurl {10.1109/bigdata59044.2023.10386678}.
{}
\bibitem{tax114}\bibalias{ref108}

Yuan Li, Lichao Sun, and Yixuan Zhang. “MetaAgents: Large Language Model Based Agents for Decision-Making
on Teaming”. In: \emph{Proceedings of the ACM on Human-Computer Interaction} 9.2 (2025), pp. 1–27.
\textsc{doi}: \nolinkurl {10.1145/3711032}.
{}
\bibitem{tax121}\bibalias{ref95}

Zahra Abbasiantaeb, Yifei Yuan, Evangelos Kanoulas, Mohammad Aliannejadi. “Let the LLMs Talk: Simulating Human-to-Human Conversational QA via Zero-Shot
LLM-to-LLM Interactions”. In: \emph{Proceedings of the 17th ACM International Conference on Web Search and
Data Mining}. ACM, 2024, pp. 8–17. \textsc{doi}: \nolinkurl {10.1145/3616855.3635856}.
{}
\bibitem{tax004}\bibalias{ref11}

Yiting Chen, Tracy Xiao Liu, You Shan, Songfa Zhong. “The emergence of economic rationality of GPT”. In: \emph{Proceedings of the National
Academy of Sciences} 120.51 (2023), e2316205120. \textsc{doi}: \nolinkurl {10.1073/pnas.2316205120}.
{}
\bibitem{tax007}\bibalias{ref17}

Daniel E. O’Leary. “Confirmation and Specificity Biases in Large Language Models: An Explorative Study”. In:
\emph{IEEE Intelligent Systems} 40.1 (2025), pp. 63–68. \textsc{doi}: \nolinkurl {10.1109/mis.2024.3513992}.
{}
\bibitem{tax008}\bibalias{ref19}

Daniel E. O’Leary. \emph{An Anchoring Effect in Large Language Models}. Tech. rep. SSRN working paper,
2025. \textsc{doi}: \nolinkurl {10.2139/ssrn.5315021}.
{}
\bibitem{tax009}\bibalias{ref20}

Simon Jerome Han, Keith J. Ransom, Andrew Perfors, Charles Kemp. “Inductive reasoning in humans and large language models”. In: \emph{Cognitive
Systems Research} 83 (2024), p. 101155. \textsc{doi}: \nolinkurl {10.1016/j.cogsys.2023.101155}.
{}
\bibitem{tax010}\bibalias{ref21}

Olivia Macmillan-Scott and Mirco Musolesi. “(Ir)rationality and cognitive biases in large language models”. In:
\emph{Royal Society Open Science} 11.6 (2024), p. 240255. \textsc{doi}: \nolinkurl {10.1098/rsos.240255}.
{}
\bibitem{tax013}\bibalias{ref27}

Cyril Chhun, Fabian M. Suchanek, and Chloé Clavel. “Do Language Models Enjoy Their Own Stories? Prompting
Large Language Models for Automatic Story Evaluation”. In: \emph{Transactions of the Association for
Computational Linguistics} 12 (2024), pp. 1122–1142. \textsc{doi}: \nolinkurl {10.1162/tacl\_a\_00689}.
{}
\bibitem{tax014}\bibalias{ref30}

Tanisha Gupta, Akarsh Tripathi, Ashutosh Kumar Dubey, Ravita Chahar. “Enhancing the imitation game: a trust-based model for distinguishing human and machine
participants”. In: \emph{Applied Intelligence} 55.6 (2025), p. 388. \textsc{doi}: \nolinkurl
{10.1007/s10489-024-06133-2}.
{}
\bibitem{tax015}\bibalias{ref33}

Thilo Hagendorff, Sarah Fabi, and Michal Kosinski. “Human-like intuitive behavior and reasoning biases emerged
in large language models but disappeared in ChatGPT”. In: \emph{Nature Computational Science} 3.10 (2023),
pp. 833–838. \textsc{doi}: \nolinkurl {10.1038/s43588-023-00527-x}.
{}
\bibitem{tax016}\bibalias{ref35}

Itay Itzhak, Gabriel Stanovsky, Nir Rosenfeld, Yonatan Belinkov. “Instructed to Bias: Instruction-Tuned Language Models Exhibit Emergent Cognitive Bias”. In:
\emph{Transactions of the Association for Computational Linguistics} 12 (2024), pp. 771–785. \textsc{doi}:
\nolinkurl {10.1162/tacl\_a\_00673}.
{}
\bibitem{tax018}\bibalias{ref41}

Simon Jerome Han, Keith Ransom, Andrew Perfors, Charles Kemp. \emph{Human-like property induction is a challenge for large language models}. 2022.
\textsc{doi}: \nolinkurl {10.31234/osf.io/6mkjy}.
{}
\bibitem{tax026}\bibalias{ref51}

Nir Fulman, Abdulkadir Memduhoğlu, and Alexander Zipf. “Distortions in Judged Spatial Relations in Large
Language Models”. In: \emph{The Professional Geographer} 76.6 (2024), pp. 703–711. \textsc{doi}: \nolinkurl
{10.1080/00330124.2024.2372792}.
{}
\bibitem{tax027}\bibalias{ref53}

Henry J. Xie, Jinghan Zhang, Xinhao Zhang, Kunpeng Liu. “Scoring with Large Language Models: A Study on Measuring Empathy of Responses in
Dialogues”. In: \emph{2024 IEEE International Conference on Big Data (BigData)}. IEEE, 2024, pp. 7433–7437.
\textsc{doi}: \nolinkurl {10.1109/bigdata62323.2024.10825836}.
{}
\bibitem{tax028}\bibalias{ref64}

David Matthew Markowitz and Jeffrey Hancock. \emph{Generative AI Are More Truth-Biased Than Humans: A
Replication and Extension of Core Truth-Default Theory Principles}. 2023. \textsc{doi}: \nolinkurl
{10.31234/osf.io/hm54g}.
{}
\bibitem{tax030}\bibalias{ref67}

Jessica Maria Echterhoff, Yao Liu, Abeer Alessa, Julian McAuley, Zexue He. “Cognitive Bias in Decision-Making with LLMs”. In: \emph{Findings of the
Association for Computational Linguistics: EMNLP 2024}. Association for Computational Linguistics, 2024,
pp. 12640–12653. \textsc{doi}: \nolinkurl {10.18653/v1/2024.findings-emnlp.739}.
{}
\bibitem{tax031}\bibalias{ref83}

Xinyi Mou, Jingcong Liang, Jiayu Lin, Xinnong Zhang, Xiawei Liu, Shiyue Yang, Rong Ye, Lei Chen, Haoyu Kuang, Xuanjing Huang, Zhongyu Wei. “AgentSense: Benchmarking Social Intelligence of Language Agents through Interactive
Scenarios”. In: \emph{Proceedings of the 2025 Conference of the Nations of the Americas Chapter of the
Association for Computational Linguistics: Human Language Technologies (Volume 1: Long Papers)}. Association
for Computational Linguistics, 2025, pp. 4975–5001. \textsc{doi}: \nolinkurl {10.18653/v1/2025.naacl-long.257}.
{}
\bibitem{tax032}\bibalias{ref88}

Tadahiro Taniguchi, Ryo Ueda, Tomoaki Nakamura, Masahiro Suzuki, Akira Taniguchi. “Generative Emergent Communication: Large Language Model is a Collective World
Model”. In: \emph{Advanced Robotics} 40.9 (2026), pp. 499–524. \textsc{doi}: \nolinkurl
{10.1080/01691864.2026.2661958}.
{}
\bibitem{tax034}\bibalias{ref123}

Daniel Martin Katz, Michael James Bommarito, Shang Gao, Pablo Arredondo. \emph{GPT-4 passes the bar exam}. 2023. \textsc{doi}: \nolinkurl
{10.2139/ssrn.4389233}.
{}
\bibitem{tax039}\bibalias{ref171}

Ryan Liu, Jiayi Geng, Joshua C. Peterson, Ilia Sucholutsky, Thomas L. Griffiths. “Large Language Models Assume People are More Rational than We Really are”. In: \emph{The
Thirteenth International Conference on Learning Representations}. 2025. \textsc{url}: \url
{https://openreview.net/forum?id=dAeET8gxqg}.
{}
\bibitem{tax040}\bibalias{al_lily2023chatgpt}
\bibalias{ref188}

Abdulrahman Essa Al Lily, Abdelrahim Fathy Ismail, Fathi M. Abunaser, Firass Al-Lami, Ali Khalifa Atwa Abdullatif. “ChatGPT and the rise of semi-humans”. In: \emph{Humanities and Social
Sciences Communications} 10.1 (2023), p. 626. \textsc{doi}: \nolinkurl {10.1057/s41599-023-02154-3}.
{}
\bibitem{tax041}\bibalias{ref197}

John J. Horton, Apostolos Filippas, and Benjamin Manning. \emph{Large Language Models as Simulated
Economic Agents: What Can We Learn from Homo Silicus?} 2023. \textsc{doi}: \nolinkurl
{10.2139/ssrn.4413859}.
{}
\bibitem{tax046}\bibalias{ref12}

Aadesh Salecha, Molly E Ireland, Shashanka Subrahmanya, João Sedoc, Lyle H Ungar, Johannes C Eichstaedt. “Large language models display human-like social desirability biases in Big Five personality
surveys”. In: \emph{PNAS Nexus} 3.12 (2024), pgae533. \textsc{doi}: \nolinkurl {10.1093/pnasnexus/pgae533}.
{}
\bibitem{tax048}\bibalias{ref16}

Ojasvi Gupta, Stefano Marrone, Francesco Gargiulo, Rajesh Jaiswal, Lidia Marassi. “Understanding Social Biases in Large Language Models”. In: \emph{AI} 6.5 (2025), p. 106.
\textsc{doi}: \nolinkurl {10.3390/ai6050106}.
{}
\bibitem{tax049}\bibalias{ref18}

Yan Tao, Olga Viberg, Ryan S Baker, René F Kizilcec. “Cultural bias and cultural alignment of large language models”. In: \emph{PNAS Nexus} 3.9
(2024), pgae346. \textsc{doi}: \nolinkurl {10.1093/pnasnexus/pgae346}.
{}
\bibitem{tax052}\bibalias{ref29}

Bojana Bodroža, Bojana M. Dinić, and Ljubiša Bojić. “Personality testing of large language models: limited
temporal stability, but highlighted prosociality”. In: \emph{Royal Society Open Science} 11.10 (2024), p. 240180.
\textsc{doi}: \nolinkurl {10.1098/rsos.240180}.
{}
\bibitem{tax056}\bibalias{ref40}

Guangyuan Jiang, Manjie Xu, Song-Chun Zhu, Wenjuan Han, Chi Zhang, Yixin Zhu. “Evaluating and inducing personality in pre-trained language models”. In:
\emph{Advances in Neural Information Processing Systems 36}. Neural Information Processing Systems
Foundation, Inc. (NeurIPS), 2023, pp. 10622–10643. \textsc{doi}: \nolinkurl {10.52202/075280-0466}.
{}
\bibitem{tax058}\bibalias{ref48}

Lorenz Sparrenberg, Tobias Schneider, Tobias Deußer, Markus Koppenborg, Rafet Sifa. “Correcting Systematic Bias in LLM-Generated Dialogues Using Big Five Personality
Traits”. In: \emph{2024 IEEE International Conference on Big Data (BigData)}. IEEE, 2024, pp. 3061–3069.
\textsc{doi}: \nolinkurl {10.1109/bigdata62323.2024.10825941}.
{}
\bibitem{tax059}\bibalias{ref52}

Wolfgang Messner, Tatum Greene, and Josephine Matalone. “From Bytes to Biases: Investigating the Cultural
Self-Perception of Large Language Models”. In: \emph{Journal of Public Policy \& Marketing} 44.3 (2025),
pp. 370–391. \textsc{doi}: \nolinkurl {10.1177/07439156251319788}.
{}
\bibitem{tax063}\bibalias{ref59}

Haijiang Liu, Yong Cao, Xun Wu, Chen Qiu, Jinguang Gu, Maofu Liu, Daniel Hershcovich. “Towards realistic evaluation of cultural value alignment in large language models: Diversity
enhancement for survey response simulation”. In: \emph{Information Processing \& Management} 62.4
(2025), p. 104099. \textsc{doi}: \nolinkurl {10.1016/j.ipm.2025.104099}.
{}
\bibitem{tax071}\bibalias{johnson2025testing}
\bibalias{ref6}

Tim Johnson and Nick Obradovich. “Testing for completions that simulate altruism in early language models”. In:
\emph{Nature Human Behaviour} 9.9 (2025), pp. 1861–1870. \textsc{doi}: \nolinkurl
{10.1038/s41562-025-02258-7}.
{}
\bibitem{tax073}\bibalias{ref36}

Patrick Schramowski, Cigdem Turan, Nico Andersen, Constantin A. Rothkopf, Kristian Kersting. “Large pre-trained language models contain human-like biases of what is right and
wrong to do”. In: \emph{Nature Machine Intelligence} 4.3 (2022), pp. 258–268. \textsc{doi}: \nolinkurl
{10.1038/s42256-022-00458-8}.
{}
\bibitem{tax081}\bibalias{ref126}

Di Zhou and Yinxian Zhang. “Political biases and inconsistencies in bilingual GPT models—the cases of the U.S.
and China”. In: \emph{Scientific Reports} 14.1 (2024), p. 25048. \textsc{doi}: \nolinkurl
{10.1038/s41598-024-76395-w}.
{}
\bibitem{tax083}\bibalias{ref71}

Meta Fundamental AI Research Diplomacy Team (FAIR). “Human-level play in the game of Diplomacy by
combining language models with strategic reasoning”. In: \emph{Science} 378.6624 (2022), pp. 1067–1074.
\textsc{doi}: \nolinkurl {10.1126/science.ade9097}.
{}
\bibitem{tax084}\bibalias{ref72}

Bahar Bateni and Jim Whitehead. “Language-Driven Play: Large Language Models as Game-Playing Agents in
Slay the Spire”. In: \emph{Proceedings of the 19th International Conference on the Foundations of Digital
Games}. ACM, 2024, pp. 1–10. \textsc{doi}: \nolinkurl {10.1145/3649921.3650013}.
{}
\bibitem{tax085}\bibalias{ref73}

Xiachong Feng, Longxu Dou, Ella Li, Qinghao Wang, Haochuan Wang, Yu Guo, Chang Ma, Lingpeng Kong. “A Survey on Large Language Model-Based Social Agents in Game-Theoretic Scenarios”. In:
\emph{Transactions on Machine Learning Research} (2025). \textsc{url}: \url
{https://openreview.net/forum?id=CsoSWpR5xC}.
{}
\bibitem{tax088}\bibalias{ref87}

Xuan Liu, Jie Zhang, Haoyang Shang, Song Guo, Chengxu Yang, Quanyan Zhu. “Exploring Prosocial Irrationality for LLM Agents: A Social Cognition View”. In: \emph{The
Thirteenth International Conference on Learning Representations}. 2025. \textsc{url}: \url
{https://openreview.net/forum?id=u8VOQVzduP}.
{}
\bibitem{tax089}\bibalias{ref89}
\bibalias{sreedhar2024simulating}

Karthik Sreedhar and Lydia Chilton. \emph{Simulating Human Strategic Behavior: Comparing Single and
Multi-agent LLMs}. arXiv. 2024. \textsc{doi}: \nolinkurl {10.48550/arXiv.2402.08189}.
{}
\bibitem{tax090}\bibalias{ref97}

Ning Bian, Hongyu Lin, Peilin Liu, Yaojie Lu, Chunkang Zhang, Ben He, Xianpei Han, Le Sun. “Influence of External Information on Large Language Models Mirrors Social Cognitive Patterns”.
In: \emph{IEEE Transactions on Computational Social Systems} 12.3 (2025), pp. 1115–1131. \textsc{doi}:
\nolinkurl {10.1109/tcss.2024.3476030}.
{}
\bibitem{tax091}\bibalias{ref109}

Tim Johnson and Nick Obradovich. \emph{Evidence of behavior consistent with self-interest and Altruism in an
artificially intelligent agent}. Tech. rep. SSRN working paper, 2023. \textsc{doi}: \nolinkurl
{10.2139/ssrn.4319609}.
{}
\bibitem{tax092}\bibalias{park2024ai}
\bibalias{ref138}

Peter S. Park, Simon Goldstein, Aidan O’Gara, Michael Chen, Dan Hendrycks. “AI deception: A survey of examples, risks, and potential solutions”. In: \emph{Patterns} 5.5
(2024), p. 100988. \textsc{doi}: \nolinkurl {10.1016/j.patter.2024.100988}.
{}
\bibitem{tax093}\bibalias{ref170}

Sonali Singh, Faranak Abri, and Akbar Siami Namin. “Exploiting Large Language Models (LLMs) through
Deception Techniques and Persuasion Principles”. In: \emph{2023 IEEE International Conference on Big Data
(BigData)}. IEEE, 2023, pp. 2508–2517. \textsc{doi}: \nolinkurl {10.1109/bigdata59044.2023.10386814}.
{}
\bibitem{tax095}\bibalias{ref70}
\bibalias{zhao2023competeai}

Qinlin Zhao, Jindong Wang, Yixuan Zhang, Yiqiao Jin, Kaijie Zhu, Hao Chen, and Xing Xie. “CompeteAI: Understanding the Competition Dynamics of Large Language Model-based
Agents”. In: \emph{Proceedings of the 41st International Conference on Machine Learning}. 2024,
pp. 61092–61107. \textsc{url}: \url {https://proceedings.mlr.press/v235/zhao24q.html}.
{}
\bibitem{tax096}\bibalias{ref74}

Reiji Suzuki and Takaya Arita. “An evolutionary model of personality traits related to cooperative behavior using
a large language model”. In: \emph{Scientific Reports} 14.1 (2024), p. 5989. \textsc{doi}: \nolinkurl
{10.1038/s41598-024-55903-y}.
{}
\bibitem{tax097}\bibalias{ref91}

I. de Zarzà, J. de Curtò, Gemma Roig, Pietro Manzoni, Carlos T. Calafate. “Emergent Cooperation and Strategy Adaptation in Multi-Agent Systems: An Extended
Coevolutionary Theory with LLMs”. In: \emph{Electronics} 12.12 (2023), p. 2722. \textsc{doi}: \nolinkurl
{10.3390/electronics12122722}.
{}
\bibitem{tax099}\bibalias{ref75}

Cristian Jimenez-Romero, Alper Yegenoglu, and Christian Blum. “Multi-agent systems powered by large language
models: applications in swarm intelligence”. In: \emph{Frontiers in Artificial Intelligence} 8 (2025), p. 1593017.
\textsc{doi}: \nolinkurl {10.3389/frai.2025.1593017}.
{}
\bibitem{tax103}\bibalias{de2025llm}
\bibalias{ref90}

Joaquim de Curto i Diaz and Irene de Zarza i Cubero. “LLM-Driven Social Influence for Cooperative Behavior in
Multi-Agent Systems”. In: \emph{IEEE Access} (2025). \textsc{doi}: \nolinkurl
{10.1109/ACCESS.2025.3548451}.
{}
\bibitem{tax105}\bibalias{ref198}
\bibalias{schoenegger2024wisdom}

Philipp Schoenegger, Indre Tuminauskaite, Peter S. Park, Rafael Valdece Sousa Bastos, Philip E. Tetlock. “Wisdom of the silicon crowd: LLM ensemble prediction capabilities rival human crowd
accuracy”. In: \emph{Science Advances} 10.45 (2024), eadp1528. \textsc{doi}: \nolinkurl
{10.1126/sciadv.adp1528}.
{}
\bibitem{tax111}\bibalias{ref98}

Huao Li, Yu Chong, Simon Stepputtis, Joseph Campbell, Dana Hughes, Charles Lewis, Katia Sycara. “Theory of Mind for Multi-Agent Collaboration via Large Language Models”. In:
\emph{Proceedings of the 2023 Conference on Empirical Methods in Natural Language Processing}. 2023,
pp. 180–192. \textsc{doi}: \nolinkurl {10.18653/v1/2023.emnlp-main.13}.
{}
\bibitem{tax112}\bibalias{ref99}

Chuanneng Sun, Songjun Huang, and Dario Pompili. “LLM-Based Multi-Agent Decision-Making: Challenges and
Future Directions”. In: \emph{IEEE Robotics and Automation Letters} 10.6 (2025), pp. 5681–5688. \textsc{doi}:
\nolinkurl {10.1109/lra.2025.3562371}.
{}
\bibitem{tax115}\bibalias{borghoff2025organizational}
\bibalias{ref128}

Uwe M. Borghoff, Paolo Bottoni, and Remo Pareschi. “An organizational theory for multi-agent interactions
integrating human agents, LLMs, and specialized AI”. In: \emph{Discover Computing} 28.1 (2025), p. 138.
\textsc{doi}: \nolinkurl {10.1007/s10791-025-09667-2}.
{}
\bibitem{tax118}\bibalias{ref77}

Taicheng Guo, Xiuying Chen, Yaqi Wang, Ruidi Chang, Shichao Pei, Nitesh V. Chawla, Olaf Wiest, Xiangliang Zhang. “Large Language Model based Multi-Agents: A Survey of Progress and Challenges”. In:
\emph{Proceedings of the Thirty-Third International Joint Conference on Artificial Intelligence}. 2024,
pp. 8048–8057. \textsc{doi}: \nolinkurl {10.24963/ijcai.2024/890}.
{}
\bibitem{tax119}\bibalias{ref93}

Antonino Ferraro, Antonio Galli, Valerio La Gatta, Marco Postiglione, Gian Marco Orlando, Diego Russo, Giuseppe Riccio, Antonio Romano, Vincenzo Moscato. \emph{Agent-Based Modelling Meets Generative AI in Social Network Simulations}.
Lecture Notes in Computer Science. 2025. \textsc{doi}: \nolinkurl {10.1007/978-3-031-78541-2\_10}.
{}
\bibitem{tax120}\bibalias{ref94}

Gian Marco Orlando, Valerio La Gatta, Diego Russo, Vincenzo Moscato. “Can Generative Agent-Based Modeling Replicate the Friendship Paradox in Social
Media Simulations?” In: \emph{Proceedings of the 17th ACM Web Science Conference 2025}. ACM, 2025,
pp. 510–515. \textsc{doi}: \nolinkurl {10.1145/3717867.3717895}.
{}
\bibitem{tax125}\bibalias{ref193}

Xinyi Mou, Xuanwen Ding, Qi He, Liang Wang, Jingcong Liang, Xinnong Zhang, Libo Sun, Jiayu Lin, Jie Zhou, Xuanjing Huang, Zhongyu Wei. \emph{From Individual to Society: A Survey on Social Simulation Driven by Large Language
Model-based Agents}. arXiv. 2024. \textsc{doi}: \nolinkurl {10.48550/arXiv.2412.03563}.
{}
\bibitem{tax128}\bibalias{ref196}

Chen Gao, Xiaochong Lan, Nian Li, Yuan Yuan, Jingtao Ding, Zhilun Zhou, Fengli Xu, Yong Li. “Large language models empowered agent-based modeling and simulation: a survey and
perspectives”. In: \emph{Humanities and Social Sciences Communications} 11.1 (2024), p. 1259. \textsc{doi}:
\nolinkurl {10.1057/s41599-024-03611-3}.
{}
\bibitem{tax130}\bibalias{ref34}

Terrence J. Sejnowski. “Large Language Models and the Reverse Turing Test”. In: \emph{Neural Computation}
35.3 (2023), pp. 309–342. \textsc{doi}: \nolinkurl {10.1162/neco\_a\_01563}.
{}
\bibitem{tax132}\bibalias{ref120}

Fabian Dvorak, Regina Stumpf, Sebastian Fehrler, Urs Fischbacher. “Adverse reactions to the use of large language models in social interactions”. In:
\emph{PNAS Nexus} 4.4 (2025), pgaf112. \textsc{doi}: \nolinkurl {10.1093/pnasnexus/pgaf112}.
{}
\bibitem{tax135}\bibalias{ref133}

Mark Steyvers, Heliodoro Tejeda, Aakriti Kumar, Catarina Belem, Sheer Karny, Xinyue Hu, Lukas W. Mayer, Padhraic Smyth. “What large language models know and what people think they know”. In: \emph{Nature
Machine Intelligence} 7.2 (2025), pp. 221–231. \textsc{doi}: \nolinkurl {10.1038/s42256-024-00976-7}.
{}
\bibitem{tax137}\bibalias{ref136}

Jorge Luis Morton. “From meaning to emotions: LLMs as artificial communication partners”. In: \emph{AI
\& SOCIETY} 41.1 (2026), pp. 171–184. \textsc{doi}: \nolinkurl {10.1007/s00146-025-02481-w}.
{}
\bibitem{tax138}\bibalias{ref147}

Vu Ngo. “Humanizing AI for trust: the critical role of social presence in adoption”. In: \emph{AI \&
SOCIETY} 41.3 (2026), pp. 1803–1819. \textsc{doi}: \nolinkurl {10.1007/s00146-025-02506-4}.
{}
\bibitem{tax139}\bibalias{ref150}

Callie Y. Kim, Christine P. Lee, and Bilge Mutlu. “Understanding Large-Language Model
(LLM)-poweredHuman-Robot Interaction”. In: \emph{Proceedings of the 2024 ACM/IEEE International
Conference on Human-Robot Interaction}. ACM, 2024, pp. 371–380. \textsc{doi}: \nolinkurl
{10.1145/3610977.3634966}.
{}
\bibitem{tax140}\bibalias{ref156}

Hari Subramonyam, Roy Pea, Christopher Pondoc, Maneesh Agrawala, Colleen Seifert. “Bridging the Gulf of Envisioning: Cognitive Challenges in Prompt Based Interactions
with LLMs”. In: \emph{Proceedings of the CHI Conference on Human Factors in Computing Systems}. ACM,
2024, pp. 1–19. \textsc{doi}: \nolinkurl {10.1145/3613904.3642754}.
{}
\bibitem{tax143}\bibalias{ref179}

Eva Paraschou, Maria Michali, Sofia Yfantidou, Stelios Karamanidis, Stefanos Rafail Kalogeros, Athena Vakali. “Ties of Trust: a bowtie model to uncover trustor-trustee relationships in LLMs”. In:
\emph{Proceedings of the 2025 ACM Conference on Fairness, Accountability, and Transparency}. ACM, 2025,
pp. 1715–1728. \textsc{doi}: \nolinkurl {10.1145/3715275.3732115}.
{}
\bibitem{tax144}\bibalias{ref180}

Oliver L. Jacobs, Farid Pazhoohi, and Alan Kingstone. “Large language models have divergent effects on
self-perceptions of mind and the attributes considered uniquely human”. In: \emph{Consciousness and Cognition}
124 (2024), p. 103733. \textsc{doi}: \nolinkurl {10.1016/j.concog.2024.103733}.
{}
\bibitem{tax145}\bibalias{ref111}

Xianzhe Fan, Qing Xiao, Xuhui Zhou, Jiaxin Pei, Maarten Sap, Zhicong Lu, Hong Shen. “User-Driven Value Alignment: Understanding Users’ Perceptions and Strategies for
Addressing Biased and Discriminatory Statements in AI Companions”. In: \emph{Proceedings of the 2025 CHI
Conference on Human Factors in Computing Systems}. ACM, 2025, pp. 1–19. \textsc{doi}: \nolinkurl
{10.1145/3706598.3713477}.
{}
\bibitem{tax146}\bibalias{ref114}

Qian Chen, Yufan Jing, Yeming Gong, Jie Tan. “Will users fall in love with ChatGPT? a perspective from the triangular theory of love”. In:
\emph{Journal of Business Research} 186 (2025), p. 114982. \textsc{doi}: \nolinkurl
{10.1016/j.jbusres.2024.114982}.
{}
\bibitem{tax147}\bibalias{ref115}

Bufang Yang, Yunqi Guo, Lilin Xu, Zhenyu Yan, Hongkai Chen, Guoliang Xing, Xiaofan Jiang. “SocialMind: LLM-based Proactive AR Social Assistive System with Human-like Perception
for In-situ Live Interactions”. In: \emph{Proceedings of the ACM on Interactive, Mobile, Wearable and
Ubiquitous Technologies} 9.1 (2025), pp. 1–30. \textsc{doi}: \nolinkurl {10.1145/3712286}.
{}
\bibitem{tax150}\bibalias{ref158}

Abeer Alessa and Hend Al-Khalifa. “Towards Designing a ChatGPT Conversational Companion for Elderly
People”. In: \emph{Proceedings of the 16th International Conference on PErvasive Technologies Related to
Assistive Environments}. ACM, 2023, pp. 667–674. \textsc{doi}: \nolinkurl {10.1145/3594806.3596572}.
{}
\bibitem{tax151}\bibalias{ref172}

Qihang He, Jiyao Wang, and Dengbo He. “The Influence of Task and Group Disparities Over Users’ Attitudes
Toward Using Large Language Models for Psychotherapy”. In: \emph{Proceedings of the Human Factors and
Ergonomics Society Annual Meeting} 68.1 (2024), pp. 1147–1152. \textsc{doi}: \nolinkurl
{10.1177/10711813241268507}.
{}
\bibitem{tax154}\bibalias{ref26}

Kobi Hackenburg, Lujain Ibrahim, Ben M. Tappin, Manos Tsakiris. “Comparing the persuasiveness of role-playing large language models and human experts
on polarized U.S. political issues”. In: \emph{AI \& SOCIETY} 41.1 (2026), pp. 351–361. \textsc{doi}:
\nolinkurl {10.1007/s00146-025-02464-x}.
{}
\bibitem{tax155}\bibalias{ref105}

Yuyang Cheng, Xingwei Qu, Tomas Goldsack, Chenghua Lin, Chung-Chi Chen. “Observing Micromotives and Macrobehavior of Large Language Models”. In:
\emph{Proceedings of the 14th International Joint Conference on Natural Language Processing and the 4th
Conference of the Asia-Pacific Chapter of the Association for Computational Linguistics}. The Asian Federation
of Natural Language Processing and The Association for Computational Linguistics, 2025, pp. 1258–1276.
\textsc{doi}: \nolinkurl {10.18653/v1/2025.ijcnlp-long.69}.
{}
\bibitem{tax156}\bibalias{ref106}

James Flamino, Mohammed Shahid Modi, Boleslaw K. Szymanski, Brendan Cross, Colton Mikolajczyk. “Testing the limits of large language models in debating humans”. In: \emph{Scientific
Reports} 15.1 (2025), p. 13852. \textsc{doi}: \nolinkurl {10.1038/s41598-025-98378-1}.
{}
\bibitem{tax157}\bibalias{ref110}

Kevin Durrheim and Michael Quayle. “Human murmuration: Group polarisation as compression in
interaction-language dynamics captured by large language models”. In: \emph{European Review of Social
Psychology} 36.2 (2025), pp. 326–365. \textsc{doi}: \nolinkurl {10.1080/10463283.2025.2499332}.
{}
\bibitem{tax162}\bibalias{ref139}
\bibalias{schmitt2024digital}

Marc Schmitt and Ivan Flechais. “Digital deception: generative artificial intelligence in social engineering and
phishing”. In: \emph{Artificial Intelligence Review} 57.12 (2024), p. 324. \textsc{doi}: \nolinkurl
{10.1007/s10462-024-10973-2}.
{}
\bibitem{tax163}\bibalias{ref141}

Yubo Shu, Haonan Zhang, Hansu Gu, Peng Zhang, Tun Lu, Dongsheng Li, Ning Gu. “RAH! RecSys-Assistant-Human: A Human-Centered Recommendation Framework With LLM
Agents”. In: \emph{IEEE Transactions on Computational Social Systems} 11.5 (2024), pp. 6759–6770.
\textsc{doi}: \nolinkurl {10.1109/tcss.2024.3404039}.
{}
\bibitem{tax165}\bibalias{ref143}

Hui Bai, Jan G. Voelkel, Shane Muldowney, johannes Christopher Eichstaedt, Robb Willer. \emph{LLM-generated messages can persuade humans on policy issues}. 2025. \textsc{doi}:
\nolinkurl {10.31219/osf.io/stakv\_v8}.
{}
\bibitem{tax166}\bibalias{ref144}

Jessy Lin, Nicholas Tomlin, Jacob Andreas, Jason Eisner. “Decision-Oriented Dialogue for Human-AI Collaboration”. In: \emph{Transactions of the
Association for Computational Linguistics} 12 (2024), pp. 892–911. \textsc{doi}: \nolinkurl
{10.1162/tacl\_a\_00679}.
{}
\bibitem{tax168}\bibalias{ref152}

Yuanning Han, Ziyi Qiu, Jiale Cheng, RAY LC. “When Teams Embrace AI: Human Collaboration Strategies in Generative Prompting in a
Creative Design Task”. In: \emph{Proceedings of the CHI Conference on Human Factors in Computing Systems}.
2024, 176:1–176:14. \textsc{doi}: \nolinkurl {10.1145/3613904.3642133}.
{}
\bibitem{tax169}\bibalias{ref153}

Maria Teresa Baldassarre, Danilo Caivano, Berenice Fernandez Nieto, Domenico Gigante, Azzurra Ragone. “The Social Impact of Generative AI: An Analysis on ChatGPT”. In:
\emph{Proceedings of the 2023 ACM Conference on Information Technology for Social Good}. ACM, 2023,
pp. 363–373. \textsc{doi}: \nolinkurl {10.1145/3582515.3609555}.
{}
\bibitem{tax171}\bibalias{ref165}

Paul Thomas, Seth Spielman, Nick Craswell, Bhaskar Mitra. “Large Language Models can Accurately Predict Searcher Preferences”. In:
\emph{Proceedings of the 47th International ACM SIGIR Conference on Research and Development in
Information Retrieval}. ACM, 2024, pp. 1930–1940. \textsc{doi}: \nolinkurl {10.1145/3626772.3657707}.
{}
\bibitem{tax172}\bibalias{ref169}

Sebastian Scholz, Alexander Lawall, and Kristina Schaaff. “The Impact of Large Language Models on IT Security
in the Corporate Environment”. In: \emph{2024 2nd International Conference on Foundation and Large Language
Models}. 2024, pp. 109–115. \textsc{doi}: \nolinkurl {10.1109/FLLM63129.2024.10852476}.
{}
\bibitem{tax173}\bibalias{ref176}

John Zhuang Liu and Xueyao Li. “How do judges use large language models? Evidence from Shenzhen”. In:
\emph{Journal of Legal Analysis} 16.1 (2024), pp. 235–262. \textsc{doi}: \nolinkurl {10.1093/jla/laae009}.
{}
\bibitem{tax174}\bibalias{ref177}

Simone Zhang, Janet Xu, and AJ Alvero. “Generative AI Meets Open-Ended Survey Responses: Research
Participant Use of AI and Homogenization”. In: \emph{Sociological Methods \& Research} 54.3 (2025),
pp. 1197–1242. \textsc{doi}: \nolinkurl {10.1177/00491241251327130}.
{}
\bibitem{tax175}\bibalias{ref178}

Marianna Ciullo. “Large Language Models: Ethics and Norms in the European Union”. In: \emph{2024 IEEE
International Conference on Metrology for eXtended Reality, Artificial Intelligence and Neural Engineering
(MetroXRAINE)}. IEEE, 2024, pp. 1065–1070. \textsc{doi}: \nolinkurl
{10.1109/metroxraine62247.2024.10795986}.
{}
\bibitem{tax181}\bibalias{ref155}

Jeongyeon Kim, Sangho Suh, Lydia B Chilton, Haijun Xia. “Metaphorian: Leveraging Large Language Models to Support Extended Metaphor Creation
for Science Writing”. In: \emph{Proceedings of the 2023 ACM Designing Interactive Systems Conference}. 2023,
pp. 115–135. \textsc{doi}: \nolinkurl {10.1145/3563657.3595996}.
{}
\bibitem{tax186}\bibalias{ref117}

Da Teng, Xiangyang Wang, Yanwei Xia, Yue Zhang, Lulu Tang, Qi Chen, Ruobing Zhang, Sujin Xie, Weiyong Yu. “Investigating the utilization and impact of large language model-based intelligent teaching
assistants in flipped classrooms”. In: \emph{Education and Information Technologies} 30.8 (2025),
pp. 10777–10810. \textsc{doi}: \nolinkurl {10.1007/s10639-024-13264-z}.
{}
\bibitem{tax187}\bibalias{ref129}

Michael Cowling, Joseph Crawford, Kelly-Ann Allen, Michael Wehmeyer. “Using leadership to leverage ChatGPT and artificial intelligence for undergraduate and
postgraduate research supervision”. In: \emph{Australasian Journal of Educational Technology} 39.4 (2023),
pp. 89–103. \textsc{doi}: \nolinkurl {10.14742/ajet.8598}.
{}
\bibitem{tax197}\bibalias{ref186}

Ahmed Tlili, Boulus Shehata, Michael Agyemang Adarkwah, Aras Bozkurt, Daniel T. Hickey, Ronghuai Huang, Brighter Agyemang. “What if the devil is my guardian angel: ChatGPT as a case study of using chatbots in
education”. In: \emph{Smart Learning Environments} 10.1 (2023), p. 15. \textsc{doi}: \nolinkurl
{10.1186/s40561-023-00237-x}.
{}
\end{thebibliography}
\end{document}